\documentclass[a4paper,11pt]{article}
\usepackage{jcappub} 
\usepackage{lineno}

\newcommand{\be}{\begin{equation}}
\newcommand{\ee}{\end{equation}}
\newcommand{\bea}{\begin{eqnarray}}
\newcommand{\eea}{\end{eqnarray}}
\newcommand{\myparagraph}[1]{\paragraph{#1}\mbox{} \vspace{0.2cm}\\ }

\newcommand{\mc}{\mathcal}
\usepackage{amsmath,amsthm,amssymb}
\usepackage{bbold}
\usepackage{mathtools}
\usepackage{slashed}
\usepackage{multicol}
\usepackage{blkarray}
\usepackage{enumerate}
\usepackage{graphicx}
\usepackage{pdfpages}
\usepackage{comment}
\usepackage{booktabs}
\usepackage{nccmath} 
\usepackage{etoolbox}
\usepackage{hyperref} 
\usepackage{array}
\usepackage{url} 
\usepackage{array}
\usepackage{epsfig,multicol,bbm}
\usepackage{colortbl}
\usepackage{color}
\usepackage{tcolorbox}
\usepackage{xcolor}
\usepackage{empheq}

\usepackage{placeins}
\usepackage{subfiles}
\usepackage{graphicx}

\definecolor{green2}{cmyk}{0, 1, 0.5, 0}
\definecolor{lightgreen}{cmyk}{0.2, 0, 0.2, 0.2}
\definecolor{lightgray}{cmyk}{0.1,0.2,0,0.1}
\definecolor{lightgray2}{cmyk}{0.4,0.4,0,0.8}
\definecolor{black}{cmyk}{1.0,1.0,1.0,1.0}
\graphicspath{{figures/}}

\usepackage{caption}
\usepackage{subcaption}

\newenvironment{aleq}
    {\begin{equation}\begin{aligned}}
    {\end{aligned}\end{equation}\ignorespacesafterend}
\usepackage{xcolor}
\title{\boldmath String Theory and the First Half of the Universe}

\author{Fien Apers,$^{a}$ Joseph P. Conlon,$^{a}$ Edmund J. Copeland,$^{b}$  Martin Mosny$^{a}$ and Filippo Revello$^{c}$}
\affiliation[a]{Rudolf Peierls Centre for Theoretical Physics,
Beecroft Building,
Clarendon Laboratory, Parks Road, University of Oxford, OX1 3PU, UK}
\affiliation[b]{School of Physics and Astronomy, University of Nottingham, Nottingham NG7 2RD, United Kingdom}
\affiliation[c]{Institute for Theoretical Physics
Utrecht University, Princetonplein 5, 3584 CC Utrecht, The Netherlands}

\emailAdd{fien.apers@physics.ox.ac.uk}
\emailAdd{ed.copeland@nottingham.ac.uk}
\emailAdd{joseph.conlon@physics.ox.ac.uk}
\emailAdd{martin.mosny@physics.ox.ac.uk}
\emailAdd{f.revello@uu.nl}

\abstract{We perform a detailed study of stringy moduli-driven cosmologies between the end of inflation and the commencement of the Hot Big Bang, including both the background and cosmological perturbations: a period that can cover half the lifetime of the universe on a logarithmic scale. Compared to the standard cosmology, stringy cosmologies motivate extended kination, tracker and moduli-dominated epochs involving significantly trans-Planckian field excursions. Conventional effective field theory is unable to control Planck-suppressed operators and so such epochs require a stringy completion for a consistent analysis. Perturbation growth in these stringy cosmologies is substantially enhanced compared to conventional cosmological histories.  The transPlanckian field evolution results in radical changes to Standard Model couplings during this history and we outline potential applications to baryogenesis, dark matter and gravitational wave production.}

\begin{document}
\maketitle
\flushbottom

\section{Introduction}

The development of the $\Lambda$CDM cosmology \cite{Planck:2018vyg} is one of the great successes of fundamental science over the last two generations (textbook accounts include \cite{Dodelson:2003ft, Mukhanov:2005sc, Weinberg:2008zzc, Baumann:2022mni}). In this
Standard Cosmology, the assumed history of the universe involves the following stages:
\begin{enumerate}
\item An early stage of inflation, terminating in reheating and a transfer of energy from the inflaton into thermalised Standard Model degrees of freedom.
\item The long period of the Hot Big Bang, with radiation domination lasting from the immediate post-inflationary epoch until matter-radiation equality.
\item A period of matter domination, in which inhomogeneities grow rapidly resulting in structure formation.
\item The modern epoch of Dark Energy domination.
\end{enumerate}
Primordial density perturbations are sourced during inflation, but do not (on any scale) grow significantly until the era of matter domination.

Although the CMB gives a strong probe of physics around 50 - 60 efolds before the end of inflation, much of the early history of the Standard Cosmology is relatively unconstrained. In particular, little constrains the equation of state of the universe between the end of inflation and the time of nucleosynthesis (a recent review is \cite{Allahverdi:2020bys}).

Although the biggest single shortcoming of string theory is surely its lack of direct connection with observation and/or experiment, the early universe offers one of the best chances to rectify this situation. In particular, the long, relatively unconstrained, history of the universe between inflation and Big Bang Nucleosynthesis, which can last as long as $\frac{t_{final}}{t_{init}} \sim 10^{30}$, offers a chance for stringy physics -- especially that associated to the moduli which are the ubiquitous low-energy footprint of compactifications -- to dominate the universe over an extended period, and perhaps, just perhaps, generate distinctive traces that can subsequently be observed.

\textit{String} cosmology (see \cite{Cicoli:2023opf} for a review) is especially relevant -- indeed, required -- when fields evolve through transPlanckian distances in field space.
This is the case in many scenarios in string models, where the 
 final vacuum in field space may be many transPlanckian distances away from the relevant locus during inflation. Such a scenario is particularly motivated in the Large Volume Scenario \cite{Balasubramanian:2005zx, Conlon:2005ki}. Here the final stabilised volume must be exponentially large in order to generate the required hierarchies of particle physics, even though inflation may have occurred at much smaller values of the volume. 
 
 However, such stringy UV completions are also required for control of several epochs considered in pure cosmological model-building.
 For example, this is true of both \emph{kination} (this name was first used in \cite{Joyce:1996cp}; see \cite{Gouttenoire:2021jhk} for a recent review) and \emph{tracker} epochs \cite{Wetterich:1987fm, Ferreira:1997hj, Copeland:1997et}. In the former, the kinetic energy of a scalar field dominates the overall energy density; in the latter, it makes up an $\mathcal{O}(1)$ fraction of the overall energy density. In both cases, this field evolves through a Planckian distance in field space approximately once per Hubble time. As any transPlanckian field excursions are associated with a breakdown of low-energy effective field theory analyses, arising from modification of any operator by terms of the form $\left( \frac{\Delta \phi}{M_P} \right)^n$, extended kination or tracker epochs can only be controlled by embedding them into a UV-complete (string theory) framework.

 A further advantage of UV embedding is that any concrete string compactification will imply correlations between very different areas of physics which could never be justified simply through the tools of, say, cosmology and quantum field theory. An easy example is provided by the volume modulus. In string compactifications, the vev of the volume modulus may simultaneously determine the scale of the UV completion, the mass of the various moduli fields, the scale of supersymmetry breaking and soft terms, the decay constant associated to the QCD axion and the tension of any potential cosmic (super)strings (as well as many of the various couplings of the Standard Model). Furthermore, concrete compactifications also permit the computation of couplings, interactions and decay modes for the various particles present in the theory.

 Epochs where the universe is dominated by the physics of string theory moduli may last a long time. String compactifications often contain light moduli $\Phi$ which come to dominate the energy density of the universe. Moduli domination begins when $H \sim m_{\Phi}$ and $t \sim \frac{1}{m_{\Phi}}$ and continues until the moduli eventually decay. As moduli are gravitationally coupled particles, their typical 
 expected lifetime is
\be
\tau_{\Phi} \sim 4 \pi \frac{M_P^2}{m_{\Phi}^3}.
\ee
The epoch of moduli domination is characterised by
\be
\frac{t_{end}}{t_{start}} \sim 4 \pi \left( \frac{M_P}{m_{\Phi}} \right)^2 .
\ee
For a modulus mass $m_{\Phi} \sim 10^6 \, \rm{GeV}$, this would give $t_{end}/t_{start} \sim 10^{25}$: an enormous period, which may also be bookended by other kination and/or tracker epochs. There are also a variety of problems/opportunities associated to moduli reheating: these include the Cosmological Moduli Problem \cite{Coughlan:1983ci, Banks:1993en, deCarlos:1993wie} and also the Moduli Dark Radiation problem of string compactifiations \cite{Cicoli:2012aq, Higaki:2012ar, Angus:2013zfa, Marsh:2014gca, Hebecker:2014gka, Acharya:2015zfk, Cicoli:2015bpq, Cicoli:2018cgu, Jaeckel:2021gah, Cicoli:2022fzy, Baer:2022fou}. See \cite{Kane:2015jia} for a review focussing specifically on the moduli-dominated epoch.

The upshot is that, instead of a radiation-dominated Hot Big Bang commencing shortly after the end of inflation, much of the stage between inflation and BBN could instead consist of a distinctively stringy cosmology where the energy density of the universe lies in compactification modes such as moduli. Furthermore, this cosmology could persist for as long as half the current lifetime of the universe (when measured on a logarithmic scale).

Our aim in this paper is to perform a detailed study of the cosmological evolution, in particular including the growth in perturbations, for the full duration of such stringy cosmologies. As extended kination and/or tracker epochs require, essentially by definition, substantially transPlanckian field excursions, we focus on the Large Volume Scenario in which the evolution of the volume modulus naturally allows for controlled large field excursions (in ways entirely consistent with the Swampland Distance Conjecture \cite{Ooguri:2006in, Ooguri:2018wrx}, as expected for any stringy completion).

Earlier work along this direction includes \cite{Brustein:1992nk, Barreiro:1998aj, Huey:2000jx, Kallosh:2004yh, Brustein:2004jp, Barreiro:2005ua, BuenoSanchez:2006epu, Conlon:2008cj, McAllister:2016vzi, Cicoli:2021fsd, Conlon:2022pnx, Rudelius:2022gbz, Cicoli:2022fzy, Apers:2022cyl, Shiu:2023nph, Shiu:2023fhb,Frey:2023jrb,Venken:2023hfa,Shiu:2023yzt,Seo:2024fki} in a stringy context and \cite{Steinhardt:1999nw, Amendola:1999dr, Barreiro:1999zs, Abramo:2001mv, Kawasaki:2001nx, Hwang:2001fb, Hwang:2001uaa, Malik:2002jb, Malquarti:2002iu, Bartolo:2003ad, Malik:2004tf, Copeland:2004qe, Pallis:2005bb, Tsujikawa:2006mw, Gomez:2008js, Redmond:2017tja, Redmond:2018xty, Gouttenoire:2021jhk, Delos:2023vfv, Khosravi:2023rhy} in the context of pure cosmology. What is new here? The previous stringy literature largely focuses solely on the background cosmology (and so neglects perturbations; partial exceptions are  \cite{Redmond:2017tja, Redmond:2018xty} and \cite{Das:2021wad}, which look at perturbation growth in a kination epoch and moduli-dominated epoch, respectively), and also tends to focus on partial epochs rather than a consistent evolution of the full cosmology. In the cosmology literature, much of the study of kination and tracker epochs treats them in the context of quintessence epochs in the current universe. Many of these cosmology studies predate the arrival of controlled moduli stabilisation scenarios; the cosmology literature also does not engage with the need for UV completion in order to control transPlanckian field excursions and furthermore lacks the periods of moduli domination and subsequent moduli reheating that are standard features of string cosmologies (an exception here is \cite{Eggemeier:2020zeg, Eggemeier:2021smj, Eggemeier:2023nyu}). Moreover, the cosmological implications of similar non-standard epochs (such as axion kination) have received significant attention recently \cite{Co:2019wyp,Co:2019jts,Chang:2019tvx,Co:2021lkc,Gouttenoire:2021wzu,Muia:2023wru}, but with no specific reference to a string theory embedding. The time is ripe for applying the full cosmological toolkit to histories of the early universe that are distinctly stringy in nature.

\begin{figure}[ht]
    \centering
    \includegraphics[width = 0.9\textwidth]{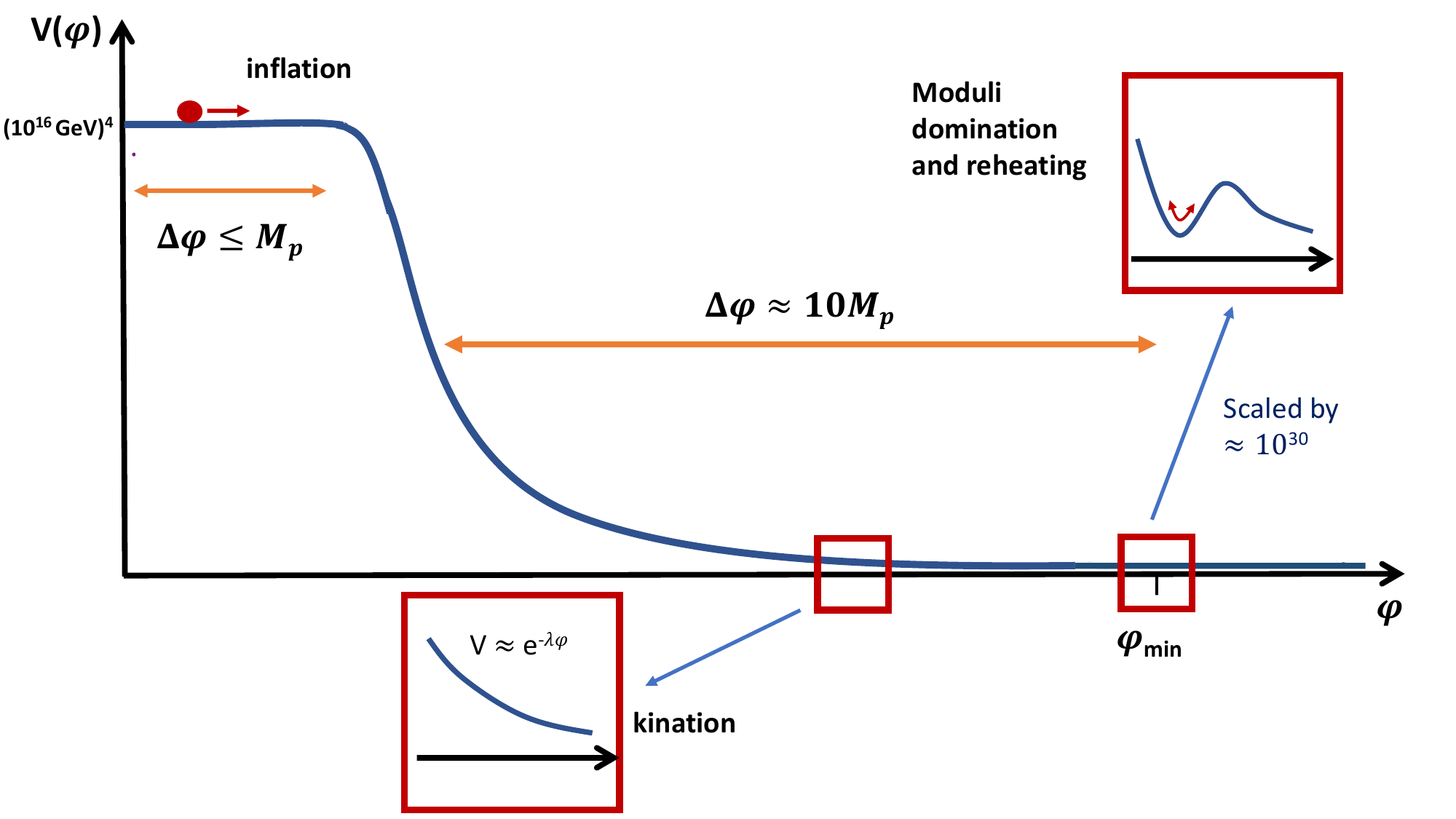} 
    \caption{A cartoon of how moduli can substantially modify the post-inflationary history of the universe. Following early inflation at very high energies, several epochs occur. We illustrate the case of a kination period followed by a tracker epoch followed by moduli domination period leading to reheating. Notice the large range of scales both in the scalar potential and the scalar field. In particular the barrier after the minimum may be 20 orders of magnitude smaller than the inflationary value ($V_{barrier}\simeq 10^{-20} V_{inflation} $). In this sense, the scale of inflation may be substantially larger than the Standard Model scale.}
    \label{ReheatingCartoon}
\end{figure}

\begin{figure}[h!]
\centering
\includegraphics[width=\textwidth]{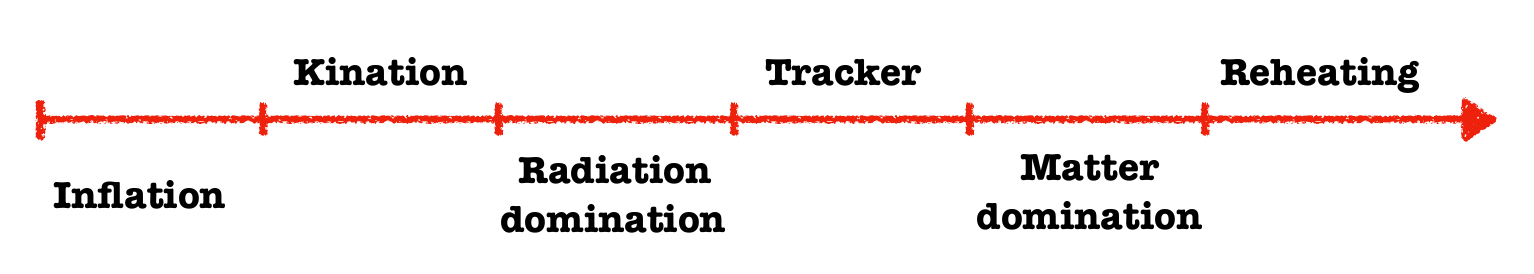}
\caption{Schematic depiction of the alternative stringy cosmological timeline.}
\label{fig:timeline}
\end{figure}

In this paper, we focus in particular on a  cosmology in which, after inflation, the volume scalar rolls down the potential, before merging into a tracker solution which guides it into the LVS minimum. As the fields approach the minimum, the volume modulus oscillates about it before settling back and eventually decaying to radiation and initiating the Hot Big Bang. This scenario is illustrated in Figure \ref{ReheatingCartoon} (taken from \cite{Cicoli:2023opf}) and the various implied epochs of the universe are
shown in figure \ref{fig:timeline}. Although the pictures suggests the volume modulus as the inflaton, this is not intrinsically key to the scenario (the volume modulus could act as a waterfall field).

The paper is structured as follows. Section \ref{BackgroundSection} describes the various epochs and describes the background cosmology. While individual treatments of the various parts are in the existing literature, this section brings these together into a unified account to cover the stringy cosmology between inflation and BBN. Section \ref{PerturbationsSection} analyses cosmological perturbations (through the framework of linear cosmological perturbation theory) generated during these epochs; cases where perturbations go non-linear are beyond the scope of this work. Section \ref{sectionNewApproaches} describes both how this cosmology motivates new approaches to old problems and also how novel physics and model-building possibilities are opened up. In section \ref{sectionConclusions} we conclude.

\section{The Background Cosmology}
\label{BackgroundSection}

This section is devoted to an analytic account of the background cosmology, before moving on to study perturbations in section \ref{PerturbationsSection}. While we do not claim originality for any of these individual epochs, there is value in bringing these together into a single coherent account.

To commence, we assume an epoch of inflation at some scale $\Lambda_{\rm{inf}}$, although we remain agnostic about the precise mechanism of inflation or the nature of the inflaton. What we do require is that, shortly after the end of inflation, the universe is rolling down the exponential potential associated to the volume modulus. This will naturally occur if the volume modulus itself is the inflaton (as in the models of \cite{Kallosh:2004yh, Conlon:2008cj, Cicoli:2015wja, Antoniadis:2020stf}); it could also occur if the volume modulus acts as a waterfall field that terminates inflation (as in the recent model of \cite{Burgess:2022nbx}).

While we do not concern ourselves here with inflationary model building, we note that as the volume modulus controls the string scale, it couples to all forms of energy density. Therefore the question (which depends on the details of inflationary model building) is not so much why the volume should be rolling at the end of inflation, but what would be capable of preventing it rolling during inflation.

\subsection{Kination}\label{kination-sub-sec}

The first epoch is a kination one associated to the volume modulus rolling down a steep exponential potential.
It follows from the equation of motion for a canonically normalised scalar field (notationally, we aim to use small $\phi$ to denote a generic scalar field and large $\Phi$ to denote the specific volume field in LVS),
\be
\ddot{\phi} + 3H \dot{\phi} + \nabla^2 \phi = -\frac{\partial V}{\partial \phi},
\ee
that, when $V$ is negligible, a kinating scalar within an FLRW cosmology evolves as
\begin{equation}
\label{darard}
\phi = \phi_0 + \sqrt{\frac{2}{3}} M_P \ln \left( \frac{t}{t_0} \right),
\end{equation}
where $\phi_0,\,t_0$ denote the value of $\phi$ and the time when kination domination begins. During this kination epoch, all energy is in the form of field kinetic energy, which evolves as
\be
\rho_{KE} \propto \frac{1}{a(t)^6}.
\ee
The logarithmic divergence in Eq. (\ref{darard}) at $t = 0$ reflects the formal divergence of the kinetic energy at the point where $a(t) = 0$.
During an epoch of pure kination, the scale factor and Hubble rate behave as
\begin{equation}\label{aandt}
a(t) = a_0 \left( \frac{t}{t_0} \right)^{1/3}, \quad \quad H=\frac{1}{3t},
\end{equation}
 where $a_0\equiv a(t_0)$.
It is convenient to switch to conformal time $\eta$, which is related to $t$ by  $d\eta = dt/a(t)$, hence
\begin{equation}\label{eta}
\eta(t) =  \frac{3}{2a_0}(t^2 t_0)^{1/3} = \eta_0  \left( \frac{t}{t_0} \right)^{2/3}, 
\end{equation}
 which implies that the field evolution of
Eq.~(\ref{darard}) becomes
\begin{equation}\label{phikinconf}
    \phi= \phi_0 + \sqrt{\frac{3}{2}} M_P \ln \left( \frac{\eta}{\eta_0} \right) ~~{\rm with}~~a(\eta) = a_0 \left(\frac{\eta}{\eta_0}\right)^{\frac{1}{2}}.
\end{equation}
Denoting derivatives with respect to $\eta$ by a prime, the conformal Hubble scale becomes
\begin{equation}
\mathcal{H} \equiv \frac{a'}{a} = \frac{1}{2 \eta}.
\end{equation}
In our context of IIB string compactification with the volume modulus as the kinating field, the extra-dimensional volume grows during kination as \cite{Conlon:2022pnx}
\be \label{volevol}
\frac{\mc{V}}{\mc{V}_0} \sim \frac{t}{t_0}.
\ee
Here $\mathcal{V}$ is the dimensionless Calabi-Yau volume, i.e. the physical volume measured in units of $l_s^6$, where $l_s = 2 \pi \sqrt{\alpha'}$.
This follows using the K\"ahler potential $K = - 3 \ln \left( T + \bar{T} \right)$, where the complexified K\"ahler modulus $T= \tau_R + i a$ and the canonically normalised field is $\frac{\Phi}{M_P} = \sqrt{\frac{3}{2}} \ln \tau_R = \sqrt{\frac{2}{3}} \ln \mc{V}$ with $\tau_R \sim \mc{V}^{2/3}$  giving the relationship between 4-cycle volumes and the overall volume.
For the LVS potential, the modulus mass (or, strictly, the second derivative of the potential), scales with the volume as $m_{\Phi} \sim M_P \mc{V}^{-3/2}$; it therefore follows that during kination the modulus mass behaves as $m_{\Phi} \sim  M_P \frac{1}{ (t/t_0)^{3/2}}$ and so $m_{\Phi} \ll H$: the field is effectively massless. 

Going back in time, one arrives at a point where the kination approximation is no longer justified as the potential energy is comparable to the kinetic one (the end of inflation). In particular, equating the two, one can approximate the beginning of kination to occur at a time $t_0$ such that
\begin{equation}\label{eq:to}
\frac{M_P^2}{3 t_0^2} \sim V (\phi_0)\sim \Lambda^4_{\rm{inf}}.
\end{equation} 
 As mentioned above, during the kination epoch, the modulus field is (effectively) massless, and from Eq.~(\ref{aandt}), we see that the comoving Hubble scale is increasing, $(aH)^{-1} \sim t^{2/3}$ and modes of the $\Phi$ field re-enter the horizon. 

Note that, provided there is an initial source of radiation or matter present, the kination epoch is time-limited; as $\rho_{kin} \sim a^{-6}$ and
$\rho_{\gamma} \sim a^{-4}$, radiation will ultimately catch up with and overtake the kinetic energy.

\subsection{The Transition from Kination to Radiation}\label{kinradtran}
As initial radiation grows relative to kinetic energy, it catches up and comes to dominate, before the fields evolve onto the tracker solution. For an analytic description, we consider the FLRW metric in the presence of a scalar field, together with some other fluid with equation of state $P= (\gamma-1) \rho$ (in practical cases of interest, this fluid will generally be radiation, with $\gamma=\frac{4}{3}$). Following standard approaches \cite{Copeland:1997et, Bahamonde:2017ize} we define the variables 

\begin{equation}\label{eq:xyv}
x = \frac{\dot{\phi}}{M_P } \frac{1}{\sqrt{6} H}, \quad \quad y = \sqrt{\frac{V(\phi)}{3}} \frac{1}{M_P H},
\end{equation}
for which the Friedmann equations can be recast as the dynamical system 
\begin{equation}\label{eq:xy2}
\left\{ \begin{aligned} 
  x'(N) &= -3x - \frac{M_P V'(\phi)}{V(\phi)}\sqrt{\frac{3}{2}} y^2 + \frac{3}{2}x \big[ 2x^2 + \gamma(1-x^2-y^2) \big], \\
  y'(N) &=   \frac{M_P\, V'(\phi)}{V(\phi)}\sqrt{\frac{3}{2}} xy + \frac{3}{2}y \big[ 2x^2 + \gamma(1-x^2-y^2) \big], \\
H'(N) &= -\frac{3}{2} H (2x^2 + \gamma(1-x^2-y^2)), \\
\phi'(N) &= \sqrt{6}  M_P x},
{ \end{aligned} \right.
\end{equation}
where the time variable is $N = \log a$ and $x'(N) = \frac{dx}{dN}$, $V'(\phi) = \frac{dV}{d\phi}$, etc. For concreteness of computations, we focus on the specific case where the kinating scalar is the volume modulus in an LVS compactification, but similar conclusions would apply more generally. In particular, stringy kination scenarios could also manifest as runaway directions for other moduli (for example, the dilaton) towards the boundary of moduli space. In the case of LVS, the scalar potential during the runaway epoch takes the form (see \cite{Balasubramanian:2005zx}, \cite{Conlon:2006gv} for derivations of the LVS potential)
\begin{equation}
V(\Phi) = M_P^4 e^{-\sqrt{\frac{27}{2}} \frac{\Phi}{M_P} }  = \Lambda^4_{\rm{inf}}  \left( \frac{\mathcal{V}_0}{ \mathcal{V}} \right)^{3},
\end{equation}
where $\frac{\Phi}{M_P} = \sqrt{\frac{2}{3}} \log \left( \mathcal{V} \right)$ is the canonically normalised volume modulus.
For more generic runaway directions, the potential has the form $V = V_0 e^{-\lambda \phi/M_P}$.\footnote{Exponential potentials arise in string theory when potentials are power-law in fields with logarithmic K\"ahler potentials. This is satisfied for the classic runaway fields, the dilaton with $K = -\ln \left( S + \bar{S}\right) $ and the volume modulus with $K = - 3 \ln \left( T + \bar{T} \right)$. Any perturbative expansion of the potential in either the string $\alpha'$ or quantum $g_s$ expansions will, by construction, give power-law potentials in $S$ or $T$ moduli.}

In this notation for generic exponential potentials, kination is represented by the (unstable) fixed point $(x,y)=(1,0)$, while the tracker solution is represented by the fixed point
\begin{equation}\label{eq:cp}
(x,y) = \left( \sqrt{\frac{3}{2}} \frac{\gamma}{\lambda},  \sqrt{\frac{3(2-\gamma)\gamma}{2 \lambda^2}}\right),
\end{equation}
which exists (and is a stable node or spiral) for $\lambda^2 > 3 \gamma$.

If the energy density of radiation (or any other fluid with $\gamma <2$ is non-zero at early times, the effects of redshift will ensure that it eventually catches up with the kinating scalar. For radiation, if we assume an initial energy fraction 
\begin{equation}\label{kinradequal}
\Omega_{\gamma}(t_0) \equiv \frac{\rho_{\gamma}(t_0)}{3 H_{\rm{inf}}^2M_P^2}=\frac{\rho_{\gamma}(t_0)}{\Lambda^4_{\rm{inf}}}= \varepsilon \ll 1,
\end{equation}
where $H_{\rm{inf}}$ is the Hubble parameter at $t_0$, the end of inflation and the start of kination, then,
radiation-kination equality is obtained at $t_{\rm{eq}} \sim t_0 \, \varepsilon^{-3/2}$, or $\eta_{\rm{eq}} \sim \eta_0/\varepsilon$. 

Under the assumption that the potential can be neglected, i.e. $y=0$, the system \eqref{eq:xy2} can be solved exactly. This represents a cosmological history which asymptotes to pure kination at early times and pure radiation at late times. If $y=0$, the first equation of \eqref{eq:xy2} becomes
\begin{equation}
    \frac{dx}{d N} = x^3-x,
\end{equation}
which can be integrated as
\begin{equation}\label{eq:x(a)}
x(a) = \frac{1}{\sqrt{1+\frac{\varepsilon}{1-\varepsilon} \left( \frac{a}{a_0}\right)^2} } \simeq \frac{1}{\sqrt{1+\varepsilon \left( \frac{a}{a_0}\right)^2} }.
\end{equation}
Radiation-kination equality is obtained at $x= \frac{1}{\sqrt{2}}$, corresponding to $\eta_{\rm{eq}} = \eta_0/\varepsilon$ $\Big( $recall from Eqs.(\ref{aandt}) and (\ref{eta}),
$\frac{\eta}{\eta_0} = \left( \frac{a}{a_0} \right)^2$ $ {\rm during \, kination} \Big)$.
The equation for $H(N)$ becomes
\begin{equation}
   \frac{dH}{d N} = -2 H -\frac{H}{1+\frac{\varepsilon}{1-\varepsilon} e^{2(N-N_0)} },
\end{equation}
and is solved by 
\begin{equation}\label{eq:Hrk}
    H(a) = H_0 \sqrt{1-\varepsilon} \left(\frac{a}{a_0}\right)^{-3} \sqrt{1+ \frac{\varepsilon}{1-\varepsilon} \left( \frac{a}{a_0}\right)^2} .
\end{equation}
In turn, integrating \eqref{eq:Hrk} and switching to conformal time, the scale factor evolves as
\begin{equation}\label{eq:aex}
    a(\eta)= a_0 \sqrt{1+ 2 \mathcal{H}_0 \left(\eta- \eta_0 \right)+ \varepsilon \mathcal{H}_0^2\left(\eta-\eta_0 \right)^2}, 
\end{equation}
with
\begin{equation}\label{eq:hex}
    \mathcal{H}(\eta) = \mathcal{H}_0\frac{1+\varepsilon \mathcal{H}_0(\eta-\eta_0)}{1+\mathcal{H}_0(\eta-\eta_0)(2+\mathcal{H}_0 \varepsilon (\eta-\eta_0))}.
\end{equation}
Therefore, in the absence of a potential the background scalar obeys the equation
\begin{equation}
    \phi''(\eta) + 2\mathcal{H}_0\frac{1+\varepsilon \mathcal{H}_0(\eta-\eta_0)}{1+\mathcal{H}_0(\eta-\eta_0)(2+\mathcal{H}_0 \varepsilon (\eta-\eta_0))} \phi'(\eta) =0,
\end{equation}
giving
\begin{equation}\label{eq:phie}
  \phi(\eta)= \phi(\eta_0)+ \sqrt{\frac{3}{2}} M_P \log \left(
  \frac{ \left( \varepsilon \mathcal{H}_0 (\eta-\eta_0) +1-\sqrt{1-\varepsilon} \right) \left( 1+\sqrt{1-\varepsilon} \right)}{\left( \varepsilon \mathcal{H}_0 (\eta-\eta_0) +1+\sqrt{1-\varepsilon} \right) \left( 1-\sqrt{1-\varepsilon} \right) } \right)
\end{equation}
with
\begin{equation}
\label{phispeed}
    \phi'(\eta) = \frac{\sqrt{6(1-\varepsilon)}\mathcal{H}_0 M_P }{1+\mathcal{H}_0(\eta-\eta_0)(2+\mathcal{H}_0 \varepsilon (\eta-\eta_0))}.
\end{equation}
We can always shift $\eta$ by a constant so that quantities diverge at $\eta=0$. This is equivalent to the condition
\begin{equation}\label{eq:1ic}
    \mathcal{H}_0 \eta_0 = \frac{1-\sqrt{1-\varepsilon}}{\varepsilon},
\end{equation}
which reduces to $ \mathcal{H}_0 \eta_0 = \frac{1}{2}$ for $\varepsilon \rightarrow 0$. We can then regard $\eta_0$ as the smallest time for which the kination approximation is valid, and treat this as the ``beginning" of kination. 

For $\eta < \eta_0$, the above formulae no longer apply as the system transitions between inflation and kination and so the potential term  (which has been neglected above) matters. Note that, having chosen our time coordinates to yield an apparent singularity at $\eta=0$, we can no longer perform shifts in $\eta$ and inflation does not necessarily happen at $\eta=0$ (but rather at some time $\eta_{\rm{inf}}$ determined by the dynamics of the transient regime).

Substituting \eqref{eq:1ic} in \eqref{eq:phie}, we obtain the much simpler expression
\begin{equation}\label{eq:phiex}
    \phi(\eta) = \phi(\eta_0) +\sqrt{\frac{3}{2}} M_P \ln \left( \frac{\left(1+\sqrt{1-\varepsilon} \right)\eta}{\left(1-\sqrt{1-\varepsilon} \right)\eta+ 2\eta_0 \sqrt{1-\varepsilon}} \right).
\end{equation}
Therefore, taking $\eta_0$ as small and considering the large $\eta$ limit in which $\eta \gg \frac{\eta_0}{\varepsilon}$, the field displacement during the kination and radiation dominated epochs asymptotes to 
\begin{equation}
\label{fuj}
    \Delta \phi = \sqrt{\frac{3}{2}} M_P \ln \left( \frac{1+\sqrt{1-\varepsilon} }{1-\sqrt{1-\varepsilon} } \right) \simeq \sqrt{\frac{3}{2}} M_P \log \left( \frac{4}{\varepsilon}\right).
\end{equation}
The finiteness of Eq. (\ref{fuj})
reflects the fact that once we enter radiation domination the scalar field slows down and effectively stops, as in the $\eta \to \infty$ limit, $\phi'(\eta) \sim \eta^{-2}$ from Eq. (\ref{phispeed}).

Although many of these expressions for the background evolution are exact for $V(\phi)=0$,
for the practical cases of interest the potential is always non-zero. We merely require $\varepsilon \ll 1$ for an extended period of kination to occur, and we can simplfy by expanding the coefficient of each $\eta^{\alpha}$ term to leading order in $\varepsilon$. Making use of Eq. \eqref{eq:1ic}, Eq. \eqref{eq:aex} can then be approximated as
\begin{equation}\label{eq:aap}
    a(\eta) = a_0 \sqrt{\frac{\eta}{\eta_0}+\frac{\varepsilon}{4}\left( \frac{\eta}{\eta_0}\right)^2 },
\end{equation}
which provides the correct limiting behaviour for kination and radiation-domination respectively,
\begin{equation}
    \frac{a(\eta)}{a_0} \underset{  \eta  \ll \frac{\eta_0}{\varepsilon}}{\simeq} \left( \frac{\eta}{\eta_0}\right)^{1/2} \quad \quad \text{and} \quad \quad
   \frac{a(\eta)}{a_0} \underset{\eta  \gg \frac{\eta_0}{\varepsilon}}{\simeq} \frac{\sqrt{\varepsilon}}{2 \eta_0}\left( \frac{\eta}{\eta_0}\right).
\end{equation}
The corresponding (conformal) Hubble scale is
\begin{equation}\label{eq:H0tr}
    \mathcal{H}(\eta)= \frac{1}{2 \eta} \frac{1+\frac{\varepsilon \eta}{2 \eta_0}}{1+\frac{\varepsilon \eta}{4 \eta_0}},
\end{equation}
which again gives the two correct asymptotic behaviours. 

These expressions provide good approximations throughout the kination and radiation regimes. Outside of these regimes, the potential energy is relevant: in particular both at the beginning of kination (as the field commences its fast-roll evolution after inflation) and also at the end of the radiation epoch, as the tracker epoch commences. 

We can make this quantitative. The effects of the potential compared to the kinetic energy of the field can be neglected as long as
\begin{equation}\label{kin-dom}
   \frac{1}{2}\dot{\phi}^2 \equiv \frac{1}{2 a(\eta)^2} \phi'^2 \gtrsim V(\phi) = V_0 e^{-\frac{\lambda \phi}{M_P}}.
\end{equation}
In this regime of kination domination, the field acts as a stiff fluid with equation of state $w=\left(\frac{\dot{\phi}^2-2 V(\phi)}{\dot{\phi}^2+2 V(\phi)}\right) \sim 1$. Substituting in, and making the approximation that $\varepsilon \ll 1$, we find that equality occurs at 

\begin{equation}\label{eq:aep}
\frac{ 3 M_P^2}{4 a_0^2 \eta_0^2 \tilde{x}^3 \left( 1 + \frac{\varepsilon \tilde{x}}{4} \right) \left( 1 + \frac{\varepsilon \tilde{x}}{2} + \frac{\varepsilon^2 \tilde{x}^2}{16} \right) } \sim V_0~ e^{-\frac{\lambda \phi_0}{M_P}}\left( \frac{4\tilde{x}}{4 + \varepsilon \tilde{x}} \right)^{-\lambda \sqrt{\frac{3}{2}}} .
\end{equation}
where $\tilde{x}=\eta / \eta_0$. 
By construction, $\tilde{x}=1$ defines the initial point at the end of inflation where the kinetic energy of the field was comparable to its potential energy. This enables us to equate
\begin{equation}
\frac{ 3 M_P^2}{4 a_0^2 \eta_0^2} \sim V_0~  e^{-\frac{\lambda \phi_0}{M_P}},
\end{equation}
a result consistent with section~\ref{kination-sub-sec} which is for a purely kination dominated period. From Eq. (\ref{eq:aep}) we can see that during the kination epoch, for $1 \lesssim \tilde{x} \lesssim \varepsilon^{-1}$, the potential energy becomes progressively more subdominant to kinetic energy (provided that  $\lambda \gtrsim \sqrt{6}$). Following this, the potential continues to grow relative to kinetic energy with equality at
\begin{equation}
    \eta_t = \eta_0 \varepsilon^{-\frac{\lambda}{6}\sqrt{\frac{3}{2}} - \frac{1}{2}},
\end{equation}
provided that $\lambda \ge \sqrt{6}$ (otherwise the potential is too flat and the potential is never subdominant). For LVS, with $\lambda = \sqrt{\frac{27}{2}}$, this gives potential-kinetic equality at (we focus on the dominant inverse powers of $\varepsilon \ll 1$) 
\begin{equation}
\label{fghy}
\tilde{x} \sim \varepsilon^{-5/4}.
\end{equation}
Note that if $\lambda \leq \sqrt{6}$, there is no solution to \eqref{eq:aep} satisfying $\varepsilon \tilde{x} \gg 1$. This corresponds to a much flatter potential, such that the potential energy does not become sub-dominant to kination as the field rolls to large values. In this case, the system transitions directly from kination to the tracker solution without undergoing an intermediate phase of radiation domination.

We can also understand the factor of $\varepsilon^{-5/4}$ in Eq. (\ref{fghy}) another way. In LVS, during kination, the volume grows as $\mc{V} \sim t$. As the potential $\sim \mc{V}^{-3}$, the potential energy falls during kination as $t^{-3} \sim a^{-9} \sim \eta^{-9/2}$.
As kinetic energy falls as $\sim \eta^{-3}$ during kination and radiation-kination equality occurs at $\eta \sim \eta_0/\varepsilon$, at kination-radiation equality the potential-kinetic ratio is $\varepsilon^{3/2} : 1$. During the radiation epoch, the potential is (effectively) frozen; as $a \sim \eta$ in this epoch, $\rho_{KE} \sim \eta^{-6}$ while $\rho_{rad} \sim \eta^{-4}$ during this period. 

The potential therefore catches up with kinetic energy at
\begin{equation}
\eta \sim \left( \frac{\eta_0}{\varepsilon} \right) \left( \varepsilon^{-3/2} \right)^{1/6} \sim \eta_0 \varepsilon^{-5/4},
\end{equation}
consistent with the above results.

The same argument implies that during the radiation epoch, the potential will catch up with radiation at
\begin{equation}
\eta \sim \left( \frac{\eta_0}{\varepsilon} \right) \left( \varepsilon^{-3/2} \right)^{1/4} \sim \eta_0 \varepsilon^{-11/8}.
\end{equation}
Taking this as the starting point of the tracker solution, we therefore have
\begin{equation}
\tilde{x}_{tracker} \sim \varepsilon^{-11/8}
\end{equation}
in LVS.

\subsection{The Tracker Solution}\label{ssc:track}

As radiation domination ends, the cosmological evolution will settle down into the tracker solution. We first discuss the tracker solution itself and then, later, the transients present as the solution settles down into the tracker.  Recently, in \cite{Copeland:2023zqz}, this type of approach to the tracker regime has been used as a possible explanation of the early dark energy postulated by many as a resolution of the Hubble tension.
The tracker solution exists as a fixed point of the evolution equations Eq. (\ref{eq:xy2}) with
\begin{equation}
\label{TrackerFractions}
(x,y) =\bigg(\sqrt{\frac{3}{2}}\frac{\gamma}{\lambda}, \sqrt{\frac{3(2-\gamma)\gamma}{2\lambda^2}}\bigg),
\end{equation}
which is stable as long as $\lambda^2>3\gamma$. This trivially satisfies the $x'$ and $y'$ equations within Eq. (\ref{eq:xy2}), while
\begin{equation}
H'(N) = -\frac{3}{2}H \gamma, \ \ \ \ \ \ \ \ \phi'(N) = \frac{3\gamma}{\lambda}  M_P.
\end{equation}
Thus, if we regard the tracker as starting with a scale factor $a_t$ at time $t_t$ the rolling scalar field has the solution 
\begin{equation}
\phi = \phi_t + \frac{3\gamma M_P}{\lambda}\log \bigg(\frac{a}{a_t}\bigg),
\end{equation}
resulting in
\begin{equation}
H = H_t \bigg(\frac{a}{a_t}\bigg)^{-\frac{3}{2}\gamma}, \qquad
a = a_t \bigg(\frac{t}{t_t}\bigg)^{2/3\gamma}, \qquad \eta(a) = \eta_t\bigg(\frac{a}{a_t}\bigg)^{3\gamma/2-1}.
\end{equation}
For a radiation background with $\gamma = 4/3$ we have $\eta(a) = a/a_t$.

As the most common types of fluid are radiation and matter, for reference we also give very explicit expressions for the field evolution within radiation and matter trackers.

\subsubsection*{Radiation Tracker}\label{ssc:rt}

In a radiation tracker epoch, the ratios of potential energy, kinetic energy and radiation remain constant. As we know that $\rho_{rad} \sim \frac{1}{a^4}$ and $V(\phi) = V_0 e^{-\lambda \phi}$, it follows that during a radiation tracker, 
\be
\label{ScalarRadTracker}
\phi = \phi_t + \frac{4 M_P}{\lambda} \ln \left( \frac{a}{a_t} \right) = \phi_t + \frac{2 M_P}{\lambda} \ln \left( \frac{t}{t_t} \right).
\ee
If we specialise to LVS and put $\lambda = \sqrt{\frac{27}{2}}$, we have
\be
\Phi = \Phi_t + \frac{4 \sqrt{2} M_P}{3 \sqrt{3}} \ln \left( \frac{a}{a_t} \right) = \Phi_t + \left( \frac{2}{3} \right)^{3/2} M_P \ln \left( \frac{t}{t_t} \right).
\ee
Note that, compared to the kination epoch, the speed of the LVS scalar field is only marginally slower in the tracker epoch (by a factor of 2/3). This also implies that the scalar field continues to move approximately one Planckian distance in field space each Hubble time (or e-folding).

During the tracker regime, the kinetic energy $\rho_{kin} = \frac{\dot{\phi}^2}{2}$ retains a constant ratio compared to both  potential energy and radiation. As
$
\dot{\phi}^2 \propto \frac{\dot{a}^2}{a^2},
$
for this to to scale the same way as $a^{-4}$, it follows that
$\dot{a}^2 \propto a^{-2}$ and so $a(t) \sim t^{1/2}$. During the tracker epoch we therefore have (also including the specialisation to LVS with $\lambda = \sqrt{\frac{27}{2}}$)
\begin{eqnarray}
a(\eta) & \propto & \eta, \\
\mathcal{H} & = & \frac{1}{\eta}, \\
\Phi(\eta) & = & \Phi_t + \frac{4 M_P}{\lambda} \ln \left( \frac{\eta}{\eta_t} \right) = \Phi_t + \frac{4 \sqrt{2} M_P}{3 \sqrt{3}} \ln \left( \frac{\eta}{\eta_t} \right).
\end{eqnarray}

\subsubsection*{Matter Tracker}\label{ssc:mt}

Although we mostly focus on the case of a radiation tracker, we also give expressions for the case of a matter tracker (as could arise with e.g. the presence of primordial black holes formed soon after inflation and providing an initial matter seed).

For a matter tracker, the ratios of potential energy, kinetic energy and matter also remain constant. As we know that $\rho_{matter} \sim \frac{1}{a^3}$ and $V(\phi) = V_0 e^{-\lambda \phi/M_P}$, it follows that during a matter tracker, 
\be
\label{ScalarMatterTracker}
\phi = \phi_t + \frac{3 M_P}{\lambda} \ln \left( \frac{a}{a_t} \right) = \phi_t + \frac{2 M_P}{\lambda} \ln \left( \frac{t}{t_t} \right).
\ee
If we specialise to LVS and put $\lambda = \sqrt{\frac{27}{2}}$, we have
\be
\Phi = \Phi_t + \sqrt{\frac{2}{3}} M_P \ln \left( \frac{a}{a_t} \right) = \Phi_t + \left( \frac{2}{3} \right)^{3/2} M_P \ln \left( \frac{t}{t_t} \right).
\ee
Somewhat surprisingly, the speed of the scalar field (as a function of time) is identical for both matter and radiation trackers.

During a matter tracker, the kinetic energy $\rho_{kin} = \frac{\dot{\phi}^2}{2}$ retains a constant ratio compared to both potential energy and matter. As
$
\dot{\phi}^2 \propto \frac{\dot{a}^2}{a^2},
$
for this to scale the same way as $a^{-3}$, it follows that
$\dot{a}^2 \propto a^{-1}$ and so $a(t) \sim t^{2/3}$. During a matter tracker we therefore have (also including the specialisation to LVS)
\begin{eqnarray}
a(\eta) & \propto & \eta^2, \\
\mathcal{H} & = & \frac{2}{\eta}, \\
\Phi(\eta) & = & \Phi_t + \frac{6 M_P}{\lambda} \ln \left( \frac{\eta}{\eta_t} \right) = \Phi_t + \frac{2 \sqrt{2} M_P}{\sqrt{3}} \ln \left( \frac{\eta}{\eta_t} \right)
\end{eqnarray}

\subsection{Approach to the (Radiation) Tracker}
\label{trackerapproach}

Although the tracker solution is a stable fixed point, the system oscillates about it as it settles down into the fixed point.
 We can interpret these as transient oscillations associated to the volume scalar, which remains massive during the tracker. As (in terms of parameters of the compactification)  $\frac{\Phi}{M_P} \sim \sqrt{\frac{2}{3}} \ln \mathcal{V}$  it follows that during the tracker itself, $\mathcal{V} \sim \left( t/t_0 \right) ^{2/3}$ with $m_{\Phi} \sim M_P \mathcal{V}^{-3/2} \sim M_P \left( t/t_0 \right)^{-1} \sim H$. We therefore expect that during the tracker, the energy density associated to away-from-equilibrium $k=0$ perturbations of the massive volume mode will behave as
\be
\rho_{\delta \Phi} \sim n_{\delta \Phi}(t) m_{\delta \Phi}(t) \sim \frac{1}{a^3(t)} \frac{1}{a^2(t)} \sim \frac{1}{a^5(t)}.
\ee
The volume scalar remains massive during the tracker with a mass remaining at a constant fraction of the Hubble scale. It follows from the scalar field equations of motion 
\be
\ddot{\phi} + \frac{3}{2t} \dot{\phi} = -\frac{\partial V}{\partial \phi},
\ee Eq. (\ref{ScalarRadTracker}) and the properties of the tracker fixed point Eq. (\ref{TrackerFractions}) in LVS that 
\be
\label{ddde}
m_{\delta \Phi}^2(t) \equiv \frac{\partial^2 V}{\partial \Phi^2} = \frac{1}{t^2} = 4 H^2.
\ee
Note that the mass of the volume perturbation, $m_{\delta \Phi}= 2H$, is actually independent of the value of $\lambda$ (even though we have specialised to LVS).

Although the volume scalar is coupled to the radiation fluid via the Einstein equations, it is instructive first to study the simpler case of a scalar with minimal contribution to the energy density, and for which the coupling to the radiation fluid can be neglected. 
 For illustration, we therefore consider the equations of motion for the $k=0$ mode of a massive scalar $\delta \phi$ with $m_{\delta \phi} = \frac{\mu}{t}$ where $\mu$ is a dimensionless constant (and so whose mass remains at a constant ratio to $H$), within a scale factor $a \sim t^{1/2}$. The equations of motion for such a scalar $\delta \phi$ are
\be
\delta \ddot{\phi} + 3H \delta \dot{\phi} = -\frac{\mu^2}{t^2} \delta \phi,
\ee
which are solved by 
\be\label{going_to_tracker}
\delta \phi(t) = t^{\alpha} \left( A \cos \left( \tilde \omega \ln t \right) + B \sin \left( \tilde \omega \ln t \right) \right),
\ee
where $\alpha = \frac{1}{2}(1-3H_t t_t) = -1/4$ for a radiation tracker (in more general trackers, $\alpha= (\gamma-2)/2\gamma$) and 
$\tilde \omega = \sqrt{\mu^2-\alpha^2}$.
The energy associated to such oscillations,
\be
\rho_{\delta \phi} = \frac{\delta \dot{\phi}^2}{2} + \frac{m_{\delta \phi}^2(t) \delta \phi^2}{2},
\ee
indeed scales as $\rho_{\delta \phi} \sim \frac{1}{a(t)^5}$ during the radiation tracker.  A structural feature of such oscillations is that they are rather slow: the oscillation period lasts around one Hubble time and so it takes a while for the system to settle down into the tracker (manifested by the logarithm inside the sinusoids). 

We now extend this to the case where the perturbed scalar field is actually that of the tracker field, and so is coupled to the background fluid. In this case, we need to consider perturbations in the full scalar field--background fluid system described by the tracker. If there were no coupling to the background, then $\mu^2 = 2(2-\gamma)/\gamma$ (which equals unity for a radiation tracker as per Eq. (\ref{ddde})) and 
\begin{equation}
\tilde \omega^2 = \frac{2-\gamma}{4\gamma^2}(9\gamma-2).
\end{equation}
To consider the full case, let the fixed point be $(x,y) =(x_0,y_0)$ and so we can write $x = x_0+f(N)$ and $y=y_0+g(N)$. Expanding the equation to first order in these functions gives a system of linear first order differential equations,
\begin{equation}
\begin{split}
\frac{d}{dN} {f \choose g} & = \begin{pmatrix} \tfrac{3}{2}(\gamma-2+ \frac{6\gamma^2}{\lambda^2}-3\frac{\gamma^3}{\lambda^2}) & 3 \sqrt{(2-\gamma)\gamma}(1-\tfrac{3}{2}\frac{\gamma^2}{\lambda^2})\\
3\sqrt{(2-\gamma)\gamma}(\frac{3\gamma}{\lambda^2}-\frac{3\gamma^2}{2\lambda^2}-\frac{1}{2}) & \frac{9\gamma^2}{2\lambda^2}(\gamma-2) \end{pmatrix} {f \choose g}, \\
& = \begin{pmatrix} \alpha & \beta \\ \gamma & \delta \end{pmatrix} { f \choose g}.
\end{split}
\end{equation}
These can be solved to give
\begin{equation}
\label{gtyu}
\begin{split}
x(N) & = x_0 + e^{\tfrac{3}{4}(\gamma-2)N}[c_1 \cos(\omega N) +\frac{1}{2\omega}[2\beta c_2 + (\alpha-\delta)c_1]\sin(\omega N))] \\
y(N) & = y_0 + e^{\tfrac{3}{4}(\gamma-2)N}[c_2 \cos(\omega N) + \frac{1}{2\omega}[2\gamma c_1 - (\alpha-\delta)c_2] \sin(\omega N))],
\end{split}
\end{equation}
where the frequency is
\begin{equation}
\omega^2 = \bigg(\frac{3\gamma}{2}\bigg)^2\tilde \omega^2 - \frac{27}{2\lambda^2}\gamma^2(2-\gamma),
\end{equation}
showing that the oscillations of the rolling scalar field within the tracker are at a  frequency slightly lower than the frequency of a subdominant scalar field. This comes from the effect that an oscillating scalar field has on the Hubble constant. This expression can then be used to solve for the constants of integration.

The oscillations in the Hubble constant behave schematically (the derivation is provided in the Appendix in section \ref{appendixtrackerapproach}) as 
\begin{equation}
\delta H = H\bigg[c_3+ A' e^{\frac{3}{4}(\gamma-2)N}\sin(\omega N + \alpha')\bigg],
\end{equation}
for some coefficients $A'$ and $\alpha'$, with initial conditions fixing $c_3$.

\subsection{Moduli domination}
\label{modulidombackground}

As the tracker solution approaches the minimum of the moduli potential, the modulus converts to a conventional massive field oscillating in a quadratic potential. These oscillations can be viewed as coherent moduli quanta, whose decays ultimately reheat the universe and initiate the Hot Big Bang. During this epoch, any high-energy radiation perturbations of the $\phi$ field will redshift to become non-relativistic and convert into matter perturbations. There is a large literature on the  physics of moduli domination and reheating in string theory (referenced in the introduction); our short summary here is largely for completeness.

If we assume a close-to-instantaneous transition to moduli domination, at time $t_{moduli}$ we can treat the scalar field potential as made up of quadratic fluctuations about some background value $\tilde \phi$ which represents the final vev of the modulus field,
\begin{equation}
V(\phi) = \frac{1}{2}m^2 (\tilde \phi - \phi)^2.
\end{equation}
Appendix \ref{appendixModuliDecay} includes a treatment that goes beyond the instantaneous-decay approximation.
Once $H < m$  we can regard the field oscillations as rapid
and treat the scalar field as a harmonic oscillator with slowly varying amplitude {\color{blue} }
\begin{equation}
\phi \approx \tilde \phi + A(a) \sin(m t).
\end{equation}
Averaging over the rapid oscillation cycles gives
\begin{equation}
\langle V\rangle = \frac{1}{4}m^2 A^2, \ \ \ \ \ \ \ \langle K\rangle = \langle \tfrac{1}{2}\dot \phi^2\rangle = \frac{1}{4}m^2 A^2,
\end{equation}
where the time average is given by $\langle V\rangle \equiv \frac{1}{T} \int_0^T V dt$, and the period of oscillation is $T$. The result corresponds to a cosmic fluid whose energy density and pressure is
\begin{equation}
\rho = \langle K\rangle +\langle V\rangle = \tfrac{1}{2}m^2 A^2, \ \ \ \ \ \ \ P = \langle K \rangle - \langle V\rangle = 0,
\end{equation}
which behaves as matter with $\gamma=1$ with the amplitude $A(a)$ decreasing as $A(a) \propto a^{-3/2}$. 

During matter domination the conformal time evolves as
\begin{equation}
\eta \propto t^{1/3} \propto a^{1/2}, \qquad \mathcal{H} = \frac{2}{\eta}.
\end{equation}
The universe remains in this moduli-dominated matter epoch until the time of moduli decay, reheating to the Standard Model and initiating the Hot Big Bang. As moduli are gravitationally coupled, the moduli life-time is long and the reheat temperature is low. For standard couplings,
 \be
  \tau_{\Phi} \sim \frac{4 \pi M_P^2}{m_{\Phi}^3}, \qquad
 T_{reheat} \sim 1 {\rm GeV} \left( \frac{M_{\Phi}}{10^6 {\rm GeV}} \right)^{3/2}.
 \ee

\subsection{Reheating}

When the moduli decay, they reheat the universe and initiate the Hot Big Bang, at which point the energy density now lies dominantly in relativistic Standard Model degrees of freedom. We then return to the Standard Cosmology. Although most of the physics of reheating can be captured by the instantaneous decay approximation, in which a cold universe filled only with moduli is instantaneously converted to the thermal Standard Model bath, this is not completely accurate as moduli decay continuously (and so, in particular, the temperature of the Standard Model sector actually decreases during reheating). This is described in e.g. \cite{Kolb:1990vq} and for completeness we also include an analysis of this in Appendix \ref{AppReheating}, in particular \ref{appendixModuliDecay}.

\subsection{Summary of Epochs and their Cosmological Evolutions}

In Table \ref{backgroundsummary} we summarise the wide ranging types of evolution the universe undergoes during the various epochs. At this point, we have still \emph{not} introduced any perturbations into the system: this table summarises the evolution of the background cosmology consisting of the scalar field together with a small initial amount of trace radiation.

\begin{table}
\begin{center}
\centering
\begin{tabular}{ | m{8em} |  m{3em}  | m{4em}| m{10em} | m{5em} | m{6em} | }
\hline
\cellcolor[gray]{0.9}  {\bf Epoch} &  \cellcolor[gray]{0.9} {\bf a(t) }&  \cellcolor[gray]{0.9} {\bf $\eta$ } &   \cellcolor[gray]{0.9} {\bf Range of $\eta$ } & \cellcolor[gray]{0.9}  $\mathcal{H} = \frac{a'(\eta)}{a(\eta)}$ & \cellcolor[gray]{0.9} {\bf PE:KE:Rad}  \\
\hline \hline
Inflation   & $e^{H_{inf}t}$ & $\sim -e^{-Ht}$ & $-\infty < \eta \lesssim 0 \sim \eta_0$ & $H_{inf}$  & $\frac{1}{2}$:$\frac{1}{2}$:$\varepsilon$ {\rm{(at end)}} \\
\hline
Kination &  $t^{1/3}$ &  $\eta \sim t^{2/3}$  & $\eta_0 \lesssim \eta \lesssim \frac{\eta_0}{\varepsilon}$ & $\frac{1}{2 \eta}$ & $\varepsilon^{3/2}:\frac{1}{2}:\frac{1}{2}$ {\rm{(at end)}} \\
\hline
 Radiation domination: PE $\leq$ KE & $t^{1/2}$   &  $\eta \propto t^{1/2}$ & $\frac{\eta_0}{\varepsilon} \lesssim \eta \lesssim \frac{\eta_0}{\varepsilon^{5/4}}$ &  $\frac{1}{\eta}$ & $\varepsilon^{1/2}$:$\varepsilon^{1/2}$:1 {\rm{(at end)}}\\
\hline
Radiation domination: PE $\ge$ KE & $t^{1/2}$   & $\eta \propto t^{1/2}$ & $\frac{\eta_0}{\varepsilon^{5/4}}\lesssim \eta \lesssim \frac{\eta_0}{\varepsilon^{11/8}} $ &  $\frac{1}{\eta}$ & $\frac{1}{2}$:$\varepsilon^{3/4}$:$\frac{1}{2}$ {\rm{(at end)}}\\
\hline
Radiation Tracker & $t^{1/2}$ & $\eta \propto t^{1/2}$ &  $ \frac{\eta_0}{\varepsilon^{11/8}} \lesssim \eta \lesssim m_{\Phi}^{-1/2}$  &  $\frac{1}{\eta}$ & $\frac{3(2-\gamma)\gamma}{2 \lambda^2}$:$\frac{3 \gamma^2}{2 \lambda^2}$:   $1-\frac{3 \gamma}{\lambda^2}$ \\
\hline
Matter domination  & $t^{2/3}$  & $\eta \propto t^{1/3}$ & $m_{\Phi}^{-1/2} \lesssim \eta \lesssim \Gamma_{\Phi}^{-1/2}$ & $\frac{2}{\eta}$ & NA   \\
\hline
Reheating to Standard Model  &  $t^{1/2}$  & $\eta \propto t^{1/2}$  & $\eta \gtrsim \Gamma_{\Phi}^{-1/2}$ &   $\frac{1}{\eta}$ &  0:0:1 {\rm{(at end)}}\\
\hline
\end{tabular}
\end{center}
\caption {A table of the various events and times within the cosmological evolution we study. Note that the tracker ends at $H \sim t^{-1} \sim m_{\Phi}$. In the table we focus on the case of LVS, where the exponential potential is characterised by $\lambda = \sqrt{\frac{27}{2}}$ and also on the case of a radiation fluid (although the relative fractions in the tracker are given for general $\gamma$ and $\lambda$).}
\label{backgroundsummary}
\end{table}

\section{Cosmological perturbations}
\label{PerturbationsSection}

We now analyse the propagation of fluctuations within this modified cosmology. We will focus in particular on fluctuations in the volume field; at the end of the cosmology, this comes to dominate the universe in the moduli epoch. As this decay, it reheats the universe and all perturbations in this field will be transferred into the early universe at the time of reheating. In our analysis of perturbations, we follow Baumann's book \cite{Baumann:2022mni} for conventions and definitions of the various gauges.

\subsection{Perturbations during the Kination Epoch}\label{kination_perturbations}

In the kination epoch, with potential neglected, the dynamics of the universe is solely that of a scalar field coupled to gravity. In this respect it is similar to the inflationary epoch, thus allowing similar formalisms to be used.

We describe the form and evolution of perturbations in both flat space gauge and also in Newtonian gauge, before moving to develop a physical understanding of these perturbations.

\subsubsection*{Flat space gauge}

In flat space gauge, for a single scalar $\phi$ coupled to gravity, the rescaled perturbations $\delta \phi_k(\eta) =  f_k(\eta) /a(\eta) $  obey the Mukhanov-Sasaki equation,
\begin{equation}\label{eq:sm1}
f''_k+\left(k^2-\frac{z''}{z}\right)f_k=0, \quad \text{with} \quad z \equiv \frac{a(\eta)\bar{\phi}'}{\mathcal{H} },
\end{equation}
where $\bar{\phi}$ refers to the background solution (in this case Eq.~(\ref{phikinconf})), with $a(\eta)$ and $\mathcal{H}$ given here in section~\ref{kination-sub-sec}.
During kination, $z = \sqrt{6} \left( \frac{\eta}{\eta_0}\right)^{1/2}$
and so the Mukhanov-Sasaki equation takes the form
\begin{equation}\label{eq:sm2}
f''_k+\left(k^2+\frac{1}{4 \eta^2}\right)f_k=0.
\end{equation}
The most general solution of \eqref{eq:sm2} is in terms of Bessel functions,
\begin{equation}
f_k(\eta) = \sqrt{\eta} \left[ C_1 J_0(k \eta)+C_2 Y_0(k \eta)\right],
\end{equation}
where the constants $C_1$ and $C_2$ are determined by the initial conditions.

As $a(\eta) \propto \eta^{1/2}$ during kination, this implies (redefining the arbitrary constants $C_1$ and $C_2$)
\begin{equation}
\delta \phi_k(\eta) = \left[ C_1 J_0(k \eta)+C_2 Y_0(k \eta)\right],
\end{equation}
and so (during the kination epoch, and again absorbing numerical factors into the arbitrary constants $C_1$ and $C_2$)
\begin{equation}\label{eq:r2}
\mathcal{R}_k(\eta)= \frac{\delta \phi_k (\eta)}{\sqrt{6}M_P} =  \left[ C_1 J_0(k \eta)+C_2 Y_0(k \eta)\right],
\end{equation}
for the gauge-invariant curvature perturbation.

\subsubsection*{Newtonian gauge}

Alternatively and equivalently, one can also use the Newtonian gauge, where $B=E=0$. This will be our preferred gauge for most of the later calculations of cosmological perturbations. 

In this case, for kination the gravitational potential $\Phi$ (which is equivalent to the Bardeen potential in Newtonian gauge) obeys the equation 
\begin{equation}
    \Phi''+\frac{3}{\eta} \Phi'- \nabla^2 \Phi =0.
\end{equation}
 Note this $\Phi$ is distinct from the LVS volume modulus $\Phi$; the difference should be clear from context.
In Fourier space, the solution for the gauge-invariant Bardeen potential $\Phi$ is given by
\begin{equation}\label{Bardeen}
    \Phi_k (\eta) = \frac{C_1}{k \eta} J_1(k \eta)  +  \frac{C_2}{k \eta}Y_1(k \eta).
\end{equation}
In Newtonian gauge, the comoving curvature perturbation is given by
\begin{equation}\label{eq:cpt}
    \mathcal{R}\equiv \Phi- \mathcal{H}v = \Phi- \frac{\mathcal{H}q}{\bar{P}+\bar{\rho}} ,
\end{equation}
with $q \equiv (\bar{P}+\bar{\rho})v $ and $T^0_{\,\,\,i}=\partial_i q$. Furthermore, for a free scalar field the anisotropic stress $\Pi$ vanishes at leading order in perturbation theory,
as 
\begin{equation}
\partial_i \partial^j \Pi =\Pi^i_{\,\,j} =  T^i_{\,\,\,j}- \delta^i_j (\bar{P}+\delta P) = \partial^i \delta \phi \partial_j \delta \phi = \mathcal{O}(\delta \phi^2)
\end{equation}
and so we can identify the two Bardeen variables $\Psi = \Phi$, given that one of the (linearised) Einstein equations implies
\begin{equation}\label{eq:psi}
    \Phi- \Psi = \frac{\Pi a^2}{M_P^2}.
\end{equation}
A further component of the Einstein equations gives 
\begin{equation}\label{eq:q}
    \frac{a^2 q}{2 M_P^2} = -\left( \Phi'+ \mathcal{H} \Phi\right),
\end{equation}
and so $q$ can be substituted back in \eqref{eq:cpt}, (recalling we are in Fourier space) giving for the comoving curvature perturbation\footnote{Using the fact that
$\frac{d Z_1(x)}{dx}= Z_0(x)-\frac{Z_1(x)}{x}$ for $Z=J,Y$.}
\begin{equation}\label{eq:Rkin}
    \mathcal{R}_k=\frac{4}{3} \Phi_k+ \frac{\Phi_k'}{3 \mathcal{H}} = \frac{2 C_1 }{3 } J_0(k \eta)+ \frac{2 C_2}{3 } Y_0(k \eta),
\end{equation}
which indeed gives the same expression as Eq. (\ref{eq:r2}) for the gauge-invariant curvature perturbation (again absorbing constant factors into $C_1$ and $C_2$, which are set by initial conditions).

\subsubsection*{Initial Conditions for Perturbations: Flat Space Gauge}

The above equations determine the propagation of perturbations throughput the whole of the kination epoch (for both sub-horizon $k \eta \gg 1$ and super-horizon $k \eta \ll 1$ scales). However, they do not specify the initial conditions for these perturbations at the start of the kination epoch (which is what will fix the constants $C_1$ and $C_2$).

In this subsection we give a discussion of initial conditions from a perspective in which the volume modulus was \emph{also} the inflaton, and so starts the kination phase with a spectrum of perturbations inherited from the  nearly scale-invariant spectrum of primordial inflationary fluctuations.  

During single-field inflation, the scalar field perturbations of the inflaton field $\varphi$, namely $\delta \varphi_k(\eta) =  f_k(\eta) /a(\eta) $ also evolve according to the Mukhanov-Sasaki equation, which in this case takes the form 
\begin{equation}\label{eq:sm111}
f''_k+\left(k^2-\frac{2}{\eta^2}\right)f_k=0.
\end{equation}
Here, the conformal time coordinate $\eta$ ranges from $-\infty$ (infinite past) to $\eta=0$ ('infinite' future, where inflation ends). Matching of inflationary and kination epochs is done by treating $(0 = \eta_{\textrm{end of inflation}}) \sim (\eta_0 = \eta_{\textrm{start of kination}})$.

Imposing suitable boundary conditions at $\eta=-\infty$, the solutions to \eqref{eq:sm111} are the \emph{Bunch-Davies} mode functions
\begin{equation}
f_k(\eta)=\frac{1}{\sqrt{2k}} \left(1-\frac{i}{k \eta} \right) e^{-i k \eta}.
\end{equation}
At the end of inflation, the inflaton perturbations are therefore of the form
\begin{equation}\label{eq:ip}
 \delta \varphi_k =\lim_{\eta \rightarrow 0}\frac{f_k(\eta)}{a(\eta)}=\frac{i H_{\rm{inf}}}{\sqrt{2k^3}},
\end{equation}
where $H_{inf}$ is the Hubble scale at the time when the perturbations are generated.
In flat space gauge, the comoving curvature perturbation at late times is then given by
\begin{equation}
\mathcal{R}_k= \lim_{\eta \rightarrow 0} \frac{f_k(\eta) }{z(\eta)} =  \left. \frac{H}{\dot{\varphi}}\right \vert_{\rm{inf}}\delta \varphi_k.
\end{equation}
where $\vert_{\rm{inf}}$ implies that this is to be evaluated at horizon crossing during inflation.
 For slow roll-inflation, this can be approximated as
 \begin{equation}\label{eq:r1}
 \mathcal{R}_k= \frac{1}{\sqrt{2 \varepsilon_V}}\frac{\delta \varphi_k}{M_P}, \quad \quad \text{where} \quad 
\varepsilon_V = \frac{M_P^2}{2} \left(\frac{V_{,\varphi}}{V} \right)^2
\end{equation}
is the slow-roll parameter evaluated at the point of horizon crossing. Assuming the transition between inflation and kination to be instantaneous,
we can treat the curvature perturbation as approximately conserved during the transient. 
Since, during kination,
\begin{equation}\label{curvpert}
\mathcal{R}_k(\eta)= \frac{\delta \phi_k (\eta)}{\sqrt{6}M_P},
\end{equation}
we can equate \eqref{eq:ip} to \eqref{eq:r2} (and its first derivatives) at $\eta = \eta_0$ to obtain the evolution of the perturbations throughout the kination phase in the flat space gauge. \begin{equation}\label{eq:pk}
\delta \phi_k (\eta)= i H_{\rm{inf}} \sqrt{\frac{3}{2\varepsilon_V k^3}} \frac{J_0(k \eta)}{J_0(k \eta_0)}.
\end{equation}
The super- and sub-horizon behaviours are approximately given by
\begin{equation}\label{supersubphi}
\delta \phi_k (\eta)=
    \begin{cases}
    i H_{\rm{inf}} \sqrt{\frac{3}{2\varepsilon_V k^3}} \quad \quad \quad \quad \text{super-horizon} \\
    i H_{\rm{inf}} \sqrt{\frac{3 }{  \pi\varepsilon_V \eta }} \frac{\cos(k \eta-\frac{\pi}{4})}{J_0(k \eta_0) k^2} \quad \text{sub-horizon} \\
    \end{cases}
\end{equation}
We note that the quantity $\delta \phi_k$ in \eqref{eq:pk} and (\ref{supersubphi}) is \emph{not} gauge invariant, and is only meaningful in a specific coordinate system (whereas the comoving curvature perturbation \eqref{curvpert} is a gauge invariant quantity; it  equals $\delta \phi_k$ in the flat space gauge, but not more generally). We can then match the initial curvature perturbations at the end of inflation and recover \eqref{eq:pk} by
choosing (for scalar field in flat space gauge)
\begin{equation}
    C_1 = \frac{3i H_{\rm{inf}}}{4 M_P} \frac{1}{J_0(k \eta_0)\sqrt{\varepsilon_V k^3}}\quad \quad \text{and} \quad \quad C_2=0.
\end{equation}

\subsubsection*{Gauge-Invariant Density Perturbations and Interpretation}

It is useful to reformulate the above analysis in terms of gauge-invariant perturbations of the fluid, in this case the kinating scalar. In general, such perturbations are encoded in the gauge invariant quantity
\begin{equation}
   \bar{\rho} \Delta \equiv \delta \bar{\rho}+\bar{\rho}' (v+B),
\end{equation}
where $\bar{\rho}$ is the background energy density of the fluid, and $\delta \rho$ the density contrast.
This is known as the comoving density contrast (as it reduces to the density contrast in the gauge where $B=v=0$). It relates to the Bardeen variable $\Phi$ through the generalised Poisson equation 
\begin{equation}\label{Phi-Delta}
    \nabla^2 \Phi= \frac{\bar{\rho} a^2}{2 M_P^2} \Delta= \frac{3 \mathcal{H}^2}{2} \Delta.
\end{equation}
 During kination (in Fourier space)
\begin{equation}\label{Phi-Delta-k}
 \Delta_k = -\frac{8k^2 \eta^2}{3}\Phi_k,
\end{equation}
with the Bardeen potential given by Eq. (\ref{Bardeen}). This gives the general form, during kination, of the density perturbations:
\begin{equation}
\label{PertKinGeneral}
\Delta_k (\eta) = A  \,k \eta \, J_1(k \eta) + B \, k \eta \,
Y_1(k \eta),
\end{equation}
where $A$ and $B$ are arbitrary constants.
In the case where the scalar $\phi$ has its initial perturbation conditions set by inflation, we obtain\footnote{An alternative derivation of \eqref{eq:ic} goes as follows (in flat space gauge). Since $R_k =- \mathcal{H}(v+B)$, the comoving density contrast for the scalar fluctuations can be rewritten as
\begin{equation}
    \Delta_k^{\phi} = \delta \rho- \bar{\rho}' \frac{\delta \phi}{\phi'}.
\end{equation}
In this coordinate system,
\begin{equation}\label{eq:ed}
    \delta \rho \equiv -\delta T^0_0= \delta \left(-\frac{1}{2}g^{00} \phi'^2 \right) = \frac{1}{a^2}\left(\phi' \delta \phi'-A\phi'^2\right),
\end{equation}
where $A =\Phi$ can either be taken from the formulas above, or obtained from the flat-space gauge Einstein equations as
\begin{equation}
    A= \frac{\varepsilon \mathcal{H}}{\phi'} \delta \phi = \frac{\phi' \delta \phi}{2 \mathcal{H}M_P^2}.
\end{equation}
Putting everything together (in Fourier space),
\begin{equation}\label{eq:ic2}
    \Delta_k = 3 \sqrt{\frac{3}{2}} \frac{M_P}{\eta} \mathcal{H}^2 \delta \phi_k' = -\frac{2 i H_{\rm{inf}} }{ M_P\sqrt{\varepsilon_V k^3}} \frac{k\eta \,J_1(k \eta)}{J_0(k \eta_0) },
\end{equation}
as in \eqref{eq:ic}. }
\begin{equation}\label{eq:ic}
    \Delta_k(\eta) = -\frac{2 i H_{\rm{inf}} }{ M_P \sqrt{\varepsilon_V k^3}} \frac{k\eta \,J_1(k \eta)}{J_0(k \eta_0) }.
\end{equation}

Let us now analyse and understand the form of the perturbations of Eqs.~(\ref{PertKinGeneral}) and (\ref{eq:ic}) a little more detail, and in particular determine how the evolution of the density contrast compares to that of the background.
The perturbations arise from the Fourier modes of a scalar field where the potential is neglected. Given the field is effectively massless during kination, can these perturbations be interpreted as radiation (which would behave as $\rho_{rad} \propto a^{-4}$ within an overall $\bar{\rho}_{total} \propto a^{-6}$ within the kination epoch)?

During this epoch, the modes fall into three categories. There are the modes which are already inside the horizon at the time kination starts (and so always have $k \eta > 1$). Then, there are modes which enter the horizon during the epoch of kination, and so start with $k\eta < 1$ and end with $k \eta > 1$. Finally, there are the modes which always have $k \eta \ll 1$ and so remain outside the horizon throughout the entire duration of kination.

We first consider the large-$k\eta$ ($k \eta \gg 1$) limit corresponding to modes always inside the horizon. In the limit of $k \eta \to \infty$, then $J_1(k \eta) \sim \frac{\sin \left(k \eta \right)}{\sqrt{k \eta} }$ and so the large-$k\eta$ behaviour of $\Delta_k$ is 
\begin{equation}
\label{growthofcontrast}
\Delta_k \sim \sqrt{k \eta} \sin \left( k \eta \right).
\end{equation}
Note that as $\lim_{k \eta \to \infty} Y_1(k \eta) \sim \frac{\cos \left( k \eta \right)}{\sqrt{k \eta}}$ similar behaviour occurs if the $Y_1$ mode is present and their linear combinations would behave 
as $\sqrt{k \eta} \, e^{ik\eta}$ and
$\sqrt{k \eta} \, e^{-ik\eta}$).

In the comoving gauge, $\Delta_k$ reduces to the density contrast $\frac{\delta \rho_k}{\bar{\rho}}$. If the Fourier modes describe radiation, we might expect the energy associated to them to grow as $\rho \sim a^{-4}$ within the $\bar{\rho} \sim a^{-6}$ kination background -- and so we would expect $\Delta_k \sim \frac{\delta \rho}{\bar{\rho}} \sim a^2 \sim \eta$. 

Clearly, this is not the behaviour of Eq. (\ref{growthofcontrast}), whose amplitude only grows as $\sqrt{k \eta}$ and also contains an oscillatory sinusoidal behaviour. How can we explain this result, given that the mass of the scalar field is far smaller than the Hubble scale, and so we can treat it as effectively massless? For the high-momentum modes well inside the Hubble radius, it is reasonable to expect the field to behave as a massless scalar field with excited Fourier modes. 

Indeed, this is how the field behaves, but the Fourier modes do not simply behave as radiation. We can understand the issue by temporarily dropping gravity and considering a single canonically normalised massless scalar field in Minkowski space,
\begin{equation}
\int d^4 x \, \partial_{\mu} \phi \partial^{\mu} \phi.
\end{equation}
The equation of motion $\partial^{\mu} \partial_{\mu} \phi = 0$ is solved by (note the explicit inclusion of the $vt$ term)
\be
\label{MinkVel}
\phi(x,t) = \phi_0 + vt + \int \frac{d^3 k}{(2 \pi)^3} \frac{1}{\sqrt{2 \omega_k}} \left( \alpha_k e^{i(kx - \omega t)} + \alpha^{*}_k e^{-i(kx - \omega t)} \right).
\ee
In the presence of a significant velocity,
the kinetic energy of the scalar field is
$$
\dot{\phi}^2 = v^2 + 2v \int \frac{d^3 k}{(2 \pi)^3} \frac{1}{\sqrt{2 \omega_k}} \left( -i \omega \alpha_k e^{i(kx - \omega t)} + i \omega \alpha^{*}_k e^{-i(kx - \omega t)} \right) + \mathcal{O}\left( \alpha_k^2 \right).
$$
As well as the dominant kinetic term, the next term (which is first-order in the mode amplitude $\alpha_k$) involves a sinusoidal cross-coupling between the velocity and the Fourier modes. 

This is consistent with what was observed in Eq. (\ref{growthofcontrast}). In the case of dynamical gravity, the velocity squared represents the dominant (kination) contribution to the energy density and the leading perturbation, linear in the scalar field perturbation, contributes to $\delta \rho$ with a sinusoidal oscillatory behaviour.

We can obtain a quantitative understanding of this by considering scalar fields inside a kinating background, where the extension of Eq. (\ref{MinkVel}) gives (as the solution to $\phi^{''} + 2\mathcal{H} \phi' = -k^2 \phi$ for high-momentum modes)
\be
\phi_k(\eta) = \phi_0 + M_P \left( \frac{2}{3} \right)^{3/2} \ln \eta + \sum_k \left( A_k \frac{e^{ik \eta}}{\sqrt{k \eta}} + A_k^{*} \frac{e^{-ik\eta}}{\sqrt{k \eta}} \right).
\ee
The energy density of a scalar field is (although note we do not use the spatially dependent $(\nabla \phi)^2$ term) as we focus on the $\eta$ dependence
\be
\label{EDen}
\frac{\dot{\phi}^2}{2} + \frac{1}{a^2} \frac{\left( \nabla \phi \right)^2}{2} \equiv \frac{1}{a^2} \frac{\phi'(\eta)^2}{2} + \frac{1}{a^2}\frac{\left( \nabla \phi \right)^2 }{2}.
\ee
As $\eta \propto a^2$ and
\be
\label{PhiD}
\phi'(\eta) = \frac{M_P}{\eta} \left( \frac{2}{3} \right)^{3/2} + 
 \sum_k \left( ik A_k \frac{e^{ik \eta}}{\sqrt{k \eta}} - ik A_k^{*} \frac{e^{-ik\eta}}{\sqrt{k \eta}} \right),
\ee
(where we have focused on subhorizon terms with $k \eta \gg 1$ and so have neglected terms suppressed by $(k \eta)^{-1}$), we see that while the leading term in $\frac{1}{a^2} \frac{\phi'(\eta)^2}{2}$ indeed gives the expected kination behaviour $\rho \propto a^{-6}$, the sinsuoidal cross-term between the velocity term and the Fourier modes gives an energy density
\be
\rho_{{\rm cross \, term}} \propto \frac{\sin \left( k \eta + \theta \right)}{\eta^{5/2}},
\ee
for some phase $\theta$, and so 
\be
\frac{\rho_{\rm{cross \, term}}}{\rho_{KE}} \sim \sqrt{k \eta} \sin \left( k \eta + \theta \right),
\ee
consistent with Eq. (\ref{growthofcontrast}). It also follows from Eqs. (\ref{EDen}) and (\ref{PhiD}) that the energy density associated purely to the Fourier modes would scale as $\rho_{Fourier} \propto a^{-4}$, consistent with the expected behaviour of radiation, although note that this term is second-order in the scalar field perturbation and so is not visible within first-order cosmological perturbation theory. Note this is an advantage of the microscopic understanding of kination in terms of a scalar field rather than simply as a cosmological fluid with $w=1$ (i.e. $\gamma=2$), as this understanding provides an insight into dynamics at higher order in perturbation theory.

This gives us a full quantitative understanding of why the self-perturbations of the massless scalar field do not appear to correspond to radiation: in the presence of a field velocity, the leading (linear) contribution of these perturbations to the energy density in the sub-horizon limit lies in their cross-coupling to the dominant velocity term, consistent with the analogous behaviour observed in the flat space limit.

Another way of rephrasing the above is as follows. The kinetic energies of the background field and the scalar perturbations scale as
\begin{aleq}\label{kination_energies}
    \rho_\phi \sim a^{-2} \phi'^2 \sim a^{-6}, \quad \rho_{\delta \phi} \sim a^{-2} \left(\delta \phi'\right)^2 \sim a^{-4},
\end{aleq}
where the kinetic energy of the scalar perturbations shows the expected radiation-like redshift. The total perturbation in the energy of the scalar field on the other hand is given by
\begin{aleq}
    \delta \rho_\phi = a^{-2}\left( \phi' \delta \phi' + \Phi \phi'^2 \right).
\end{aleq}
Each of these contributions scale as follows
\begin{aleq}
    a^{-2}\left( \phi' \delta \phi'\right) \sim a^{-5}, \quad a^{-2}\left(\Phi \phi'^2 \right)\sim a^{-9}.
\end{aleq}
The first contribution is the dominant one, and so the final 
behaviour,
\begin{aleq}
    \rho_{\delta \phi} \sim a^{-2}\left( \phi' \delta \phi'\right) \sim \sqrt{\rho_\phi \cdot \rho_{\delta \phi}} \sim a^{-5},
\end{aleq}
averages the the scalings of the background and the perturbations in \eqref{kination_energies}.

We next consider the super-horizon limit of the self-perturbations. Taking $k \eta \ll 1$, then $J_1(k \eta) \propto k \eta$. On super-horizon scales it therefore follows that
\be 
\label{ogyu}
\Delta_k \propto \left( k \eta \right)^2 \sim a^4,
\ee
resulting in a fast growth of the density contrast for superhorizon modes (which again does not correspond to radiation like behaviour). We can also understand this behaviour from the perspective of the underlying scalar field dynamics. Each $\Delta_k$ term is a Fourier mode and so also associated to spatial dependence of the scalar field. The gradient term in the scalar field energy of Eq. (\ref{EDen}), $\frac{\left( \nabla \phi \right)^2}{2 a^2}$, behaves as $a^{-2}$. During the kination epoch, spatially separated Hubble patches with misaligned vevs for the scalar fields, $\phi_{{\rm patch} \, 1} - \phi_{{\rm patch} \, 2} \neq 0$ maintain this misalignment. This super-horizon gradient energy behaves as $a^{-2}$ and so grows relative to the kination background as $a^4$, precisely as observed in Eq. (\ref{ogyu}), as long as the modes remain outside the horizon.

Note this differs from the behaviour of scalar fields oscillating about a quadratic potential (i.e. matter), where the overall amplitude of the field is constantly decreasing and so any spatial misalignment reduces with time.

It is an interesting question (but one beyond the scope of this work) to study the endpoint of a pure kination epoch (and also a pure scalar field on an exponential potential), if the self-perturbations of the scalar field are allowed to grow until they become non-linear.
As such a endpoint would involve $\mathcal{O}(1)$ density contrasts, it can only be studied numerically and requires going beyond the analytic treatment here.

\subsection{Perturbations of other fields during the Kination Epoch}

During the kination epoch, the volume scalar dominates the evolution. However, in any string model there are many other fields. These include massless fields (for example, the volume axion which partners the volume mode) and also other massive moduli such as the complex structure moduli. One of the distinctive aspect of \emph{string} cosmologies is that they fix the masses and couplings of such fields as functions of the dominant fields. Given that the presence of such fields is also an unavoidable feature of string cosmologies, it is important to understand their behaviour.

Naively, one would expects the energy densities of perturbations associated to massless and massive fields to redshift as $\rho \sim a^{-4}$ (radiation) and $\rho \sim a^{-3}$ (matter) respectively. However, in string theory things are more subtle: the kinetic terms and masses for such fields depend on the volume, and so are time-dependent throughout the kination epoch, which will alter expectations based on having canonical kinetic energy terms.

\subsubsection{The Volume Axion}

One additional field that will always be present in LVS cosmologies is the volume axion, the axionic partner of the volume modulus. In presence of the volume axion, the equations of motion \eqref{eq:xy2} need to be modified to account for its background evolution, which is inextricably linked to that of the saxion. We will assume that the initial velocity of the axion (denoted by $\chi$ to avoid confusion with the scale factor $a(t)$) initially satisifies $\dot{\chi}(t) \ll \dot{\phi}(t)$, being $\chi$ an approximately flat direction of the potential. In that case, its Lagrangian is
\be
L = \int d^4 x  \sqrt{-g} \frac{1}{\mc{V}^{4/3}} \partial_{\mu} \chi \partial^{\mu} \chi
\ee
(the non-trivial kinetic term follows straightforwardly from the standard K\"ahler potential $K = - 3 \ln \left( T + \bar{T} \right)$). 

We assume that we can treat the volume evolution as a `slow' background process against which we study `fast' modes of the axion. In particular, using Eq.~(\ref{aandt}) and Eq.~(\ref{volevol}), we have $\mc{V} \sim t$ and $a(t) \sim t^{1/3}$ leading to an action of the form
\be
L = \int d^4 x   \frac{1}{t^{1/3}}\partial_{\mu} \chi \partial^{\mu} \chi.
\ee
The resulting equation of motion is $\partial_{\mu} \left( t^{-1/3} \partial^{\mu} \chi \right) = 0$, giving
\be
\ddot{\chi} - \frac{1}{3t} \dot{\chi} + \frac{\nabla^2 \chi}{t^{2/3}} = 0,
\ee
to which the zero mode solutions are (in accordance with \cite{Revello:2023hro})
\be
\label{zeromodeaxion}
\chi(t) = A + B t^{4/3}.
\ee
When $\dot{\chi}(t) \sim \dot{\phi(t)}$, the above approximation is no longer valid, as the axion kinetic energy is then sufficient to backreact on the volume modulus. However, it was shown in \cite{Revello:2023hro} that for the LVS potential, with $\lambda= \sqrt{\frac{27}{2}}$, the late time solutions of the full, non-linear equations of motion are still the tracker solutions described in Section \ref{ssc:rt} (for background radiation and matter respectively), with $\dot{\chi}(t)=0$.

To study perturbations in $\chi$, the conformal time equation is (using $\eta \sim a^2$ and $\mc{V} \sim \eta^{3/2}$ during kination)
$\partial_{\mu} \left( \frac{1}{\eta} \partial^{\mu} \chi \right) = 0$, giving (in Fourier space),
\be
 \chi''_k - \frac{1}{\eta}  \chi'_k + k^2 \chi_k =0.
\ee
The corresponding mode solutions are
\be
\chi_k = A \left( k \eta \right) J_1( k \eta) + 
B \left( k \eta \right) Y_1 (k \eta).
\ee
which, consistent with Eq. (\ref{zeromodeaxion}), reduces to
\be
\chi = A + B \eta^2
\ee
in the $k \to 0$ superhorizon limit. For subhorizon modes with $k \eta \gg 1$, 
\be
\chi_k \sim \alpha \sqrt{k \eta} \sin \left( k \eta + \beta \right),
\ee
with arbitrary integration constants $\alpha$ and $\beta$. The energy density associated to the scalar field is (note the non-trivial volume terms present),
\be 
\rho_{\chi} = \frac{1}{\mc{V}^{4/3}} \frac{\dot{\chi}^2}{2} + \frac{1}{\mc{V}^{4/3}} \frac{(\nabla \chi)^2}{2 a^2}.
\ee
Using $a \sim \eta^{1/2}$, $\mc{V} \sim \eta^{3/2}$ and $k \eta \gg 1$, the mode energy density behaves as
\be
\rho_{\chi_k} \sim \frac{\vert \alpha \vert^2}{\eta^2} \propto a^{-4},
\ee
thus confirming that, despite the time-dependent kinetic term, the energy associated to the volume axion indeed redshifts as radiation. This is particularly relevant for the cosmology, as it shows that there is a new source of radiation present which has not been put in by hand, but always arises from the fluctuations of the volume axion. This has important consequences for the forthcoming cosmology of the system as such radiation provides a means to reach a tracker solution.

\subsubsection{Massive Moduli and other Massive Fields}

As well as the modes of the volume axion, there are also other massive modes present in the compactification. In the case of LVS, the simplest examples are those corresponding to all the other moduli (i.e. the residual K\"ahler moduli and complex structure moduli in addition to the rolling volume modulus). The masses of such fields are time-dependent; for example, the complex structure moduli $U_i$ have masses $m_U(t) \sim \frac{M_P}{\mc{V}}$ (and so $ m_U(t) \sim \frac{M_P}{(t/t_0)}$ during the kination epoch).

This suggests that energy density stored in such massive modes will not redshift as the conventional $\rho_{m} \sim n(t) m_0 \sim a^{-3}$, where $n(t)$ is the number density and $m_0$ the time-independent mass, but will rather reflect the time-dependent masses (and so behave as e.g. $\rho_{U}(t) \sim n(t) m_U(t) \sim a^{-6}$ for complex structure moduli).

We now verify this through study of the equations for massive scalar fields in a kinating background. The Lagrangian is 
\be
L = \int d^4 x \sqrt{-g} \left( \frac{1}{2} \partial_{\mu} U \partial^{\mu} U -  V(U) \right)
\ee
As the K\"ahler potential for Calabi-Yau compactifications factorises as $K(U, \bar{U}) + K(T + \bar{T})$, the kinetic terms for the complex structure moduli have no time dependence. Focussing on the leading mass term, the Lagrangian is then 
\be
L = \int d^4 x \sqrt{-g} \left( \frac{1}{2} \partial_{\mu} U \partial^{\mu} U -  \frac{\mu_U^2}{ 2t^2} U^2 \right),
\ee
where for simplicity we treat $U$ as a real scalar field and we use $m_U \sim \frac{\mu_U}{t}$ for some fixed constant $\mu_U$.

The equation of motion for $U$ (restricting to the zero mode) is
\be
\ddot{U} + 3 H \dot{U} = - \frac{\mu_U^2}{t^2} U,
\ee
which is solved by
\be
U(t) = \tilde{\alpha} \cos \left( \mu_U \ln t + \tilde{\beta} \right) 
\ee
where $\tilde{\alpha}$ and $\tilde{\beta}$ are arbitrary integration constants.
The energy density in this scalar field is then
\be
\frac{\dot{U}^2}{2} + \frac{\mu_U^2 U^2}{2 t^2} \propto \frac{1}{t^2} = \frac{1}{a^6},
\ee
consistent with earlier arguments involving particle number densities.

As this energy density of $a^{-6}$ is a similar scaling to kination, it does not grow relative to the kinating background. Furthermore, we also expect little initial energy density in such modes in the post-inflationary epoch, and therefore we do not consider them further.

\subsection{Perturbations including a Radiation (Matter) Background}\label{addingrad}

The analysis of cosmology perturbations becomes more involved as soon as background radiation (or matter) ``catches up'' with the kinating scalar and becomes relevant. We first describe this for a radiation fluid but also give the formulae for a matter background in section \ref{uffjj}.

As radiation catches up, not only does the background evolution first change from purely kination type behaviour to radiation domination and then tracker-like behaviour as described  in section~\ref{kinradtran}, but we now have to also study the coupled evolution for the perturbations in the radiation, the kinating scalar field, as well as in the metric. We will first write down the general equations that describe such a system, and then specialise to the radiation domination and tracker epochs.  In this section, we will exclusively make use of the Newtonian gauge $B=E=0$. As a reminder, the metric is of the form
\begin{equation}
    ds^2=  - a^2(\eta)(1+2 \Psi)d\eta^2+ a^2(\eta)(1-2 \Phi)(dx_1^2+dx_2^2+dx_3^2).
\end{equation}
Given the extensive appearance of the gravitational potential $\Phi$ in this section, here we will exclusively use lower-case $\phi$ for the scalar field, even when considering a specialisation to the LVS potential.

We will describe two separate formalisms (giving the same answers) for the computations applying for both the kination-radiation transition and also for the tracker epoch, one the formalism based on \cite{Baumann:2022mni} and the other (described in appendix \ref{entropyappendix}) an entropy formalism based on \cite{Bartolo:2003ad}.

In the tracker epoch, the presence of a potential for $\phi$ cannot be neglected, and is in fact needed to ensure the precise balancing of all energy sources. In particular, the energy densities stored in radiation, kinetic and potential energy are all of the same order of magnitude. For the scalar field, the general expression for the density fluctuations is now
\begin{equation}\label{eq:rhophi}
    \delta \rho \equiv -\delta T^0_0= \delta \left(-\frac{1}{2}g^{00} \phi'^2 +V(\phi) \right) = \frac{1}{a^2}\left(\phi' \delta \phi'-\Phi \phi'^2\right) +\frac{dV(\phi)}{d \phi} \delta \phi,
\end{equation}
whereas the pressure perturbations are 
\begin{equation}\label{eq:Pphi}
    \delta P \equiv \delta T^i_{\,\,i} = \delta \left(-\frac{1}{2}g^{00} \phi'^2 -V(\phi) \right) = \frac{1}{a^2}\left(\phi' \delta \phi'-\Phi \phi'^2\right) -\frac{dV(\phi)}{d \phi} \delta \phi.
\end{equation}
We can formally treat the potential and kinetic energy contributions to be two separate fluids, although there is energy transfer between the two and only their total energy-momentum density is conserved. They are characterised by a constant equation of state (with $w_p=-1$ and $w_k =1$ respectively), while perturbations in radiation have $w_r=\frac{1}{3}$.
 
 Both fluids do not have anisotropic stress, so $\Pi =0$ and $\Psi = \Phi$ from the linearised Einstein equations in Newtonian gauge. The fluctuations in the field $\delta \phi$ obey the perturbed Klein-Gordon equation,
\begin{equation}\label{eq:c1}
    \delta \phi''+ 2 \mathcal{H}\delta \phi'-\nabla^2 \delta \phi+\frac{d^2V(\phi)}{d \phi^2} a^2 \delta \phi-4 \phi' \Phi'+2\frac{dV(\phi)}{d \phi} a^2 \Phi =0,
\end{equation}
which is implied by the conservation of the stress energy tensor for a scalar field. Since fluctuations in the scalar field and radiation are not directly coupled, the latter evolve as
\begin{equation}\label{eq:c2}
    \delta_r^{\prime \prime}-\frac{1}{3} \nabla^2 \delta_r-\frac{4}{3} \nabla^2 \Phi-4 \Phi^{\prime \prime} = 0,
\end{equation}
where $\delta_r= \frac{\delta \rho_r}{\rho_r}$, $\rho_r$ being the background energy density in radiation, and $\delta \rho_r$ the fluctuations in that energy density.
They are complemented by the $G^0_{\,\,0}$ and $G^i_{\,\,i}$ Einstein equations, given respectively by
\begin{equation}\label{eq:e1}
    \nabla^2 \Phi-3 \mathcal{H}\left(\Phi^{\prime}+\mathcal{H} \Phi\right)=\frac{a^2}{2 M_P^2} \left( \delta \rho_k+ \delta \rho_p+\delta \rho_r \right)
\end{equation}
\begin{equation}\label{eq:e2}
    \Phi^{\prime \prime}+3 \mathcal{H} \Phi^{\prime}+\left(2 \mathcal{H}^{\prime}+\mathcal{H}^2\right) \Phi= \frac{a^2}{2 M_P^2} \left( \delta \rho_k - \delta \rho_p+ \frac{1}{3}\delta \rho_r\right).
\end{equation}
Together, Eqs \eqref{eq:c1}-\eqref{eq:e2} are a closed system of four differential equations in three unknowns ($\Phi, \delta \phi, \delta \rho_r$), and seemingly over-determined. However, they are not all independent, since the Einstein equations imply the conservation of total energy-density; together with either one of \eqref{eq:c1} or \eqref{eq:c2} they imply the other of the two.

For an exponential potential, $V(\phi)= V_0 e^{- \lambda \phi/M_P}$, \eqref{eq:c1} reduces to (in Fourier space)
\begin{equation}\label{eq:m1}
 \delta \phi_k''+ 2 \mathcal{H}\delta \phi_k'+k^2 \delta \phi_k+ \lambda^2 a^2 \frac{V(\phi)}{M_P^2} \delta \phi_k = 4 \phi' \Phi_k'+2 \lambda  a^2 \frac{V(\phi)}{M_P}\Phi_k,   
\end{equation}
whereas a linear combination of \eqref{eq:e1} and \eqref{eq:e2} gives 
\begin{equation}\label{eq:m2}
    \Phi_k''+ 4 \mathcal{H}\Phi_k'+ 2\left( \mathcal{H}'+\mathcal{H}^2 + \frac{k^2}{6}\right) \Phi_k= \frac{1}{3 M_P^2} \left( \phi' \delta \phi_k' -\Phi_k \phi'^2+2 \lambda a^2 \frac{V(\phi)}{M_P}  \delta \phi_k \right).
\end{equation}
These two equations are then a closed system for the perturbations $\left( \delta \phi_k, \Phi_k\right)$, which only depend on the background evolution for $\phi(\eta)$. In the following, we will look at their solution during the radiation dominated and tracker epochs respectively. In what follows for convenience we drop the $k$ subscripts in the Fourier mode equations, unless there is a need to reinstate them, for example in Appendix ~\ref{pert-vol-axion}.

\subsubsection{Transition from kination to radiation domination}\label{ssc:ktr}
We now give an analytic solution of the equations for the evolution of the perturbations during the kination-radiation epoch (neglecting the potential). As before, we assume $\varepsilon \ll 1$, and expand the coefficients of each $\eta^{\alpha}$ term to lowest order in $\varepsilon$. In this epoch, it is convenient to define the ratio of the energy density in radiation compared to that in the scalar field, through the variable $y$ \footnote{The variable $y$ here should not be confused with the one defined in Eq.~\eqref{eq:xyv}.} 
\begin{aleq}\label{yepsilona}
    y = \dfrac{\rho_r}{\rho_\phi} =\dfrac{a^2}{a^2_{eq}} = \varepsilon a^2.
\end{aleq}
From the definition of $y$ in Eq.~(\ref{yepsilona}),

it follows that
\begin{equation}
    \frac{dy}{d \eta} = 2 y \mathcal{H} \quad \quad \quad \text{and} \quad  \quad \quad \mathcal{H}=\frac{\varepsilon }{2 \eta_0} \frac{\sqrt{1+y}}{y},
\end{equation}
where we  have taken $\mathcal{H}$ from Eq.~\eqref{eq:H0tr}. In particular, by differentiating the latter with respect to conformal time, one obtains
\begin{eqnarray}
    \mathcal{H}' = -\mathcal{H}^2 \left(1+\frac{y}{1+y} \right).
\end{eqnarray}
By the chain rule, the derivatives of a generic function $F(\eta)$ with respect to $\eta$ can be expressed as
\begin{equation}\label{eq:371}
    F'(\eta) = 2 y \mathcal{H} \frac{\mathrm{d}F}{\mathrm{d} y} \quad \quad  F''(\eta) = 4y^2 \mathcal{H}^2 \frac{\mathrm{d}^2 F}{\mathrm{d}y^2}+\frac{2y^2 \mathcal{H}^2}{1+y} \frac{\mathrm{d}F}{\mathrm{d} y}.
\end{equation}
Then, the system Eqs.~\eqref{eq:m1}-\eqref{eq:m2} can be recast as 
\begin{equation}\label{eq:krad}
    \begin{cases}
     2y \frac{\mathrm{d}^2 \Phi}{\mathrm{d}y^2}  +\left( 4+ \frac{y}{1+y}\right) \frac{\mathrm{d}\Phi}{\mathrm{d} y} + \frac{2 \eta_0^2 k^2}{3 \varepsilon^2} \frac{y}{1+y} \Phi = \frac{\sqrt{6}}{3 M_P\sqrt{1+y} } \frac{\mathrm{d}\delta \phi}{\mathrm{d} y} \\[0.5cm]
     2 y \frac{\mathrm{d}^2 \delta \phi}{\mathrm{d}y^2}  +\left( 2+ \frac{y}{1+y}\right) \frac{\mathrm{d} \delta \phi}{\mathrm{d} y}+ \frac{2 \eta_0^2 k^2}{\varepsilon^2}  \frac{y}{1+y} \delta \phi = \frac{4\sqrt{6} M_P}{ \sqrt{1+y}}  \frac{\mathrm{d}\Phi}{\mathrm{d} y}.
    \end{cases}
\end{equation}
Thanks to an accidental cancellation (and to the fact that $V(\phi)\equiv 0$), the only linear terms in $\Phi$ and $\delta \phi$ are those coming from their spatial gradient, proportional to $k^2$.

\myparagraph{Long wavelength modes}
For long wavelenghts $k \eta \ll 1$, the system reduces to
\begin{equation}
    \begin{cases}
     2y \frac{\mathrm{d}^2 \Phi}{\mathrm{d}y^2}  +\left( 4+ \frac{y}{1+y}\right) \frac{\mathrm{d}\Phi}{\mathrm{d} y} = \frac{\sqrt{6}}{3 M_P \sqrt{1+y}} \frac{\mathrm{d}\delta \phi}{\mathrm{d} y} \\[0.5cm]
     2 y \frac{\mathrm{d}^2 \delta \phi}{\mathrm{d}y^2}  +\left( 2+ \frac{y}{1+y}\right) \frac{\mathrm{d} \delta \phi}{\mathrm{d} y} = \frac{4\sqrt{6}M_P}{ \sqrt{1+y}} \frac{\mathrm{d}\Phi}{\mathrm{d} y}.
    \end{cases}
\end{equation}
Notice how, as a consequence of the accidental cancellations in Eq.~\eqref{eq:krad}, it reduces to a first order system in terms of $\frac{\mathrm{d} \Phi}{\mathrm{d} y}$ and $\frac{\mathrm{d} \delta \phi}{\mathrm{d} y}$. In particular, it can be turned into a second order ODE for $ q(y) \equiv \frac{\mathrm{d} \Phi}{\mathrm{d} y} $, namely
\begin{equation}\label{eq:qd}
 2 y (1 + y) \frac{\mathrm{d}^2 q}{\mathrm{d} y}  + (8 + 11 y) \frac{\mathrm{d} q}{\mathrm{d} y}+10 q=0.
\end{equation}
The general solution to Eq.~\eqref{eq:qd} is
\begin{equation}\label{eq:qs}
\begin{aligned}
q(y) =& -\frac{9 c_1 \left(y \left(\sqrt{y+1}-3\right)+4 \left(\sqrt{y+1}-1\right)\right)}{24 y^3 \sqrt{y+1}}\\ & - \frac{8 c_3 \left(y\sqrt{y+1} -3 y+4 \sqrt{y+1}-4\right)+9 c_4 (3 y+4)}{24 y^3 \sqrt{y+1}},
\end{aligned}
\end{equation}
where the integration constants (only two of them are independent) have been chosen to match with Eq.~\eqref{eq:phigamma} with $\gamma=\frac{4}{3}$. Indeed, Eq.~\eqref{eq:qs} can be integrated as
\begin{equation}
    \Phi(y) =  \frac{8 c_3 \left(2 y^2+y-2 \sqrt{y+1}+2\right)+9 c_1 \left(y-2 \sqrt{y+1}+2\right)+18 c_4 \sqrt{y+1}}{24 y^2}.
\end{equation}
For an adiabatic solution that remains finite as $y\rightarrow0$ (kination domination), we need
\begin{aleq}
    c_1=c_4=0
\end{aleq}
yielding
\begin{aleq}
    \frac{\Phi( y\rightarrow\infty)}{\Phi( y\rightarrow 0)} = \frac{8}{9},
\end{aleq}
signifying that the gravitational potential drops by $8/9$ during the kination-radiation transition. In Appendix \ref{Largescalepotential}, we derive a general expression for the large-scale evolution of the gravitational potential during a transition involving arbitrary cosmic fluids characterized by parameters  $\gamma_1$ and $\gamma_2$, along with the general magnitude by which the potential decreases. We also have
\begin{equation}
    \mathcal{R}= \Phi- \frac{y+1}{2y+3} \left( \Phi + 2 y \frac{\mathrm{d} \Phi}{\mathrm{d} y}\right)  = \frac{3 c_1}{8 (2 y+3)}+c_3.
\end{equation}
One can also solve for $\delta \phi$ to obtain\footnote{The constant $c_2$ is different from the one defined in the Appendix, whereas $c_1,c_3,c_4$ are the same.}
\begin{equation}
    \delta \phi (y) = \frac{\left(9 c_1+8 c_3\right) (y-2) \sqrt{y+1}+2 \left(9 c_1+8 c_3-9 c_4\right)}{4 \sqrt{6} y^2}+c_2,
\end{equation}
solving the system completely. Finally,
from Eq.~\eqref{eq:rhophi}
\begin{equation}
    \delta_{\phi} = 2\frac{\delta \phi'}{\phi'}- 2\Phi= \frac{-3 c_1 \left(y^2-3 y+6 \sqrt{y+1}-6\right)+8 c_3 \left(-y^2+y-2 \sqrt{y+1}+2\right)+18 c_4 \sqrt{y+1}}{4 y^2}
\end{equation}
and
\begin{equation}
\begin{aligned}
   \delta_{r} & = -\frac{\delta_{\phi}}{y}- \frac{2(y+1)}{y} \left( \Phi + 2y \frac{\mathrm{d} \Phi}{\mathrm{d} y} \right)\\ &=
   \frac{8 c_3 \left(-y^2+y-2 \sqrt{y+1}+2\right)+9 c_1 \left(y-2 \sqrt{y+1}+2\right)+18 c_4 \sqrt{y+1}}{6 y^2}.
   \end{aligned}
\end{equation}

\myparagraph{Short wavelength modes}
For short (arbitrary) wavelengths, one needs to solve the full system Eq.~\eqref{eq:krad}, for which we could not find any analytic solutions unless $y \gg 1$. In that case, the system reduces to 
\begin{equation}
\label{ftyg}
    \begin{cases}
     2y \frac{\mathrm{d}^2 \Phi}{\mathrm{d}y^2}  + 5 \frac{\mathrm{d}\Phi}{\mathrm{d} y}+ \frac{2 \eta_0^2 k^2}{3 \varepsilon^2} \Phi = \frac{\sqrt{6}}{3 M_P \sqrt{y}} \frac{\mathrm{d}\delta \phi}{\mathrm{d} y} \\[0.5cm]
     2y \frac{\mathrm{d}^2 \delta \phi}{\mathrm{d}y^2}  + 3\frac{\mathrm{d} \delta \phi}{\mathrm{d} y} +  \frac{2 \eta_0^2 k^2}{ \varepsilon^2}  \phi = \frac{4\sqrt{6} M_P}{ \sqrt{y}}  \frac{\mathrm{d}\Phi}{\mathrm{d} y}.
    \end{cases}
\end{equation}

To obtain the asymptotic form of the solution, we can assume  
\begin{equation}
   \frac{\Phi(x)}{\delta \phi(x)} \sim \mathcal{O}\left( x^{-b}\right) \quad \quad \quad \text{with} \quad \quad b >0,
\end{equation} 
and verify it a posteriori. If the above holds, for large $y$ one can neglect the source term in the second equation of \eqref{ftyg}, and solve it as
\begin{equation}\label{eq:phiz}
    \delta \phi = \frac{c_1}{\sqrt{y}} \cos \left(\frac{2 \eta_0 k\sqrt{y}}{\varepsilon} \right) + \frac{c_2}{\sqrt{y}} \sin \left(\frac{2 \eta_0 k\sqrt{y}}{\varepsilon} \right),
\end{equation}
where the $c_i$'s will always denote arbitrary constants. Substituting \eqref{eq:phiz} back into the first equation in \eqref{ftyg}, the latter can be solved  for large $y$, as
\begin{equation}
    \Phi = \frac{(\sqrt{6} \pi c_2+c_3)}{y} 
    \cos \left(\frac{2 \eta_0 k\sqrt{y}}{\sqrt{3}\varepsilon} \right)+\frac{c_4}{y}  \cos \left(\frac{2 \eta_0 k\sqrt{y}}{\sqrt{3}\varepsilon} \right)+ \mathcal{O}\left(\frac{1}{y^2} \right).
\end{equation}
Since $\sqrt{y} \simeq \frac{\varepsilon \eta}{2 \eta_0} $ for late times, the asymptotic behaviour reduces to
\begin{equation}
    \Phi \simeq \frac{c_3'}{a^2(\eta)} \cos \left( \frac{k \eta}{\sqrt{3}} \right)+ \frac{c_4'}{a^2(\eta)} \sin \left( \frac{k \eta}{\sqrt{3}} \right)
\end{equation}
and
\begin{equation}
     \delta \phi  \simeq \frac{c_1'}{a(\eta)} \cos \left( k \eta \right)+ \frac{c_2'}{a(\eta)} \sin \left( k \eta \right).
\end{equation}
As expected, this matches with the behaviour of perturbations during the radiation era \cite{Baumann:2022mni}. Since the perturbations in radiation are solely determined by the evolution of $\Phi$ (from \eqref{eq:c2}), they also behave as during an epoch of radiation domination, namely
\begin{equation}
    \Delta_r \simeq c_5 \cos \left( \frac{k \eta}{\sqrt{3}} \right)+  c_6 \sin \left( \frac{k \eta}{\sqrt{3}} \right).
\end{equation}

\subsubsection{Transition from kination to matter domination}
\label{uffjj}
A similar computation for the perturbations can be carried out for the transition from kination to matter domination. In analogy with the previous subsection, we introduce the variable
\begin{aleq}\label{ymat}
    y = \dfrac{\rho_m}{\rho_\phi} = \dfrac{a^3}{a^3_{eq}}= \varepsilon a^3,
\end{aleq}
which satisfies\footnote{The second equality is just a rewriting of the first Friedmann equation.}
\begin{equation}
    \frac{dy}{d \eta} = 3 y \mathcal{H} \quad \quad \quad \text{and} \quad  \quad \quad \mathcal{H}=\frac{\varepsilon^{2/3} }{2 \eta_0} \frac{\sqrt{1+y}}{y^{2/3}}.
\end{equation}
Following the same prescription as \eqref{eq:371}, we see that
\begin{equation}
    \mathcal{H}'=-\mathcal{H}^2 \frac{y+4}{2y+2},
\end{equation}
and derivatives of a generic function $G(\eta)$ with respect to $\eta$
can be expressed as
\begin{equation}
     G'(\eta) = 3 y \mathcal{H} \frac{\mathrm{d}G}{\mathrm{d} y} \quad \quad  G''(\eta) =9 \mathcal{H}^2y^2 \frac{\mathrm{d}^2 G}{\mathrm{d} y^2}+3y \mathcal{H}^2 \frac{5y+2}{2y+2} \frac{\mathrm{d}G}{\mathrm{d} y}.
\end{equation}
For matter perturbations, equation \eqref{eq:e2} is replaced by
\begin{equation}\label{eq:newe2}
    \Phi^{\prime \prime}+3 \mathcal{H} \Phi^{\prime}+\left(2 \mathcal{H}^{\prime}+\mathcal{H}^2\right) \Phi= \frac{a^2}{2 M_P^2} \left( \delta \rho_k - \delta \rho_p \right)
\end{equation}
Together with \eqref{eq:m2}, \eqref{eq:newe2} gives a system that can be expressed as
\begin{equation}\label{eq:kmat}
    \begin{cases}
     3y \frac{\mathrm{d}^2 \Phi}{\mathrm{d}y^2}  + \frac{11y+8}{2y+2} \frac{\mathrm{d}\Phi}{\mathrm{d} y} = \frac{\sqrt{6}}{2 M_P\sqrt{1+y} } \frac{\mathrm{d}\delta \phi}{\mathrm{d} y} \\[0.5cm]
     3 y \frac{\mathrm{d}^2 \delta \phi}{\mathrm{d}y^2}  +\frac{9y+6}{2y+2} \frac{\mathrm{d} \delta \phi}{\mathrm{d} y}+ \frac{4 k^2 \eta_0^2}{3 \varepsilon^{4/3}} \frac{y^{1/3}}{y+1} \delta \phi = \frac{4\sqrt{6} M_P}{ \sqrt{1+y}}  \frac{\mathrm{d}\Phi}{\mathrm{d} y}.
    \end{cases}
\end{equation}

\myparagraph{Long wavelength modes}
For long wavelength modes, $k \eta \ll 1$, and \eqref{eq:kmat} reduces to
\begin{equation}
    \begin{cases}
     3y \frac{\mathrm{d}^2 \Phi}{\mathrm{d}y^2}  + \frac{11y+8}{2y+2} \frac{\mathrm{d}\Phi}{\mathrm{d} y} = \frac{\sqrt{6}}{2 M_P\sqrt{1+y} } \frac{\mathrm{d}\delta \phi}{\mathrm{d} y} \\[0.5cm]
     3 y \frac{\mathrm{d}^2 \delta \phi}{\mathrm{d}y^2}  +\frac{9y+6}{2y+2} \frac{\mathrm{d} \delta \phi}{\mathrm{d} y} = \frac{4\sqrt{6} M_P}{ \sqrt{1+y}}  \frac{\mathrm{d}\Phi}{\mathrm{d} y}.
    \end{cases}
\end{equation}
As in the case of the kination to radiation transition, it can be turned into a system of two first order equations, and with the substitution $r(y) \equiv \frac{{\rm{d}}\Phi(y)}{{\rm{d}}y}$ it is equivalent to the ODE
\begin{equation}
    6 y (1 + y)\frac{{\rm{d}}^2 r(y)}{{\rm{d}}y^2} + (20 + 29 y) \frac{{\rm{d}}r(y)}{{\rm{d}}y} + 
  22 r(y)=0.
\end{equation}
The general solution is given by
\begin{equation}
\begin{aligned}
    & \quad \quad \quad \quad \quad \quad \quad \quad \quad \quad  r(y)= \frac{8 \left(c_1+3 c_3 (y+2)-\frac{c_4 \sqrt{y+1} (5 y+8)}{y^{4/3}}\right)}{48 y (y+1)}\\
    & +\frac{\frac{1}{5} (5 y+8) \left(\, _2F_1\left(\frac{5}{6},1;\frac{7}{3};-y\right) (3 c_3 (y-4)-5 c_1)\right)-18 c_3 (y+1) \, _2F_1\left(\frac{5}{6},2;\frac{7}{3};-y\right)}{48 y (y+1)},
    \end{aligned}
\end{equation}
where the constants have been chosen to match with the notation in Appendix \ref{ssc:c1}. The latter can be integrated to give
\begin{equation}
    \Phi(y)= -\frac{1}{24} (y+1) (5 c_1+6 c_3) \, _2F_1\left(1,\frac{11}{6};\frac{7}{3};-y\right)+\frac{c_1}{3}+c_3+\frac{c_4 \sqrt{y+1}}{y^{4/3}},
\end{equation}
matching the results obtained with the entropy formalism (see in particular  \eqref{eq:phigamma} in the Appendix) for $\gamma=1$. For an adiabatic solution that does not blow up for $y\rightarrow 0$, we require $c_1 =c_4=0$ (See Appendix \ref{ssc:c1} for details), so that
\begin{aleq}
    \frac{\Phi( y\rightarrow\infty)}{\Phi( y\rightarrow 0)} = \frac{4}{5}.
\end{aleq}
Analogously to the case of radiation domination, this can then be used to compute the form of all the other perturbations ($\mathcal{R},\delta_{m},\delta_{\phi}$...), although we will not report here the explicit expressions.

\myparagraph{Short wavelength modes}
As for the case of transitioning from kination to radiation domination, in this case where we transition to matter domination, in the short wavelength limit, no analytic solution exists for this system. However, approximate solutions can be obtained for $y \gg 1$. In that case, the equations reduce to
\begin{equation}
    \begin{cases}
     3y \frac{\mathrm{d}^2 \Phi}{\mathrm{d}y^2}  + \frac{11}{2} \frac{\mathrm{d}\Phi}{\mathrm{d} y} = \frac{\sqrt{6}}{2 M_P\sqrt{y} } \frac{\mathrm{d}\delta \phi}{\mathrm{d} y} \\[0.5cm]
     3 y \frac{\mathrm{d}^2 \delta \phi}{\mathrm{d}y^2}  +\frac{9}{2} \frac{\mathrm{d} \delta \phi}{\mathrm{d} y}+ \frac{4 k^2 \eta_0^2}{3 \varepsilon^{4/3} y^{2/3}}  \delta \phi = \frac{4\sqrt{6} M_P}{ \sqrt{1+y}}  \frac{\mathrm{d}\Phi}{\mathrm{d} y}.
    \end{cases}
\end{equation}
At leading order in $y$ (where we can neglect the source term for the second equation), they are solved by
\begin{equation}
    \Phi(y)= c_1+\frac{c_2}{y^{5/6}} +\mathcal{O}\left(\frac{1}{y}\right) 
\end{equation}
and
\begin{equation}
\delta \phi(y) = \frac{c_3}{y^{1/3}} \sin \left(\frac{4 k \eta_0 y^{1/6}}{\varepsilon^{2/3}} \right)+  \frac{c_4}{y^{1/3}} \cos \left( \frac{4 k \eta_0 y^{1/6}}{\varepsilon^{2/3}}  \right)+ \mathcal{O} \left(\frac{1}{\sqrt{y}}\right).
\end{equation}
This matches with the behaviour during matter domination, since $y \sim a^3$ implies
\begin{equation}
    \Phi(a) = c_1+ c_2 a^{-5/2}
\end{equation}
for the Bardeen potential and $\Delta_m \sim a$ for the matter perturbations.

\subsection{Perturbations within Tracker Epochs}

Here we consider perturbations within tracker epochs, where the background solution is given by the tracker solutions described in Section \ref{ssc:track}. As in section \ref{ssc:track}, our primary consideration is a radiation tracker but for completeness we also give formulae for a matter tracker.

\subsubsection{Perturbations about the Radiation Tracker}
During the tracker epoch, the potential cannot be neglected anymore. Using the values of the kinetic and potential energy densities for the radiation tracker (described in section \ref{ssc:track}), Eqs.~\eqref{eq:m1}-\eqref{eq:m2} reduce to the system
\begin{equation}\label{eq:f1}
\begin{cases}
    \delta \phi''+\frac{2}{\eta}\delta \phi'+\left(k^2+\frac{4}{\eta^2} \right)\delta \phi= \frac{8 M_P }{\lambda \eta} \left(2 \Phi'+\frac{\Phi}{\eta}\right) \\[0.5cm]
\Phi''+\frac{4}{\eta}\Phi'+\frac{\Phi}{3}\left(k^2+\frac{16}{ \lambda^2 \eta^2}\right) = \frac{4}{3 M_P\lambda \eta} \left( \delta \phi' +\frac{2 \delta \phi}{\eta}\right).\\
\end{cases}
\end{equation}
The explicit dependence on the wave number can be eliminated by the following substitution (where we have reintroduced the subscript $k$ for clarity)\footnote{The variable $x$ here should not be confused with the one defined in Eq.~\eqref{eq:xyv}.}
\begin{equation}
    x\equiv k \eta \quad \quad \delta \tilde{\phi} \equiv \frac{\delta \phi_k (k \eta)}{M_P} \quad \quad \tilde{\Phi}(x) \equiv \Phi_k (k \eta).
\end{equation}
Then, Eq. \eqref{eq:f1} can be recast as
\begin{equation}\label{eq:f1m}
\begin{cases}
    x^2 \delta\tilde{\phi}''(x)+2x\delta\tilde{\phi}'(x)+\left(x^2+4 \right) \delta\tilde{\phi}(x)= \frac{8 }{\lambda } \left(2 x\tilde{\Phi}'(x)+\tilde{\Phi}(x)\right) \\[0.5cm]
x^2 \tilde{\Phi}''(x)+4 x \, \tilde{\Phi}'(x)+\frac{\tilde{\Phi}(x)}{3}\left(x^2+\frac{16}{ \lambda^2 }\right) = \frac{4}{3 \lambda } \left( x  \delta \tilde{\phi} '(x) + 2  \delta \tilde{\phi} (x) \right),\\
\end{cases}
\end{equation}
where primes now denote derivatives with respect to $x$. Although the system of \eqref{eq:f1}-\eqref{eq:f1m} is hard to treat analytically, we can determine its behaviour in the super- and sub-horizon limits respectively.

\myparagraph{Long wavelength modes}
In the super-horizon limit, $k \eta \ll 1$ or equivalently $x\ll1$, we can neglect the $x^2 \delta \tilde{\phi}$ and $x^2 \tilde{\Phi}$ terms in Eq.~(\ref{eq:f1m}), which then reduce to homogeneous equations of degree two, and we can try a power-law ansatz of the form
\begin{equation}
     \delta \tilde{\phi} (x) = x^{\alpha} \quad \quad  \tilde{\Phi} (x) = Kx^{\beta}.
\end{equation}
Substituting in the above, we find that solutions only exist if $\alpha = \beta$ and 
\begin{equation}
    \begin{cases}
          (4 + \alpha + \alpha^2) \lambda -8 K (1 + 2 \alpha) =0 \\
          (16  + 9 \alpha  \lambda^2 + 
 3 \alpha^2  \lambda^2)K - 4  (2 +\alpha) \lambda=0,
    \end{cases}
\end{equation}
which admits the following exponents as a solution:
\begin{equation}\label{alfa}
    \alpha = \left \{0,-3,\frac{-1\pm \sqrt{\frac{64}{\lambda^2}-15}}{2} \right\} .
\end{equation}
For $\lambda = \sqrt{\frac{27}{2}}$ there are no growing modes, and the dominant mode is the constant one which represents a shift along the tracker solution.
 For $2<\lvert \lambda \rvert <\sqrt{64/15}$, the last two modes (which are non-adiabatic, see Appendix \ref{entropyappendix}) are polynomially decaying. These are the values for $\lambda$ for which the tracker is a stable node. For $\lvert \lambda \rvert>\sqrt{64/15}$ on the other hand (when the tracker is a stable spiral), the last two values for $\alpha$ are imaginary and we get an oscillating solution

\begin{aleq}
    \delta \phi(\eta)= c_a \eta^{-\frac{1}{2}} \cos\left( \frac{\sqrt{15-\frac{64}{\lambda^2}}}{2}\log \left( \eta \right) \right) + c_b \eta^{-\frac{1}{2}} \sin\left( \dfrac{\sqrt{15-\frac{64}{\lambda^2}}}{2} \log \left( \eta \right) \right),
\end{aleq}
where $c_a$ and $c_b$ are integration constants.
These oscillatory perturbations represent exactly the transient settling down of the volume modulus in the approach to the tracker, at the frequencies discussed earlier in section~\ref{trackerapproach}.

\myparagraph{Short wavelength modes}

For $k \eta \gg 1$, or equivalently $x\gg 1$ one can expand \eqref{eq:f1m} in powers of $x$ and only keep the fastest growing power in each equation, amounting to 
\begin{equation}\label{eq:trsw}
\begin{cases}
    x^2 \delta \tilde{\phi}''(x)+2x\delta\tilde{\phi}'(x)+x^2 \delta\tilde{\phi}(x)= \frac{8 }{\lambda } \left(2 x\tilde{\Phi}'(x)+\tilde{\Phi}(x)\right) \\[0.5cm]
x^2 \tilde{\Phi}''(x)+4 x \, \tilde{\Phi}'(x)+\frac{x^2}{3}\tilde{\Phi}(x) = \frac{4}{3 \lambda } \left( x  \delta \tilde{\phi} '(x) + 2  \delta \tilde{\phi} (x) \right).\\
\end{cases}
\end{equation}
However, the system Eq.~\eqref{eq:trsw} still does not admit an exact solution. To proceed, we shall first assume that, asymptotically, 
\begin{equation}
   \frac{\tilde{\Phi}(x)}{\delta \tilde{\phi}(x)} \sim \mathcal{O}\left( x^{-b}\right) \quad \quad \quad \text{with} \quad \quad b>0,
\end{equation} 
to be verified a posteriori. Then, in a large $x$ expansion the source term can be neglected in the first (but not in the second) equation in Eq.~\eqref{eq:trsw}. The homogeneous equation for $\delta \tilde{\phi} (x)$ can then be solved by
\begin{equation}\label{eq:fhom}
    \delta \tilde{\phi} (x) = \frac{c_1 \cos(x)+ c_2 \sin(x)}{x},
\end{equation}
with $c_1$ and $c_2$ arbitrary constants. Using Eq.~\eqref{eq:fhom} as a source term, the second equation in Eq.~\eqref{eq:trsw} is solved by
\begin{equation}\label{eq:philw}
\begin{aligned}
    \tilde{\Phi}(x)= 
    \frac{c_3 \sin \left(\frac{x}{\sqrt{3}}\right)+c_4 \cos \left(\frac{x}{\sqrt{3}}\right)}{x^2} +
\frac{8 \pi  c_1 \cos \left(\frac{x}{\sqrt{3}}\right)+2 c_1 \sin (x)-2 c_2 \cos (x)}{\lambda  x^2}+ \mathcal{O}\left( \frac{1}{x^3} \right)
\end{aligned}
\end{equation}
where the first term is the solution to the homogeneous equation for $\tilde{\Phi}(x) $ without any sources. Indeed, for $\lambda \rightarrow \infty$ the two equations should decouple, and the homogeneous term is the only one that survives in that limit. From Eq.~\eqref{eq:cpt}, and using the Einstein equations \eqref{eq:psi}-\eqref{eq:q} in Newtonian gauge, the curvature perturbation can be expressed as\footnote{Notice that this expression is different from \eqref{eq:Rkin}, because the background is different.}
\begin{equation}\label{eq:Rt}
    \mathcal{R}(\eta) = \frac{3}{2}\Phi+\frac{\Phi'}{2 \mathcal{H}}.
\end{equation}
To arrive at \eqref{eq:Rt}, we have used results about the background radiation tracker in Section \ref{ssc:track}, and surprisingly all dependence on $\lambda$ drops out. Then, the subhorizon curvature perturbation becomes, at leading order in $x$,
\begin{equation}
    \mathcal{R}(\eta)= \frac{c_3 \cos \left(\frac{k \eta}{\sqrt{3}}\right)-c_4 \sin \left(\frac{k \eta}{\sqrt{3}}\right)}{\sqrt{3} k \eta}+\frac{6 c_1 \cos (k \eta)+6 c_2 \sin (k \eta)-8 \sqrt{3} \pi  c_1 \sin \left(\frac{k \eta}{\sqrt{3}}\right)}{3 \lambda  k \eta}
    \end{equation}
One can also define an overall comoving density contrast Eq.~(\ref{Phi-Delta}) for the whole fluid from the total stress-energy tensor, which is equal to 
\begin{equation}
\begin{aligned}
    \Delta=&-\frac{2}{3}(k \eta)^2 \Phi (\eta)= -\frac{2}{3}  \left(c_4 \cos \left(\frac{k \eta}{\sqrt{3}}\right)+c_3 \sin \left(\frac{k \eta}{\sqrt{3}}\right)\right) \\ &+\frac{4 c_2 \cos (k \eta)-4 c_1 \left(\sin (k \eta)+4 \pi  \cos \left(\frac{k \eta}{\sqrt{3}}\right)\right)}{3 \lambda }
    \end{aligned}
\end{equation}
If we want to track the evolution of the perturbations in the individual components, we can use Eqs \eqref{eq:e1} and \eqref{eq:e2} to express
\begin{equation}
    \delta_r \equiv \frac{\delta \rho_r }{\rho_r} = \frac{3\lambda^2}{2(\lambda^2-4)}\left( \Phi'' \eta^2 +4 \Phi \frac{4-\lambda^2}{\lambda^2}-(k \eta)^2\Phi-\frac{4}{ \lambda}\delta \phi' \eta \right).
\end{equation}
Plugging in the expressions obtained above, one gets
\begin{equation}
    \delta_r =\frac{4 (c_4 \lambda +8 \pi  c_1) \cos \left(\frac{k \eta}{\sqrt{3}}\right)}{3 \lambda }-\frac{4}{3} c_3 \sin \left(\frac{k \eta}{\sqrt{3}}\right)
\end{equation}
A non-trivial check is provided by the fact that the sinusoidal terms in $k \eta$, present in both $\Phi(\eta)$ and $\delta \phi ( \eta)$, precisely cancel out in $\delta_r$: this is because $\delta_r$ must satisfy Eq. \eqref{eq:c2}, where the source terms are subleading.

Overall, the perturbations behave very similarly as during an epoch of radiation domination, with corrections parameterised by $1/\lambda$. In particular, both the scalar field and radiation perturbations behave as one would expect during a radiation dominated epoch, both oscillating at their proper frequencies. However, the gravitational potential (Bardeen variable) and the overall density contrast feel the effect of both components, and are the sum of two oscillatory contributions with the two different proper frequencies. In any case, all the perturbed quantities have the same power law scaling with $k \eta$ (or $a(\eta)$) that one would expect during radiation domination, so these are only $\mathcal{O}(1)$ differences.

\myparagraph{Behaviour of the sub-horizon scalar perturbations in the radiation tracker}
We can understand why the perturbations should redshift as radiation in a similar way to our earlier discussion in Section \ref{kination_perturbations} for a kination epoch. The solution to the equation $\phi'' + 2 \mathcal{H} \phi' +k^2 \phi = -a^2 \frac{dV(\phi)}{d\phi}$ during the tracker, (where $a(\eta) = \eta$ and $\frac{dV}{d\phi} = -\frac{4}{\eta ^4 \lambda }$) can be written as 
\begin{aleq}
    \phi(\eta) = \phi_t + \frac{4}{\lambda} \log\left( \eta \right) + \sum_k a_{\pm k} \frac{e^{\pm ik\eta}}{(k \eta)},
\end{aleq}
where $\phi_t$ is a constant of integration determined by the initial conditions. For sub-horizon scales ($k \eta \gg 1$) we have 
\begin{aleq}
    \phi'(\eta) = \frac{4}{\lambda \eta} + \sum_k (\pm ik) a_{\pm k} \frac{e^{\pm ik\eta}}{(k \eta)},
\end{aleq}
and we see that both the velocity-term and the Fourier modes have the same dependence on $\eta$. Hence all contributions to $a^{-2} \phi'(\eta)^2$ will have the same radiation-like $a^{-4}$ dependence on the scale factor in contrast to the case of kination). The contributions to the amplitude of the energy density of the scalar perturbations go like

\begin{aleq}
     \frac{1}{a^2} \left( \phi' \delta \phi'\right) \sim a^{-4}, \quad \frac{1}{a^2} \left( \Phi \phi'^2 \right)\sim  a^{-6}, \quad \frac{dV(\phi)}{d \phi} \delta \phi \sim a^{-5},
\end{aleq}
so the leading contribution indeed behaves like radiation. 

\subsubsection{Perturbations about the Matter Tracker}

We now also consider the case where the background solution is the matter tracker described in section \ref{ssc:mt}, as may arise
if e.g. primordial black holes are formed within the kination epoch. The perturbations in this case can still be effectively described using the equations \eqref{eq:e1}, \eqref{eq:e2},\eqref{eq:m1}, where we now substitute a matter contribution for that of radiation contribution. As in the previous discussion, we proceed to calculate the perturbations across various epochs, considering both sub- and super-horizon limits.

The equations for the gravitational potential and the scalar perturbations in the matter tracker take the form
\begin{equation}\label{mattertrackersysten}
\begin{cases}
   \Phi''+ \frac{6}{\eta} \Phi' + \frac{18}{\lambda^2\eta^2} \Phi = \frac{3}{\lambda \eta M_p}\left( \frac{3}{\eta}\delta \phi+ \delta \phi' \right), \\[0.5cm]
\delta \phi'' + \frac{4}{\eta} \delta \phi' + k^2 \delta \phi = \frac{12}{\lambda \eta} M_p \left( 2 \Phi' + \frac{3}{\eta} \Phi \right).\\
\end{cases}
\end{equation}
\myparagraph{Matter tracker, long wavelengths}
Similarly as for the radiation tracker the system \eqref{mattertrackersysten} has an exact power-law solution in the superhorizon limit ($k=0$), which reads
\begin{aleq}
    \delta \phi(\eta) = \eta^\alpha, \quad \Phi(\eta) = K \eta^ \alpha,
\end{aleq}
for a constant $K$, where $\alpha$ is given by
\begin{aleq}\label{eq:mattertrackercoefs}
    \alpha =  \left\{0,-\frac{5}{2},\frac{3 \left( \lambda \pm \sqrt{24-7 \lambda ^2} \right)}{4 \lambda }\right\}.
\end{aleq}
The first solution is the leading constant adiabatic mode, which represents a shift along the tracker, and there is also a second adiabatic decaying mode. The last two modes are non-adiabatic. As long as $\lvert \lambda \rvert >\sqrt{3}$, which is the condition for a tracker to exist, these modes will be decaying as well. More precisely, when $\sqrt{3}< \lvert \lambda \rvert <\sqrt{24/7}$ (when the tracker is a stable node \cite{Copeland:1997et}), the modes are power-law decaying. On the other hand, when $ \lvert \lambda \rvert >\sqrt{24/7}$ (the tracker is a stable spiral \cite{Copeland:1997et}), these modes are decaying oscillations on the approach of the tracker, given by
\begin{aleq}
    \delta \phi(\eta) = c_a \eta^{-\frac{3}{2}} \cos \left( \frac{3 \sqrt{7 \lambda^2-24}}{2 \lambda} \log  \eta \right) +  c_b \eta^{-\frac{3}{2}} \sin \left( \frac{3 \sqrt{7 \lambda^2-24}}{2 \lambda} \log  \eta \right).
\end{aleq}
where $c_a$ and $c_b$ are integration constants.

The amplitude of the energy density in each of the modes arising from Eq.~\eqref{eq:mattertrackercoefs} is given by
\begin{aleq}
    \rho_{\delta \phi_1} \sim a^{-3}, \quad \rho_{\delta \phi_2} \sim a^{-8}, \quad \rho_{\delta \phi_{3,4}} \sim a^{-9/2}
\end{aleq}
where $\rho_{\delta \phi_1}$ corresponds to the solution with $\lambda_1$ etc..

\myparagraph{Matter tracker, short wavelengths}
For the sub-horizon limit $k\eta\gg1$,
we consider the Ansatz solution
\begin{aleq}
    \Phi(\eta) = A (k\eta)^{-a}, \quad \delta \phi(\eta) = B (k\eta)^{-b} f(k\eta),
\end{aleq}
for constants $A,B,a$ and $b$, where $f$ is a trigonometric function of amplitude equal to $1$, and we assume that $a<b$.
For $k\eta\gg1$, the first equation is solved if
\begin{aleq}
    a = \frac{5}{2}\pm\frac{\sqrt{25 \lambda ^2-72}}{2 \lambda }.
\end{aleq}
The second equation is solved as well for large $k\eta$ if 
\begin{aleq}
    b = 2, \quad f(k\eta) =\cos\left(k \eta\right) \quad \text{or} \quad f(k\eta) =\sin\left(k \eta\right).
\end{aleq}
In summary 
\begin{aleq}\label{mattertracker1}
    \Phi(\eta) &= A_1 \eta ^{\frac{1}{2} \left(\frac{\sqrt{25 \lambda ^2-72}}{\lambda }-5\right)} + A_2 \eta ^{\frac{1}{2} \left(-\frac{\sqrt{25 \lambda ^2-72}}{\lambda }-5\right)}\\
    &= A_1 (k\eta)^{\frac{\sqrt{\frac{59}{3}}}{2}-\frac{5}{2}}+A_2 (k\eta)^{-\frac{5}{2}-\frac{\sqrt{\frac{59}{3}}}{2}} \approx A_1 (k\eta)^{-0.28} + A_2 (k\eta)^{-4.72},
\end{aleq}
where we substituted $\lambda = \sqrt{27/2}$ in the second line, 
and for the scalar field 
\begin{aleq}\label{mattertracker2}
    \delta \phi(\eta) =  \dfrac{B_1\cos\left( k \eta \right)+B_2\sin\left( k \eta \right)}{\left( k \eta \right)^2}.
\end{aleq}
Hence, just as in the radiation tracker, the amplitude of these perturbations goes like $1/a$. These sub-horizon solutions \eqref{mattertracker1}, \eqref{mattertracker2} are essentially solutions to the homogenized system \eqref{mattertrackersysten}, meaning that far into the matter tracker the influence of the perturbations of the scalar field, which is mimicking the matter background, on the gravitational potential can be neglected.\\
The curvature perturbation defined in Eq.~(\ref{eq:cpt}) then equals 
\begin{aleq}\label{curvpertmattertrack}
    \mathcal{R}(\eta) &=\left( -\frac{A_1 \sqrt{25 \lambda ^2-72} }{6 \lambda }+\frac{5}{6} A_1\right) (k\eta) ^{\frac{\sqrt{25 \lambda ^2-72}}{2 \lambda }-\frac{5}{2}}+\left(\frac{A_2 \sqrt{25 \lambda ^2-72}}{6 \lambda }+\frac{5}{6}
   A_2 \right)(k\eta) ^{-\frac{\sqrt{25 \lambda ^2-72}}{2 \lambda }-\frac{5}{2}}.
\end{aleq}
The equation for the matter perturbations is \cite{Baumann:2022mni}
\begin{aleq}
    \delta_m'' + \mathcal{H} \delta_m' = 3 \Phi''+3 \mathcal{H} \Phi' -k^2 \Phi,
\end{aleq}
leading to 
\begin{aleq}
    \delta_m \approx  \frac{k^2 \lambda ^2 \eta ^{\frac{\sqrt{25 \lambda ^2-72}}{2 \lambda }-\frac{1}{2}}}{18-6 \lambda ^2} A_1,
\end{aleq}
by substituting the leading mode of $\Phi$ in \eqref{mattertracker1}
for large $k\eta$. Note that the tracker solution only exists for $\lambda^2> 3\gamma=3$, such that this is well defined.
\myparagraph{Behavior of sub-horizon perturbations in the matter tracker}
From \eqref{mattertracker1}, we observe that the leading order time evolution of the comoving density contrast behaves as
\begin{aleq}
    \Delta \sim \mathcal{H}^{-2} \Phi \sim \eta^2 \Phi \sim \eta ^{\frac{1}{2} \left(\frac{\sqrt{25 \lambda ^2-72}}{\lambda }-1\right)} .
\end{aleq}
In the limit $\lambda \rightarrow \infty$, we recover matter-like behavior,
\begin{aleq}
    \Delta \sim \eta^2 \sim a,
\end{aleq}
while for $\lambda = \sqrt{27/2}$, we obtain
\begin{aleq}
\label{zsd}
    \Delta \sim \eta^{\sqrt{\frac{59}{3}}+\frac{1}{6} \left(-3-\sqrt{177}\right)} \approx \eta^{1.72} \sim a^{0.86},
\end{aleq}
indicating that the growth of perturbations is slightly suppressed compared to the matter domination case.

\myparagraph{Behaviour of the scalar perturbations in the matter tracker}
The scalar equation of motion $\phi'' + 2 \mathcal{H} \phi' +k^2 \phi = -a^2 \frac{dV}{d\phi}(\phi)$ in the matter tracker is solved by
\begin{aleq}
    \phi(\eta) = \phi_t + \frac{6}{\lambda} \log\left( \eta \right) + \sum_k a_{\pm k} \frac{e^{\pm ik\eta}}{(k \eta)^2},
\end{aleq}
where $\phi_t$ is a constant of integration, leading to 
\begin{aleq}
    \phi'(\eta) = \frac{6}{\lambda \eta} + \sum_k (\pm ik) a_{\pm k} \frac{e^{\pm ik\eta}}{(k \eta)^2}.
\end{aleq}
As expected, the leading term in $a^{-2} \phi'(\eta)^2$ scales as $a^{-3}$, the contribution from Fourier modes scales as $a^{-4}$, and there is a cross-term scaling as $a^{-7/2}$. Denoting $x = \frac{1}{2}\left(5-\frac{\sqrt{25 \lambda^2-72}}{\lambda}\right)$ (where $x \approx 0.28$ for $\lambda = \sqrt{27/2}$), the contributions to the energy density in the scalar perturbations, $\delta \rho_\phi = a^{-2}\left( \phi' \delta \phi' - \Phi \phi'^2\right) + \left(dV(\phi)/d\phi\right) \delta \phi$, show the following scaling behaviors
\begin{aleq}
     \frac{1}{a^2} \left( \phi' \delta \phi'\right) \sim a^{-\frac{7}{2}}, \quad \frac{1}{a^2} \left( \Phi \phi'^2 \right)\sim  a^{-3-\frac{x}{2}}, \quad \frac{dV(\phi)}{d \phi} \delta \phi \sim a^{-4}
\end{aleq}
and we see that the first contribution indeed scales as the cross-term in $a^{-2} \phi'(\eta)^2$. In this case however, as the gravitational potential does not decay as rapidly as the scalar perturbations for $\lambda > 3/\sqrt{2}$, the leading contribution to the amplitude of the energy density comes from the second term,
\begin{aleq}
    \delta \rho_\phi \sim a^{-3-\frac{x}{2}} \approx a^{-3.14},
\end{aleq}
with $\lambda = \sqrt{27/2}$.  Note that, as this gives the energy density perturbation in the scalar field component, it differs from the behaviour of Eq. (\ref{zsd}) which refers to the perturbation in the overall energy density (including the matter).

\subsection{Perturbations during Moduli domination}
As the field reaches the bottom of the potential, we enter an era of moduli domination (described in section \ref{modulidombackground}).
As the epoch of moduli domination is, in effect, an early matter epoch, standard results apply. We briefly include these standard results for completeness.
During this epoch, the background energy density
\begin{aleq}
    \rho \sim a^{-3}
\end{aleq}
falls as in a matter-dominated phase and the perturbations hence behave accordingly.
The equation for the gravitational potential is thus given by
\begin{aleq}
    \Phi''+ \frac{6}{\eta} \Phi' =0,
\end{aleq}
with solution
\begin{aleq}
    \Phi = c_1 + c_2 \eta^{-5},
\end{aleq}
with $\eta \propto a^{1/2}$ and $c_1$ representing the dominant, constant contribution.
The comoving density contrast grows linearly with the scale factor,
\begin{aleq}
    \Delta \sim \eta^2 \sim a.
\end{aleq}
The effects of long moduli-dominated epochs on early structure formation are studied in e.g. \cite{Das:2021wad, Eggemeier:2020zeg, Eggemeier:2021smj}.
At the end of moduli domination, the universe reheats to the Hot Big Bang and we re-enter the standard cosmology and the standard account of the growth of cosmological perturbations.

\subsection{Summary of the evolution}

To close off this section, we report here the final results for the evolution of perturbations during the various epochs that we considered, in order of appearance. Although we have also obtained analytic results for some of the transients connecting these different epochs, these will be omitted here for conciseness. We will be particularly interested in highlighting differences (and similarities) with respect to a standard cosmological history. 
All results that are gauge dependent will be reported in the Newtonian gauge, defined at the beginning of this section.\\

\noindent{\bf{Kination}}\\
During kination, the scale factor is proportional to $a(\eta) \sim \sqrt{\eta}.$ The gravitational potential evolves as
\begin{equation}\label{Bardeen2}
    \Phi (\eta) = \frac{C_1}{k \eta} J_1(k \eta)  +  \frac{C_2}{k \eta}Y_1(k \eta),
\end{equation}
from which the short and long wavelength limits can be extracted straightforwardly. Assuming kination to have taken place right at the end of inflation, the coefficients $C_1$ and $C_2$ can be fixed as
\begin{equation}
    C_1 = \frac{3i H_{\rm{inf}}}{4 M_P} \frac{1}{J_0(k \eta_0)\sqrt{\varepsilon_V k^3}}\quad \quad \text{and} \quad \quad C_2=0.
\end{equation}
Then, the density contrast of the field perturbations can be expressed as
\begin{equation}
    \Delta(\eta) = -\frac{2 i H_{\rm{inf}} }{ M_P \sqrt{\varepsilon_V k^3}} \frac{k\eta \,J_1(k \eta)}{J_0(k \eta_0) }.
\end{equation}
On large scales, it grows as $\Delta \sim a(\eta)^4$, much faster than the behaviour of the density contrasts\footnote{Denoted as $\Delta_r$ and $\Delta_m$ for radiation and matter perturbations respectively.} during radiation or matter domination epochs ($\Delta_{r,m} \sim a(\eta)^2$ and $\Delta_{r,m} \sim a(\eta)$ respectively). On small scales, 
$\Delta \sim a(\eta) \sin (k \eta)$, still faster than perturbations during a radiation dominated epoch.\\

\noindent{\bf{Radiation Tracker}}\\
 The behaviour of perturbations in the radiation tracker (where $a(\eta) \sim \eta$) is very similar to those during a radiation dominated epoch. For super-horizon scales, the gravitational potential $\Phi$ is constant, while on sub-horizon scales it evolves as
\begin{equation}\label{eq:sumrt}
    \Phi(\eta) \simeq \frac{ C_3 \sin \left( \frac{k \eta}{\sqrt{3}}\right)+ C_4 \sin \left( k \eta +C_5 \right)}{a( \eta)^2},
\end{equation}
where the coefficient $C_4$ is proportional to $1/\lambda$. In the limit $\lambda \rightarrow \infty$, \eqref{eq:sumrt} correctly reduces to the case of a radiation dominated epoch, but in general it will be the sum of two oscillatory contributions with different frequencies, corresponding to fluctuations in the radiation fluid and the scalar field. The coefficients $C_3,C_4$ and $C_5$ are in general model-dependent, as they depend on the specific seed of radiation that leads to the tracker (and the initial perturbations in the radiation fluid). The comoving density contrast behaves as
\begin{equation}
    \Delta \simeq C_3 \sin \left( \frac{k \eta}{\sqrt{3}}\right)+ C_4 \sin \left( k \eta +C_5 \right),
\end{equation}
on sub-horizon scales, while fluctuations in radiation ($\delta_r$) and the scalar field ($\delta_{\phi}$) only oscillate at their specific frequencies. On super-horizon scales, $\Delta \sim a^2(\eta)$, as during radiation domination.
 \\

\noindent{\bf{Matter Tracker}}\\
On large (super-horizon) scales, the matter tracker (with $a(\eta)\sim \eta^2$) behaves as a matter dominated epoch: the gravitational potential $\Phi$ is approximately constant and perturbations grow as $\Delta \sim a$. However, the matter tracker exhibits an anomalous growth of perturbations with respect to its simpler counterpart of a matter dominated epoch on sub-horizon scales. This is a crucial difference with respect to a radiation tracker, with potentially interesting implications. Indeed, for $k \eta \ll 1$, 
\begin{equation}
    \Phi \simeq C_6 \, a(\eta)^{\frac{1}{4} \left(\frac{\sqrt{25 \lambda ^2-72}}{\lambda }-5\right)}
\end{equation}
and
\begin{equation}
    \Delta \simeq C_7 \, a(\eta)^{\frac{1}{4} \left(\frac{\sqrt{25 \lambda ^2-72}}{\lambda }-1\right)},
\end{equation}
where $\lambda$ is the exponent of the potential. For $\lambda \rightarrow \infty$, these reduce to a constant $\Phi$ and $\Delta \sim a(\eta)$, as during matter domination. \\

\noindent{\bf{Moduli domination}}\\
During moduli domination, perturbations behave as they would during a standard matter domination epoch. Perturbations grow linearly with the scale factor in both the super- and sub-horizon limits, \emph{i.e.} $\Delta \sim a(\eta)$, whereas $\Phi$ is constant.

\section{Particle Cosmology: Old Problems and New Approaches}
\label{sectionNewApproaches}

The thrust of this paper is to emphasise the substantial differences between stringy cosmologies and the standard picture of the early post-inflationary universe.
This cosmology also motivates novel approaches to some of the important questions in particle cosmology/astroparticle physics. These are the focus of this section; various aspects are outlined here with full explorations left for future work.

To recap, in the Standard Cosmology, reheating occurs rapidly at the end of inflation, directly transitioning into a radiation-dominated Hot Big Bang. The energy density quickly takes the form of relativistic and thermalised Standard Model degrees of freedom which subsequently cool.

It is standard in stringy cosmologies that there is a long period of moduli (matter) domination prior to reheating through their decay. 
The evolution through kination and tracker epochs offers something new and compelling. Additional radiation (or matter) is an essential part of the tracker epoch, in order to catch the rolling modulus and prevent overshoot. However, the nature of this seed radiation (or matter) is relatively unspecified. 

There are various candidates for seed radiation: these include
\begin{itemize}
    \item Gravitons (or gravitational waves), for example those emitted by cosmic strings or other natural sources.
    \item Radiation modes associated to the volume axion which partners the universal volume modulus.
    \item Relativistic degrees of freedom associated to the Standard Model itself, such as the gauge bosons and chiral fermions including quarks and leptons (which are all either massless or effectively massless).
\end{itemize}
The most plausible candidate for seed matter would be (small) primordial black holes formed shortly after the end of inflation, either directly or through the growth of density contrast during the kination epoch.

During the tracker solution, although the universe is not fully radiation dominated, the radiation fraction is large (as per Eq. (\ref{TrackerFractions})). This realises a scenario where the universe starts on the tracker solution (with a significant fraction of its energy in thermalised Standard Model degrees of freedom), before passing into moduli (matter) domination and then returning into a radiation-dominated epoch with energy in thermalised Standard Model degrees of freedom.

The last of these candidates then results in a cosmology 
in which the universe first consists principally of a thermalised Standard Model Hot Big Bang plasma, then enters a period of matter domination which dilutes this, before re-commencing the thermal Hot Big Bang at a much lower temperature.

However, there is something rather unusual about the Standard Model plasma that appears here.

\subsection{The Standard Model: but not as we know it}
\label{SMsubsec}

In string theory, all couplings of the Standard Model are set by the moduli vevs; there are no true coupling `constants'. This is true of the gauge couplings; it is true of Yukawa couplings; it is true of the self-interactions present in the Higgs potential. As one simple example of this, for non-Abelian gauge groups realised by 7-branes wrapping a 4-cycle $\Sigma_i$, the tree-level gauge coupling is given by 
\begin{equation}
    \frac{2 \pi}{g_{YM}^2} = \int_{\Sigma_i} e^{-\varphi} \sqrt{g},
\end{equation}
where $\varphi$ is the string dilaton with $g_s \equiv e^{-\varphi}$ the string coupling and $g$ is the 10-dimensional string frame metric pulled back onto the worldvolume of the 7-brane.
In principle, many of these compactification moduli may have a role in setting the Standard Model couplings. However, the presence of this dependence is clearest for the volume modulus (which is the rolling field in the LVS tracker). The volume sets the foundational scale of the theory (the string scale) and all e.g. gauge couplings, as above, depend on volumes of cycles. As in a string cosmology all physics ultimately originates in string physics, the string scale (and thus the volume) will, either directly or indirectly, enter all physical couplings.\footnote{One example of indirect dependence is the way the volume determines, via the string scale $M_s$, the ultraviolet scale from which any coupling $y$ starts to run to a low-energy scale $\mu$. Using $\beta$ to denote a renormalisation group coefficient, the volume enters the low-energy couplings as$$y(\mu) = y(M_s) + \beta \ln \left( \frac{M_s}{\mu} \right) \equiv y\left( \frac{M_P}{\sqrt{\mathcal{V}}} \right) + \beta \ln \left( \frac{M_s}{\mu} \right),$$ and so low-energy couplings will always implicitly depend on $\mathcal{V}$ even if there is no explicit dependence.}
We can therefore schematically write the Standard Model Lagrangian in terms of the rolling field $\Phi$ (the dimensionless quantity is of course $\Phi/M_P$) as
\be
\mc{L}_{SM} = \sum_i \frac{1}{g_i^2(\Phi)} {\rm Tr} \left( F_{i,\mu \nu} F^{i, \mu \nu} \right) + \sum_i K_f(\Phi) \bar{\psi} \gamma^i \partial_i \psi + K_h(\Phi) \partial_{\mu} H \partial^{\mu} H^{*} + \sum_i Y_{ab}(\Phi) H \bar{\psi}^a \psi^b,
\ee
where $K_f$ and $K_h$ are the kinetic terms.

Throughout kination, the tracker solution and the approach to the minimum, the volume modulus $\Phi$ evolves through transPlanckian distances in field space. As an example, a post-inflationary evolution in the volume from $\mathcal{V} \sim 100 \, l_s^6 $ to $\mathcal{V} \sim 10^{10} \, l_s^6$ (as would be reasonable in LVS) would involve a field excursion $\Delta \Phi \sim \sqrt{\frac{2}{3}} \ln \left( 10^8 \right) M_P \sim 15 M_P$. During this evolution, although the particle content may remain that of the Standard Model,
 the actual couplings present within the theory change and the Lagrangian is certainly not that of the Standard Model.

This evolution continues until the modulus reaches its minimum.
In the final approach to the minimum, as the tracker settles down, the volume modulus approaches the minimum, overshoots it and then settles back to start oscillating about the minimum and enter moduli domination. For an LVS potential of the form (restricting to the light volume modulus and integrating out heavy fields; more detailed derivations of this as the form of the LVS potential can be found in \cite{Conlon:2008cj,Cicoli:2023opf})
\be
V(\Phi) = V_0 e^{-\sqrt{\frac{27}{2}}\Phi} \left( 1 - \alpha \Phi^{3/2} \right) + \varepsilon e^{-\sqrt{6} \Phi},
\ee
(where $\alpha \propto g_s^{-3/2}$ is the part of LVS that depends on flux quanta and results in exponentially large volumes and the $\varepsilon$ term is the more \emph{ad hoc} uplift potential included to ensure a Minkowski minimum), numerical evolution shows that the field overshoots the minimum by around $0.2 M_P$ before it settles down to oscillate about the minimum.

One consequence of this evolution is that for any aspects of its Lagrangian where the Standard Model itself is close-to-critical as a result of some sort of tuning or accidental cancellation among more fundamental parameters, during the approach of the modulus to the minimum the Standard Model will at some point pass through precise criticality and cross over to the other side of the critical point, before returning to its final value.

We enumerate various places where this behaviour can be relevant.
\begin{itemize}
\item
{\bf The size of the Higgs mass} -- If, as appears plausible,  the Higgs mass arises as a consequence of some fine-tuning, then away from the minimum it would attain its natural, larger, value. In this case, the Higgs mass would scan over many values during the cosmological evolution, and only be small at the actual minimum of the potential.
\item
{\bf The sign of the Higgs quartic} -- the $\lambda h^4$ term is close to marginal within the Standard Model and the Standard Model parameters, evolved to high scales, 
appear very close to critical \cite{Degrassi:2012ry, Buttazzo:2013uya}. 
As the volume modulus runs, it can scan through the possible values of $\lambda$: in particular, including negative values of $\lambda$. In this case, during the evolution these instabilities in the Higgs potential would be triggered, causing the Higgs potential to be temporarily unstable and the Higgs to run away to large vevs (before eventually reverting to the current vev as the final modulus vacuum is reached).\footnote{See \cite{Laverda:2024qjt} for a recent analysis of instabilities in the Higgs potential induced by a phase of kination for the inflaton.}
\item 
{\bf Colour and charge breaking (CCB) minima} -- in phenomenological applications of the MSSM, one potential danger is the need to avoid colour and charge-breaking minima: soft terms, e.g. the trilinear A-terms, that cause the preferred minimum of the theory to lie at a point that directly breaks charge and colour. In normal studies of supersymmetry, these minima are problematic as they are incompatible with the observed universe. 

However, in the case that the Standard Model (or MSSM) parameters are scanned during the cosmological evolution, such minima may represent an opportunity rather than a problem. During the evolution of the volume modulus, the Standard Model sector may pass through epochs in which its vacuum is a CCB minimum carrying large $B$ and $L$ number and in which $SU(3)$ and $U(1)_Y$ are badly broken, before returning to the `standard' minimum preserving colour and charge, as the moduli settle down. Such a generation of $B$ or $L$ may be phenomenologically useful, as we now discuss.
\end{itemize}

\subsection{Dark Matter, Baryogenesis and Dilution}
\label{baryosubsec}

If stable matter particles (of whatever nature) are produced prior to the moduli epoch, their abundance will be diluted during the moduli dominated epoch. The dilution factor can be computed as follows. 

Suppose the number density of some species $\psi$ is denoted by $n_{\psi}$. We use $t_{moduli}$ to denote the start of moduli domination and $t_{reheat}$ to denote the end. Then, 
\begin{equation}
n_{\psi}(t_{reheat}) = n_{\psi}(t_{moduli}) \left( \frac{a (t_{moduli})}{a(t_{reheat})} \right)^3,
\end{equation}
and so as $a(t) \sim t^{2/3}$ during modulus (matter) domination, 
\begin{equation}
n_{\psi}(t_{reheat}) = n_{\psi}(t_{moduli}) \left( \frac{t_{moduli}}{t_{reheat}} \right)^2.
\end{equation}
At the point where the tracker reaches the minimum of the potential and transitions to moduli domination, we have $H \sim m_{\Phi}$. Given $t_{reheat} \sim \Gamma_{\Phi}^{-1}$, we have
\begin{equation}
n_{\psi}(t_{reheat}) = n_{\psi}(t_{moduli}) \left( \frac{\Gamma_{\Phi}}{m_{\Phi}} \right)^2.
\end{equation}

If we specialise to the case where the background radiation in the tracker is that of the Standard Model, then at the transition an $\mathcal{O}(1)$ fraction of the energy density is in Standard Model radiation. The temperature of this radiation bath is
\begin{equation}
T^4 \sim 3 H^2 M_P^2, \qquad T \sim \left( M_P m_{\Phi} \right)^{1/2} \sim 10^{12} \, {\rm GeV} \left( \frac{m_{\Phi}}{10^6 \, {\rm GeV}} \right)^{\frac{1}{2}},
\end{equation}
and so the entropy density of the Standard Model is
\begin{equation}
s_{moduli} \sim T^3 \sim M_P^{3/2} m_{\Phi}^{3/2}.
\end{equation}
The moduli dominated epoch lasts until $t \sim \Gamma_{\Phi}^{-1}$, and so $H_{reheat} \sim \Gamma_{\Phi}$. After moduli complete their decays, the reheat temperature of the universe is then
\begin{equation}
    T_{reheat} \sim \Gamma_{\Phi}^{1/2} M_P^{1/2} \qquad {\rm with} \qquad s_{reheat} \sim \Gamma_{\Phi}^{3/2} M_P^{3/2},
\end{equation}
and so
\begin{equation}
s_{reheat} = s_{moduli} \left( \frac{\Gamma_{\Phi}}{m_{\Phi}} \right)^{3/2}. 
\end{equation}
It follows that
\begin{equation}
\frac{n_{\psi}\left( t_{reheat} \right)}{s\left( t_{reheat} \right)}  = 
\frac{n_{\psi}\left( t_{moduli} \right)}{s\left( t_{moduli} \right)} \left( \frac{\Gamma_{\Phi}}{m_{\Phi}} \right)^{1/2}
\end{equation}
and so the dilution factor in abundance is $y=\left( \frac{\Gamma_{\Phi}}{m_{\Phi}} \right)^{1/2}$. For conventional moduli decays, with $\Gamma \sim \frac{m_{\Phi}^3}{M_P^2}$, the resulting dilution factor is
\begin{equation}
y_{dilution} \sim \frac{m_{\Phi}}{M_P} \sim 10^{-12} \left( \frac{m_{\Phi}}{10^6 \,{\rm GeV}} \right).
\end{equation}
However, note there are also scenarios where moduli decays can be much faster than the naive rate of $\Gamma_{\Phi} \sim \frac{m_{\Phi}^3}{M_P^2}$. For example, \cite{Cicoli:2022fzy} provides a scenario where, in the presence of a finely-tuned Standard Model Higgs mass, the rate of moduli decays is determined by the untuned mass. This results in a decay rate (using the relation between $m_{3/2}$ and $\mathcal{V}$ in LVS)
\begin{equation}
\Gamma_{\Phi} \sim \frac{m_{3/2}^4 \alpha_{loop}^2}{m_{\Phi} M_P^2} \sim \left( \alpha_{loop} \mc{V} \right)^2 \frac{m_{\Phi}^3}{M_P^2} \gg \frac{m_{\Phi}^3}{M_P^2},
\end{equation}
where $\alpha_{loop} \sim \frac{1}{16\pi^2}$ is a loop factor.
In this case, the dilution factor is 
\begin{equation}
y_{dilution} \sim \left( \alpha_{loop} \mathcal{V} \right) \frac{m_{\Phi}}{M_P} \sim 10^{-12} \left( \alpha_{loop} \mathcal{V} \right) \left( \frac{m_{\Phi}}{10^6 \,{\rm GeV}} \right),
\end{equation}
which is significantly enhanced compared to the standard estimate.

There are two physics problems where large dilution factors may be particularly interesting. These are the origins of 
\begin{itemize}
    \item the baryon asymmetry of the universe,
    \item the dark matter content of the universe.
\end{itemize} 
The baryon asymmetry is parametrised by the baryon-to-entropy ratio,
$n_B/s \sim 10^{-9}$. There are, broadly, two approaches to obtaining the baryon-to-entropy ratio. In the first, the baryon asymmetry is directly generated at the small, currently measured level (example mechanisms are electroweak baryogenesis \cite{Trodden:1998ym, Morrissey:2012db} or baryogenesis via leptogenesis \cite{Davidson:2008bu}). In the second, an $\mc{O}(1)$ baryon asymmetry is initially generated, which is subsequently reaches the observable value through dilution. As moduli cosmologies naturally lead to considerable dilution, the second approach appears the most promising within stringy models.

The second approach is normally associated to Affleck-Dine baryogenesis \cite{Affleck:1984fy}, in which an Affleck-Dine field (a susy D-flat direction carrying baryon number) acquires a vev during inflation. This field experiences a torque about the origin from the trilinear A-terms, which result in an eventual non-zero baryon number. As the Affleck-Dine field directly carries baryon number, the Affleck-Dine mechanism can result in very large baryon numbers; an epoch of dilution is therefore beneficial for reducing the baryon asymmetry to be consistent with observed values. Given the presence of large dilution factors and low reheating temperature, the Affleck-Dine mechanism is probably the most appealing in a stringy context
(although baryogenesis has not been extensively studied in a stringy context). A treatment of Affleck-Dine baryogenesis in the context of LVS is \cite{Allahverdi:2016yws}, while also see \cite{Akita:2017ecc} and \cite{Kane:2019nes} for other stringy approaches to baryogenesis.

The cosmology outlined in this paper offers novel and beneficial features in terms of Affleck-Dine baryogenesis. During the long evolution to the minimum, CCB minima may be present and indeed the preferred vacuum of the Standard Model -- thus triggering a vev for the Affleck-Dine field -- even though CCB vacua may be absent in the final minimum. This mitigates some of the difficulties of Affleck-Dine model-building as it is no longer necessary for the Affleck-Dine vev to be generated during inflation. Instead, the vev may be generated during the transPlanckian post-inflationary evolution towards the final minimum, in a period when the Standard Model couplings are very different from their final values in today's minimum.

The cosmology described in this paper offer a further clear general advantage for baryogenesis. As the tracker approaches, overshoots and eventually settles down in the minimum, the changing time-dependent couplings define an explicit arrow of time when restricted to the Standard Model sector, thereby violating Lorentz symmetry, CPT and ensuring that the 3rd Sakharov condition is satisfied.

In addition to baryogenesis, a large dilution factor may also be relevant for understanding the origin of the dark matter content of the universe. The similarities in the baryon ($\Omega_b \sim 0.05$) and dark matter ($\Omega_{dm} \sim 0.25$) energy densities have long been striking. In cases where the dark matter mass is comparable to the baryonic mass, the number densities are also comparable, and in this case a universal dilution factor from the moduli epoch could sit behind both the dark matter number density and the baryon number density.

\subsection{Primordial Legacies from Gravitational Waves}

The holy grail of work in string cosmology are observational legacies that offer distinctive signatures of stringy epochs or behaviour. There are several difficulties in obtaining these. One is that of scales: in terms of Hubble horizons, post-inflationary epochs correspond to scales that are much smaller than CMB scales today. In terms of density perturbations, such galactic and sub-galactic scales are highly contaminated by non-linear and baryonic effects. The other is that of particle lifetimes: even if they once dominated the universe, moduli and other stringy particles eventually decay and are no longer present today.

For these reasons, one of the most promising approaches to such stringy legacies are through relativistic particles that survive in the present universe. Prominent examples include relic spectra of either axion-like particles (for example, through the notion of a Cosmic Axion Background arising from moduli decays \cite{Cicoli:2012aq, Higaki:2012ar, Higaki:2013qka, Conlon:2013txa, Angus:2013sua, Marsh:2014gca, Evoli:2016zhj, Ayad:2019hrj}; also see \cite{Schiavone:2021imu} which considers an early population of Primordial Black Holes as a source of relativistic axion-like particles) or gravitational waves through non-linear structures (e.g. \cite{Antusch:2017flz, Das:2021wad, Cicoli:2022sih, Bhattacharya:2023ztw}) or via axions (e.g. see the recent paper \cite{Dimastrogiovanni:2023juq}).

The cosmology described in this paper gives several mechanisms to produce and enhance primordial gravitational wave signatures.
The simplest is that of direct enhancement through a kination epoch, where gravitational waves represent one candidate for the radiation that grows relative to the kination background, and in principle could constitute the radiation portion of the tracker solution with gravitational waves an $\mc{O}(1)$ part of the energy density of the universe. Such gravitational waves could be formed, for example, from a high-tension ($G\mu_{then} \gg 10^{-7}$) cosmic string network (bounds on the string tension today which restrict $G \mu_{today} \lesssim 10^{-7}$ are not applicable as the string tension will also evolve with the evolution of the volume modulus). The difficulty is to ensure that this spectrum is not diluted to unobservable levels during the period of moduli domination preceding reheating: this may require the existence of `fast' moduli decay channels as per the ideas of \cite{Cicoli:2022fzy}.

Other mechanisms exploit the rapid growth of small scale structure that is present within kination, matter tracker and moduli domination epochs. In each of these, sub-horizon density perturbations grow with powers of the scale factor. If these go non-linear (and note the initial magnitude of these perturbations can be significantly larger than those present on CMB scales) they can form non-linear structures, and potentially primordial black holes, at very early scales. If formed at any stage prior to moduli domination, the density of such PBHs will increase relative to the background.

Indeed, PBHs are a candidate to source a matter tracker during the modulus evolution and, as matter dominates over radiation, may come to form an $\mc{O}(1)$ fraction of the universe during a long kination and tracker epoch.

A primordial black hole of mass $M_{PBH}$ has a lifetime
$$
\tau_{Hawking} \sim \frac{M_{PBH}^3}{M_P^4},
$$
and typically forms with a horizon mass at time of formation,
$$
M_{PBH} \sim \frac{M_P^2}{H_{formation}}.
$$
With a standard modulus lifetime $\tau \sim \frac{M_P^2}{m_{\Phi}^3}$, it follows that PBHs formed as above prior to the epoch of moduli domination, $H > m_{\Phi}$, will evaporate prior to moduli decay.

This does not mean that such early PBHs would be observationally irrelevant. There are two ways such early PBHs can lead to gravitational waves. One is through their evaporation and direct emission of gravitons: as early PBHs can survive deep into the moduli epoch, effects of dilution will be greatly diminished. 
The other is through mergers of PBHs and emission of gravitational waves from the merger. During the long epoch of moduli/matter domination, all matter (including PBHs) will clump, which can (as with the present universe) lead to mergers of black holes, albeit on much smaller scales.

The other clear origin of gravitational waves would be those emitted from networks of cosmic (super)strings \cite{Copeland:2003bj} (see \cite{Gouttenoire:2019kij} for an analysis also focusing on kination epochs). One effect of the kination and tracker epochs is that the tension of such cosmic strings may vary through the cosmic history: in particular, tensions $G \mu \gg 10^{-7}$ would be possible at earlier stages in this evolution. This motivates a largely unstudied scenario in which the string tension varies appreciably during the cosmic evolution, emitting gravitational waves as it does so.

In principle, the form of the gravitational wave spectrum could provide evidence for the non-standard evolution of string cosmologies due to the distinctive nature of the various epochs the system passes through; however, we defer a detailed analysis of it to further work.

\section{Conclusions and Open Directions}
\label{sectionConclusions}

The epoch between the end of inflation and nucleosynthesis is poorly constrained observationally and also one where string models strongly motivate changes to the equations of state. In this paper we have performed a detailed analysis of both the background cosmology and perturbations to this cosmology for a string cosmology across this epoch. This analysis reveals many striking features. These include the amplification of normally small effects (such as background radiation or gravitational waves) during a long kination epoch, a detailed understanding of perturbations during matter and radiation trackers, and the phenomenological opportunities offered by changes in Standard Model coupling constants during these eras.\\

\noindent Directions of future work in this area include:
\begin{itemize}
\item 
In this paper, we have studied perturbations separately in each of the different eras the string cosmology passes through. It would be good to extend this to a single unified analytical or numerical treatment which matches the perturbations across all of these epochs and carries the cosmology all the way from the end of inflation to match into BBN and $\Lambda$CDM. Such a treatment would also require an understanding of the initial spectrum of perturbations within the radiation (or matter) seed that catches up with the kinating scalar to bring the cosmology onto the tracker solutions.
    \item 
    In the approach to the tracker, we have assumed that the seed radiation/matter is separate to the rolling scalar field, and that self-perturbations in the scalar field remain subdominant. In Eq. (\ref{growthofcontrast}), we saw that scalar field self-perturbations do indeed grow relative to the background. With the analytic perturbative techniques of this paper, we are unable to study what happens if these perturbations reach the level of the background with an $\mc{O}(1)$ density contrast. However, it is a well-defined problem in numerical GR, which should be accessible with tools such as \cite{Andrade:2021rbd}, to determine the end-point of a kinating scalar with self-perturbations on an exponential potential. What happens? Do the self-perturbations end up forming black holes? Does the system end up in a tracker in which the self-perturbations of the scalar field can act as radiation? 
    \item
The development and analysis of vanilla test particle physics models for the scenarios described in sections \ref{SMsubsec} and \ref{baryosubsec}, in which the effects of explicit time dependence of MSSM couplings on the vacuum structure of the Standard Model is studied. In particular, this would enable a more quantitative study of the prospects of baryogenesis in this fashion. 

\item 
A detailed study of the gravitational wave spectrum this non-standard cosmology gives rise to, arising either from the evolution of cosmic string networks or through formation and decay/merger of primordial black holes: with the aim of determining any unusual spectral features that could give distinctive signatures of the stringy cosmological evolution, for example the correlation between gravitational wave production and cosmic microwave background fluctuations induced by a network of cosmic superstrings \cite{Avgoustidis:2011ax,Charnock:2016nzm}.

\end{itemize}

In summary, the era between inflation and BBN offers one of the most promising opportunities for string cosmology in terms of chances to connect string theory to observations.

\acknowledgments

FA is supported by the Clarendon Scholarship in partnership with the Scatcherd European Scholarship, Saven European Scholarship, and the Hertford College Peter Howard Scholarship. JC acknowledges support from the STFC consolidated grants ST/T000864/1 and ST/X000761/1. MM is supported by the St John’s College Graduate Scholarship in partnership with STFC.
The research of FR is partly supported by the Dutch Research Council
(NWO) via a Start-Up grant and a Vici grant. 
EJC is supported by STFC Consolidated Grants [ST/T000732/1 and ST/X000672/1] and by a Leverhulme Research Fellowship [RF- 2021 312].
For the purpose of Open Access, the authors have applied a CC BY public copyright licence to any Author Accepted Manuscript version arising from this submission. We thank Michele Cicoli, Anish Ghoshal, Eugene Lim, Swagat Mishra, Sirui Ning, Fernando Quevedo and Gonzalo Villa for discussions. EJC would like to thank the members of Oxford's Rudolf Peierls Centre for Theoretical Physics for their kind hospitality during the period this work was being undertaken.

\appendix

\section{Additional Details on the Background Cosmology}
\label{AppReheating}

This section contains extra details on the background cosmology.

\subsection{Oscillations about the Tracker Solution}
\label{appendixtrackerapproach}

Here we give further details about the oscillations of the background about the tracker solution, as the system settles down into the tracker (i.e. section \ref{trackerapproach}).

The result of Eq. (\ref{gtyu}) can be rewritten in a simpler form as
\begin{equation}
\begin{split}
x(N) & = x_0+e^{\frac{3}{4}(\gamma-2)N}\eta_1 \sin(\omega N+ \tilde \alpha_1), \\
y(N) & = y_0+e^{\frac{3}{4}(\gamma-2)N}\eta_2 \sin(\omega N + \tilde \alpha_2),
\end{split}
\end{equation}
where $(x_0,y_0)$ are the tracker values. Here 
\begin{equation}
\begin{split}
\eta_1 = \sqrt{c_1^2 + \frac{1}{4\omega^2}[2\beta c_2 + (\alpha-\delta)c_1]^2}, \\
\eta_2 = \sqrt{c_2^2 + \frac{1}{4\omega^2}[2\gamma c_1 - (\alpha-\delta)c_2]^2}.
\end{split}
\end{equation}
Meanwhile the angular shifts are given by
\begin{equation}
\tan \alpha_1 = \frac{2\omega c_1}{2\beta c_2 + (\alpha-\delta)c_1}, \ \ \ \ \ \ \ 
\tan \alpha_2 = \frac{2\omega c_2}{2\gamma c_1 - (\alpha-\delta)c_2}.
\end{equation}
From $x(N)$ we can also get the expression for the scalar field as
\begin{equation}
\phi(N) = \phi_0 + \sqrt 6 M_P x_0 N + \frac{\sqrt 6 M_P\eta_1}{\sqrt{\omega^2+\frac{9}{16}(\gamma-2)^2}}e^{\frac{3}{4}(\gamma-2)N}\sin(\omega N + \tilde \alpha_1 + \alpha'),
\end{equation}
where
\begin{equation}
\tan \alpha' = -\frac{4}{3}\frac{\omega}{\gamma-2}.
\end{equation}

Lastly, we can also calculate how the Hubble constant behaves during these tracker oscillations by expanding its equation around the tracker solution $H = H_0 e^{-3\gamma N/2} + \delta H$, then at linear order this gives
\begin{equation}
\delta H' = -\frac{3}{2}\gamma \delta H - 3\sqrt{\frac{3}{2}}\frac{\gamma}{\lambda}H_0 e^{-3\gamma N/2}[(2-\gamma)(x(N)-x_0) - \sqrt{(2-\gamma)\gamma} (y(N)-y_0)].
\end{equation}
Inserting our expressions for the perturbations yields a differential equation of the form
\begin{equation}
\delta H' = -\frac{3}{2}\gamma \delta H -3\sqrt{\frac{3}{2}}\frac{\gamma}{\lambda}H_0 e^{-3(\gamma+2)N/4}[(2-\gamma)\eta_1 \sin(\omega N+\tilde \alpha_1)-\sqrt{(2-\gamma)\gamma}\eta_2 \sin(\omega N+\tilde \alpha_2)].
\end{equation}
This can be solved directly to give
\begin{equation}
\delta H = H \bigg[ c_3 + \frac{12}{9(\gamma-2)^2+16 \omega^2}\sqrt{\frac{3}{2}}\frac{\gamma}{\lambda}H_0e^{-3(\gamma+2)N/4}h(N)\bigg],
\end{equation}
where $c_3$ is an integration constant and $h(N)$ is a trigonometric function given by
\begin{equation}
\begin{split}
h(N) & = 3(2-\gamma)[(2-\gamma)\eta_1 \sin(\omega N + \tilde \alpha_1)-\sqrt{(2-\gamma)\gamma}\eta_2 \sin(\omega N + \tilde \alpha_2)] \\
 & \ \ \ \ \ \ \ \ + 4\omega[(2-\gamma)\eta_1 \cos(\omega N + \tilde \alpha_1)-\sqrt{(2-\gamma)\gamma}\eta_2 \cos(\omega N + \tilde \alpha_2)].
\end{split}
\end{equation}
Through appropriate trigonometric manipulations the Hubble constant can be re-expressed in a simpler form which schematically behaves as
\begin{equation}
\delta H = H\bigg[c_3+ A' e^{\frac{3}{4}(\gamma-2)N}\sin(\omega N + \alpha')\bigg],
\end{equation}
for some coefficients $A'$ and $\alpha'$.

\subsection{More Details on Reheating}
\label{appendixModuliDecay}

Here we give, for completeness, an account of reheating that goes beyond the instantaneous decay approximation.

In an initially completely matter dominated universe where the matter particles start to decay into pairs of photons at time $t_0$, the total number of matter particles in a volume $V$ is given by $N = N_0 e^{-\alpha (t-t_0)}$, so then the energy density of the radiation component is
\begin{equation}
\rho_\gamma = \frac{N_\gamma}{V}\langle E_\gamma(t)\rangle = \frac{2N_0(1-e^{-\alpha (t-t_0)})}{V_0}\frac{a_0^3}{a^3}\langle E_\gamma(t)\rangle, 
\end{equation}
where $\langle E_\gamma(t)\rangle$ is the average energy of the photon bath at a particular time.

The energy of a photon can be related to the scale factor through $E_\gamma(\lambda) = 2\pi/\lambda = ma_c/2a$ where $a_c$ is the scale factor at which the photon was created. The average energy density is then given by
\begin{equation}
\langle E_\gamma(t)\rangle = \frac{1}{n_{\gamma,\text{tot}}} \int_{4\pi/m}^{\frac{4\pi a(t)}{ma(t_0)}} d\lambda \ \tilde n(t,\lambda)E_\gamma(\lambda),
\end{equation}
where $\tilde n_\gamma(t,\lambda)d\lambda$ is the total number of photons per unit volume in the wavelength interval $d\lambda$, while the total number density is
\begin{equation}
n_{\gamma,\text{tot}} = \int_{4\pi/m}^{\frac{4\pi a(t)}{ma(t_0)}} d\lambda \ \tilde n_\gamma(t,\lambda) = 2n_0 (1-e^{-\alpha (t-t_0)}),
\end{equation}
where $n_0 = N_0/V_0$ is the initial number density of the matter particles.

To rewrite the average energy in a more useful form, consider photons originally created with a wavelength of $\lambda = 4\pi/m$. Over time the change in their energy is due to the redshifting of the wavelength with
\begin{equation}
d\lambda = \frac{4\pi}{m}\frac{a(t)}{a_c(t_c)} - \frac{4\pi}{m}\frac{a(t)}{a_c(t_c+dt)} \approx \frac{4\pi}{m}\frac{a H_c}{a_c} dt,
\end{equation}
providing a relation between $d\lambda$ and $dt$. Since $\dot n_\gamma dt|_c$ photons are initially created in a time interval $dt$, then this can be related to the number density per unit wavelength interval as
\begin{equation}
\tilde n_\gamma d\lambda = \dot n_\gamma(a_c)\frac{m}{4\pi H_c}\frac{a_c d\lambda}{a(t)}.
\end{equation}
This formula states that the number density over a wavelength interval is given by the number of photons created per unit time, together with a $dt\rightarrow d\lambda$ conversion factor, along with stretching of the wavelength interval by a factor of $a_c/a$.

The average energy of the radiation bath can now be written through a change of integration variables $da_c/d\lambda = -ma_c^2/4\pi a$ as
\begin{equation}
\langle E_\gamma(t)\rangle = \frac{1}{n_{\gamma,\text{tot}}} \frac{m}{2a(t)} \int_{a(t_0)}^{a(t)}da_c \frac{\dot n_\gamma(a_c)}{H_c} = \frac{1}{n_{\gamma,\text{tot}}}\frac{m}{2a(t)}\int_{t_0}^{t} dt_c \ a(t_c) \dot n_\gamma(t_c).
\end{equation}
Writing the production rate as
\begin{equation}
\dot n_\gamma(t_c) = 2\alpha n_0 e^{-\alpha (t_c-t_0)}.
\end{equation}
then the Hubble equation can be written as
\begin{equation}
H^2 = \frac{H_0^2a_0^3}{a^3}e^{\alpha t_0}\bigg(e^{-\alpha t}+ \frac{\alpha}{a}\int_{t_0}^t dt_c a(t_c) e^{-\alpha t_c}\bigg),
\end{equation}
where $H_0$ is the initial Hubble constant at $t_0$. This can be differentiated to give a differential equation for $a(t)$
\begin{equation}
\ddot a + \frac{\dot a^2}{a} = \frac{H_0^2a_0^3}{2a^2}e^{-\alpha (t-t_0)}.
\end{equation}

\section{Cosmological perturbations for general equation-of-state parameters}
\subsection{Large-scale evolution of gravitational potential through transitions between cosmic fluids}\label{Largescalepotential}
In this Appendix, we derive a general expression for the large-scale evolution of gravitational potential through transitions between cosmic fluids characterized by parameters $\gamma_1$ and $\gamma_2$.
We consider the large-scale Einstein equation
\begin{equation}
    - 3 \mathcal{H} ( \Phi' + \mathcal{H} \Psi) = 4\pi G a^2 (\rho_1 \delta_1 + \rho_2 \delta_2).
\end{equation}
Define
\begin{equation}
    y \equiv \dfrac{\rho_1}{\rho_2} \sim a^{-3(\gamma_1-\gamma_2)},
\end{equation}
and assume adiabaticity
\begin{align}
    \dfrac{\delta_1}{\gamma_1} = \dfrac{\delta_2}{\gamma_2} 
\end{align}
 to obtain
\begin{equation}
    - 3 \mathcal{H} ( \Phi' + \mathcal{H} \Phi) = 4\pi G a^2 \rho_2 \delta_2 \left( 1+ \dfrac{1}{y} \dfrac{\gamma_1}{\gamma_2}\right) = \dfrac{3}{2} \mathcal{H}^2 \dfrac{y}{y+1}   \delta_2 \left( 1+ \dfrac{1}{y} \dfrac{\gamma_1}{\gamma_2} \right) ,
\end{equation}
where we substituted the Friedman equation $8\pi G \rho_2/3 = (\mathcal{H}^2 y)/(a^2(y+1))$. This turns into
\begin{equation}
    -3(3(\gamma_1-\gamma_2)y \dfrac{d\Phi}{dy}+ \Phi) = \dfrac{3}{2} \dfrac{y}{y+1} \left( 1+ \dfrac{1}{y} \dfrac{\gamma_1}{\gamma_2} \right) \delta_2
 \end{equation}
 and with $\delta_2' \approx 3 \gamma_2 \Phi'$,
 \begin{align}
     &-\left(\gamma _1 \left(9 \gamma _1-6 \gamma _2+2\right)+\left(6 \gamma _1-3 \gamma _2+2\right) \gamma _2 y^2+2 \left(6 \gamma _1^2-3 \gamma _2 \gamma _1+\gamma _1+\gamma
   _2\right) y\right) \Phi '(y)\\
   &-6 \left(\gamma _1-\gamma _2\right) y (y+1) \left(\gamma _1+\gamma _2 y\right) \Phi ''(y)+2 \left(\gamma _2-\gamma _1\right) \Phi (y)=0.
 \end{align}
The general solution is
\begin{align}
    \Phi(y) = \frac{2 \left(\gamma _1-\gamma _2\right) \left(3 \gamma _1-2 \sqrt{y+1} \, _2F_1\left(\frac{1}{2},\frac{3 \gamma _1+2}{6 \gamma _1-6 \gamma _2};\frac{3 \gamma _1+2}{6 \gamma _1-6 \gamma _2}+1;-y\right)+2\right)}{3 \gamma _1+2}
\end{align}
Then
\begin{align}
    \lim_{y \rightarrow 0} \Phi(y) = \frac{6 \gamma _1 \left(\gamma _1-\gamma _2\right)}{3 \gamma _1+2}
\end{align}
and
\begin{align}
    \sqrt{y+1} \, _2F_1\left(\frac{1}{2},b; b+1, -y\right) = b \sqrt{y+1}\int^1_0 \dfrac{t^{b-1}}{(1+ty)^{1/2}} \rightarrow b \int^1_0 t^{b-3/2} = \dfrac{b}{b-1/2}
\end{align}
as $y\rightarrow\infty$ so that
\begin{align}
    \lim_{y \rightarrow \infty} \Phi(y)  = \frac{6 \left(\gamma _1-\gamma _2\right) \gamma _2}{3 \gamma _2+2}.
\end{align}
Hence
\begin{align}\label{eq:dropPhi}
    \lim_{y \rightarrow \infty} \Phi(y)/\Phi(0)  = \frac{\gamma _2\left(3 \gamma _1+2\right) }{\gamma _1 \left(3 \gamma _2+2\right)}.
\end{align}
As a different way to check this result,
during an epoch dominated by a single fluid with EoS parameter $\gamma_i$, the Bardeen variable $\Phi$ and curvature perturbation in the super-horizon limit are related by \cite{Baumann:2022mni}
\begin{equation}
    \mathcal{R}   \simeq \frac{3 \gamma_i+2}{3 \gamma_i} \Phi \quad \quad \quad \text{for} \quad \quad  k \eta \ll 1.
\end{equation}
Assuming adiabaticity, the curvature pertubation on super-horizon scales is conserved, and one can deduce that in the transition between epochs dominated by different fluids $\Phi$ drops exactly as predicted by \eqref{eq:dropPhi}.
\subsection{Cosmological perturbations for $\gamma$-trackers}
In this section, we discuss perturbations in tracker regimes, characterized by an equation-of-state parameter $\gamma$. For super-horizon limits (long wavelength modes), the discussion is the same for all values of $\gamma$, while for the sub-horizon limits (short wavelength modes), we need to distinguish between the cases $\gamma>1$ and $\gamma<1$. The limiting case $\gamma=1$ is described in the main text.

For a $\gamma$-tracker, the scale factor goes like
\begin{aleq}
    a(\eta) = a_0 \eta ^{\frac{2}{3 \gamma -2}},
\end{aleq}
and the evolution of the scalar field is described by
\begin{aleq}
    \phi = \phi_t + \frac{6\gamma}{(3\gamma-2) \lambda} \log\left(\eta\right),
\end{aleq}
and the scalar potential $V$ is expressed as
\begin{aleq}
    V = V_0 e^{-\lambda \phi} = -\frac{18 (\gamma -2) \gamma  \eta ^{\frac{6 \gamma }{2-3 \gamma }}}{(2-3 \gamma )^2 \lambda ^2}.
\end{aleq}

The Einstein equations, within a tracker background, take the form
\begin{equation}
    \nabla^2 \Phi-3 \mathcal{H}\left(\Phi^{\prime}+\mathcal{H} \Phi\right)=\frac{a^2}{2 M_P^2} \left( \delta \rho_\phi+\delta \rho_\gamma \right)
\end{equation}
\begin{equation}
    \Phi^{\prime \prime}+3 \mathcal{H} \Phi^{\prime}+\left(2 \mathcal{H}^{\prime}+\mathcal{H}^2\right) \Phi= \frac{a^2}{2 M_P^2} \left( \delta P_\phi+ (\gamma-1)\delta \rho_\gamma\right),
\end{equation}
where $\delta \rho_\phi$ and $\delta P_\phi$ are given by \eqref{eq:rhophi}, \eqref{eq:Pphi}:
\begin{aleq}
    \delta \rho_\phi &= \frac{1}{a^2}\left(\phi' \delta \phi'-\Phi \phi'^2\right) +\frac{dV(\phi)}{d \phi} \delta \phi\\
    &= \frac{6 \gamma  \eta ^{\frac{6 \gamma }{2-3 \gamma }} \left((3 \gamma -2) \eta  \lambda  \delta \phi '(\eta )+3 (\gamma -2) \lambda  \delta \phi (\eta )-6 \gamma  \Phi
   (\eta )\right)}{(2-3 \gamma )^2 \lambda ^2},
\end{aleq}
\begin{aleq}
    \delta P_\phi &= \frac{1}{a^2}\left(\phi' \delta \phi'-\Phi \phi'^2\right) -\frac{dV(\phi)}{d \phi} \delta \phi\\
    &= -\frac{6 \gamma  \eta ^{\frac{6 \gamma }{2-3 \gamma }} \left((2-3 \gamma ) \eta  \lambda  \delta \phi '(\eta )+3 (\gamma -2) \lambda  \delta \phi (\eta )+6 \gamma  \Phi
   (\eta )\right)}{(2-3 \gamma )^2 \lambda ^2}.
\end{aleq}
By combining the Einstein equations, we can eliminate the dependence on $\delta \rho_\gamma$, and we also consider the Klein-Gordon equation \eqref{eq:c1}, to obtain a system of second order equations for $\Phi(\eta)$ and $\delta \phi (\eta)$,
\begin{equation}\label{gentrackersysten}
\begin{cases}
\Phi''(\eta)+\frac{6 \gamma  \Phi '(\eta )}{\left(3 \gamma -2\right) \eta }+\Phi (\eta ) \left((\gamma -1) k^2-\frac{18 (\gamma -2) \gamma ^2}{(3 \gamma -2)^2 \eta ^2 \lambda ^2}\right) = -\frac{3 (\gamma -2) \gamma  \left((3 \gamma -2) \eta  \delta \phi '(\eta )+3 \gamma  \delta \phi (\eta )\right)}{(3 \gamma -2)^2 \eta ^2 \lambda }.\\[0.5cm]
  \delta \phi ''(\eta ) +\frac{4 \delta \phi '(\eta )}{ \left(3 \gamma  -2 \right) \eta }+\delta \phi (\eta ) \left(k^2-\frac{18 (\gamma -2) \gamma }{(3 \gamma -2)^2 \eta ^2}\right)=\frac{12 \gamma  \left(2 (3 \gamma -2) \eta  \Phi '(\eta )-3 (\gamma -2) \Phi (\eta )\right)}{(3 \gamma-2 )^2 \eta ^2 \lambda }.\\
\end{cases}
\end{equation}
\subsubsection{Long wavelength modes}
In the limit $k \rightarrow 0$, the system \eqref{gentrackersysten} can be solved analytically. When $\lambda^2>24 \gamma^2/(9\gamma-2)$, which is the condition for the tracker solution to be a stable spiral \cite{Copeland:1997et}, the solution is given by
\begin{aleq}
    \Phi(\eta) = c_1+c_2 \eta ^{\frac{3 \gamma +2}{2-3 \gamma }} +\gamma \eta ^{\frac{3 (\gamma -2)}{2 (3 \gamma -2)}}\left( c_3  \cos\left(\omega \log \eta \right) + c_4 \sin\left(\omega \log \eta \right)\right),
\end{aleq}
\begin{aleq}
    \delta \phi(\eta) = \frac{2c_1}{\lambda}-3\gamma c_2 \eta ^{\frac{3 \gamma +2}{2-3 \gamma }}  + \frac{4(3\gamma-2)\lambda \omega}{9(\gamma-2)}\eta ^{\frac{3 (\gamma -2)}{2 (3 \gamma -2)}} \left(   c_3 \cos\left(\omega \log \eta \right) + c_4  \sin\left(\omega \log \eta \right) \right),
\end{aleq}
where we defined 
\begin{aleq}
    \omega = \frac{3 \sqrt{-(3 \gamma -2)^2 (\gamma -2) \left((9 \gamma -2) \lambda ^2-24 \gamma ^2\right)}}{2 (3 \gamma -2)^2 \lambda }.
\end{aleq}

When $\lambda^2<24 \gamma^2/(9\gamma-2)$, on the other hand, all solutions are power-law
\begin{aleq}\label{tracker21}
    \Phi(\eta) = c_1+c_2 \eta ^{\frac{3 \gamma +2}{2-3 \gamma }} +\gamma \eta ^{\frac{3 (\gamma -2)}{2 (3 \gamma -2)}}\left( c_3  \eta^\omega + c_4 \eta^{-\omega}\right),
\end{aleq}
\begin{aleq}\label{tracker22}
    \delta \phi(\eta) = \frac{2c_1}{\lambda}-3\gamma c_2 \eta ^{\frac{3 \gamma +2}{2-3 \gamma }}  + \frac{4(3\gamma-2)\lambda \omega}{9(\gamma-2)}\eta ^{\frac{3 (\gamma -2)}{2 (3 \gamma -2)}} \left(   c_3 \eta^\omega + c_4 \eta^{-\omega} \right),
\end{aleq}
where
\begin{aleq}
    \omega = \frac{3 \sqrt{(3 \gamma -2)^2 (\gamma -2) \left((9 \gamma -2) \lambda ^2-24 \gamma ^2\right)}}{2 (3 \gamma -2)^2 \lambda }.
\end{aleq}
We note that a special case of \eqref{tracker21}, \eqref{tracker22}  for $\gamma=0$ with $\Phi(\eta) = c_1+c_2 \eta$ and $\delta \phi(\eta) = (2\lambda/3) c_3\eta^3 + (2c_1/\lambda+2\lambda c_4/3) $.

The solutions corresponding to the coefficients $c_3$ and $c_4$ are non-adiabatic, and for $\gamma >2/3$, these are decaying.

\subsubsection{Small wavelength modes for trackers with $\gamma >1$}
We consider an Ansatz solution to Eqs. \eqref{gentrackersysten} of the form
\begin{aleq}
    \delta \phi(\eta) =(k\eta)^{\frac{-2}{3\gamma-2}} f(k \eta), \quad \Phi(\eta) = (k\eta)^{\frac{-3 \gamma}{3\gamma-2}} g(k \eta),    
\end{aleq}
where $f(k \eta)$ and $g(k \eta)$ are assumed to be trigonometric functions. Plugging this Ansatz into the equation for the scalar perturbations, and keeping the leading terms as $k \eta \gg 1$, we obtain
\begin{aleq}
    f''(k \eta) + f(k \eta) =0,
\end{aleq}
so that we can take
\begin{aleq}\label{gentracker_scalar}
    \delta \phi(\eta) = A_1 (k\eta)^{\frac{-2}{3\gamma-2}} \cos\left(k \eta \right).
\end{aleq}
The leading terms to both $\delta \phi$ and $\delta \phi'$ go as $1/a$, and hence the intrinsic kinetic energy in the scalar field perturbations $a^{-2}\delta \phi'(\eta)^2 \sim a^{-4}$ will behave like radiation in all trackers.\\
Similarly, keeping the leading terms in the equation for the gravitational potential assuming that $\gamma \neq 1$, we obtain
\begin{aleq}
    (2-3 \gamma )^2 \eta  k \lambda  \left(g''(k\eta)+(\gamma -1) g(k \eta)\right)=3 A_1 (\gamma -2) \gamma  \sin (k \eta)
\end{aleq}
which is solved by
\begin{aleq}
    g(k \eta) = A_1 \frac{3 \gamma \sin (k \eta)}{(3 \gamma-2)\lambda} + A_2 \cos\left( k \sqrt{\gamma-1}\eta\right),
\end{aleq}
as long as $\gamma >1$. We thus conclude that
\begin{aleq}
    \Phi(\eta) = (k\eta)^{\frac{-3 \gamma}{3\gamma-2}} \left[ A_1 \frac{3 \gamma \sin (k \eta)}{(3 \gamma-2)\lambda} + A_2 \cos\left( k \sqrt{\gamma-1}\eta\right)\right],
\end{aleq}
consists of a superposition of both modes characteristic to the fluid, and those following the scalar field.
\paragraph{Behaviour of the energy density of the scalar perturbations}
To understand how the scalar perturbations redshift, we consider the solution to the scalar equation $\phi'' + 2 \mathcal{H} \phi' +k^2 \phi = -a^2 V'(\phi)$,
\begin{aleq}
    \phi(\eta) = \phi_t + \frac{6 \gamma}{(3\gamma-2)\lambda} \log\left( \eta \right) + \sum_k a_{\pm k} \frac{e^{\pm ik\eta}}{(k \eta)^{\frac{2}{3\gamma-2}}},
\end{aleq}
and for sub-horizon scales we have 
\begin{aleq}
    \phi'(\eta) = \frac{6 \gamma}{(3\gamma-2)\lambda}\frac{1}{\eta}+ \sum_k (\pm ik) a_{\pm k} \frac{e^{\pm ik\eta}}{(k \eta)^{\frac{2}{3\gamma-2}}},
\end{aleq}
and we see that both the velocity-term and the Fourier modes have the same dependence on $\eta$. Hence the leading contribution to $a^{-2} \phi'(\eta)^2$ will scale like $a^{-3 \gamma}$, while the Fourier modes go as $a^{-4}$, and the cross-term mixes these behaviors: $a^{-\frac{4+3 \gamma}{2}}$.The contributions to the energy density of the scalar perturbations $\delta \rho_\phi = a^{-2}\left( \phi' \delta \phi' - \Phi \phi'^2\right) + \left(dV(\phi)/d\phi\right) \delta \phi$ go like
\begin{aleq}
     \frac{1}{a^2} \left( \phi' \delta \phi'\right) \sim a^{-\frac{4+3 \gamma}{2}}, \quad \frac{1}{a^2} \left( \Phi \phi'^2 \right)\sim  a^{-\frac{9 \gamma}{2}}, \quad \frac{dV(\phi)}{d \phi} \delta \phi \sim a^{-1-3\gamma},
\end{aleq}
and we see that the first contribution matches with the behavior of the cross-term, and this contribution is the leading one (as $\gamma >1$).
\subsubsection{Small wavelength modes for trackers with $0<\gamma < 1$}
Here, we consider the Ansatz
\begin{aleq}
    \Phi(\eta) = f(k\eta) e^{\sqrt{1-\gamma}k \eta}, \quad \delta \phi(\eta) =  g(k\eta) e^{\sqrt{1-\gamma}k \eta},
\end{aleq}
where we assume that
\begin{aleq}
    f(k \eta) \gg f'(k \eta),  f''(k \eta) , g(k\eta) , g'(k \eta) , g''(k \eta) \,
\end{aleq}
in the limit for large $k \eta$. In this limit the equation for the gravitational potential decouples from the scalar perturbations, leading to
\begin{aleq}
    f(k\eta) = \left(\sqrt{1-\gamma } (3 \gamma -2) k \eta\right)^{\frac{3 \gamma }{2-3 \gamma }}c_1.
\end{aleq}
Substituting this in the equation for the scalar perturbations, we obtain
\begin{aleq}
    g(k\eta) &= -\frac{24 \gamma  \left(\left(\sqrt{1-\gamma } (3 \gamma -2) k \eta\right)^{\frac{2}{2-3 \gamma }}-3 \gamma  \left(\sqrt{1-\gamma } (3 \gamma -2) k \eta\right)^{\frac{3 \gamma }{2-3 \gamma }}\right)}{(2-3 \gamma )^2 (\gamma -2) \lambda  k \eta^2}c_1\\
    &+e^{-\sqrt{1-\gamma } (k \eta)} (c_2 \cos (k \eta)+c_3 \sin (k \eta)).
\end{aleq}
\subsubsection{Small wavelength modes for trackers with $\gamma = 0$}
For this value of $\gamma$, the equations \eqref{gentrackersysten} decouple, and we can solve without the need to go to an asymptotic limit,
\begin{aleq}
    \Phi(\eta) =c_1 e^{k\eta}+c_2 e^{-k \eta},
\end{aleq}
\begin{aleq}
    \delta \phi(\eta) = (c_3 k \eta+c_4) \cos (k \eta)+(c_4 k \eta-c_3) \sin (k \eta).
\end{aleq}

\subsubsection{Perturbations of the volume axion}\label{pert-vol-axion}
The equation for the axion perturbations is
\begin{aleq}
    \chi_k''(\eta) + \left[ \frac{2a'}{a}+\frac{f'}{f}\right] \chi_k'(\eta) + k^2 \chi_k(\eta) = 0,
\end{aleq}
where in a tracker background
\begin{aleq}
    a(\eta) = \eta^{\frac{2}{3\gamma-2}}, \quad f(\eta) = \exp\left(\sqrt{\frac{8}{3}\frac{\Phi}{M_p}}\right) \sim \eta ^{-\frac{4 \sqrt{6} \gamma }{(3 \gamma -2) \lambda }},
\end{aleq}
so that this equation takes the form 
\begin{aleq}
    k^2 \chi_k(\eta )+\chi_k''(\eta )+\frac{4 \left(\lambda -\sqrt{6} \gamma \right) \chi_k'(\eta )}{(3 \gamma -2) \eta  \lambda }=0.
\end{aleq}
The solutions are
\begin{aleq}
    \chi_k(\eta) = \eta^{-\frac{3 \gamma  \lambda +4 \sqrt{6} \gamma -6 \lambda }{4 \lambda -6 \gamma  \lambda }} \left(c_1 J_{\alpha}(k \eta
   )+c_2 Y_{\alpha}(k \eta )\right),
\end{aleq}
with
\begin{aleq}
    \alpha = -\frac{\sqrt{96 \gamma ^2+9 (\gamma -2)^2 \lambda ^2+24 \sqrt{6} (\gamma -2) \gamma  \lambda }}{2 (3 \gamma -2) \lambda }.
\end{aleq}
The kinetic energy scales then scales like radiation
\begin{aleq}
    \rho = \dfrac{f(\eta) \chi_k'(\eta)^2}{a^(\eta)} \sim a^{-4}.
\end{aleq}
\section{Entropy formalism}
\label{entropyappendix}
We can alternatively express the second-order equations \eqref{eq:m1} and \eqref{eq:m2} as a set of four first-order equations, employing the formalism introduced in \cite{Bartolo:2003ad}. In this approach we keep the gravitational potential $\Phi$ as one of the variables. Additionally, we introduce the gauge-invariant curvature perturbation $\mathcal{R}$, along with the gauge-invariant entropy perturbations $\mathcal{S}$ and $\Gamma$. For convenience we drop the subscript $k$ in what follows for $\Phi,\mathcal{R},\mathcal{S}$ and $\Gamma$, but it is important to remember that the quantities are all being evaluated for a particular $k$ in Fourier space. $S$, denotes the relative entropy perturbation and is  defined as
\begin{aleq}\label{entropy1}
    \mathcal{S} =  \dfrac{3 \mathcal{H}\gamma_r\gamma_\phi \Omega_r}{\gamma} \left( \dfrac{\delta \rho_\phi}{\rho_\phi'}-\dfrac{\delta \rho_r}{\rho_r'}\right)= \Omega_r \frac{\gamma_\phi \delta_r- \gamma_r \delta_\phi}{\gamma},\\
\end{aleq}
where $\gamma = \Omega_\phi \gamma_\phi + \Omega_r \gamma_r$, where the equation of state $\gamma_r = 1+w_r$ etc..., and the intrinsic entropy perturbation of the scalar field is
\begin{aleq}
   \Gamma =  \dfrac{3 \mathcal{H} \gamma_\phi c_\phi^2}{1-c_\phi^2}\left( \dfrac{\delta \rho_\phi}{\rho_\phi'}-\dfrac{\delta P_\phi}{P_\phi'}\right),
\end{aleq}
where the adiabatic sound speed is $c_\phi^2 = P_\phi'/\rho_\phi'$, in terms of the background scalar field pressure and energy density. The resulting first-order system in terms of these variables is then expressed as \cite{Bartolo:2003ad}
\begin{aleq}\label{eq:system}
    \dfrac{d \mathbf{v}}{dN} = \left( M_0 + k^2 \mathcal{H}^{-2}M_1 + k^4 \mathcal{H}^{-4} M_2\right) \mathbf{v}, \quad \mathbf{v} = \left( \Phi, \mathcal{R}, \mathcal{S}, \Gamma\right)^T,
\end{aleq}
with $N = \log(a)$, and the matrices are defined as follows

\begin{aleq}
    M_0 = \left(
\begin{array}{cccc}
 -\frac{3 \gamma }{2}-1 & \frac{3 \gamma }{2} & 0 & 0 \\
 0 & 0 & \frac{\Omega _\phi \left(\omega _r-c_\phi^2\right)}{\gamma } & \frac{\left(1-c_\phi^2\right) \Omega _r}{\gamma } \\
 0 & 0 & \frac{3 \gamma _r \Omega _r \left(\omega _r-c_\phi^2\right)}{\gamma }+3 \left(\omega _\phi-\omega _r\right) & \frac{3 \left(1-c_\phi^2\right) \gamma _r \Omega _r}{\gamma } \\
 0 & 0 & -\frac{3 \gamma }{2} & 3 \left(\omega _\phi-\frac{\gamma }{2}\right) \\
\end{array}
\right),
\end{aleq}
\begin{aleq}
    M_1 = \left(
\begin{array}{cccc}
 0 & 0 & 0 & 0 \\
 -\frac{2 c^2}{3 \gamma } & 0 & 0 & 0 \\
 0 & 0 & \frac{1}{3} & \frac{1}{3} \\
 0 & -\gamma _\phi & -\frac{1}{3} & -\frac{1}{3} \\
\end{array}
\right), \quad M_2 = \left(
\begin{array}{cccc}
 0 & 0 & 0 & 0 \\
 0 & 0 & 0 & 0 \\
 \frac{2 \gamma _\phi}{9 \gamma } & 0 & 0 & 0 \\
 -\frac{2 \gamma _\phi}{9 \gamma } & 0 & 0 & 0 \\     
 \end{array}
\right),
\end{aleq}

with
\begin{aleq}\label{speedofsound}
    c^2 = \dfrac{\left(\rho_\phi + P_\phi \right) c_\phi^2 + \left( \rho_r + P_r \right) c_r^2}{\left(\rho_\phi + P_\phi \right) + \left( \rho_r + P_r \right)}.
\end{aleq}
This formalism is mainly useful for efficiently solving superhorizon limits ($k=0$) and identifying which are the adiabatic modes. We illustrate this in a couple of examples.
\subsection{Example 1: Transition from kination to fluid domination}\label{ssc:c1}
We consider the transition from kination to some fluid with parameter $\gamma$ and introduce the variable
\begin{aleq}
    y = \dfrac{\rho_\gamma}{\rho_\phi} =\dfrac{a^{3(2-\gamma)}}{a^{3(2-\gamma)}_{eq}} .
\end{aleq}
and consider the system
\begin{aleq}
    3(2-\gamma)y\dfrac{d \mathbf{v}}{dy} =  \dfrac{d \mathbf{v}}{dN} = 
    \left( M_0 + k^2 \mathcal{H}^{-2}M_1 + k^4 \mathcal{H}^{-4} M_2\right) \mathbf{v} \equiv \mathcal{M}\mathbf{v}, \quad \mathbf{v} = \left( \Phi, \mathcal{R}, \mathcal{S}, \Gamma\right)^T.
\end{aleq}
The matrix $M_0$ governs the large-scale evolution ($k=0$), and because

\begin{aleq}
    \Omega_\phi(y) = \frac{1}{1+y}, \quad \Omega_r(y) = \frac{y}{1+y},
\end{aleq}
this is given by
\begin{aleq}
   M_0 = \left(
\begin{array}{cccc}
 -\frac{3 (\gamma  y+2)}{2 (y+1)}-1 & \frac{3 \gamma  y+6}{2 y+2} & 0 & 0 \\
 0 & 0 & \frac{\gamma -2}{\gamma  y+2} & 0 \\
 0 & 0 & -\frac{6 (\gamma -2)}{\gamma  y+2} & 0 \\
 0 & 0 & -\frac{3 (\gamma  y+2)}{2 (y+1)} & -\frac{3 (\gamma -2) y}{2 (y+1)} \\
\end{array}
\right).
\end{aleq}
This system has an exact solution equal to
\begin{aleq}
    & S(y) = \frac{c_1 y}{\gamma  y+2}, \quad \Gamma(y) = \sqrt{2} c_2 \sqrt{y+1}-\frac{c_1}{\gamma -2}, \quad \mathcal{R}(y) =\frac{c_1}{6 \gamma +3 \gamma ^2 y}+c_3,
\end{aleq}
\begin{aleq}\label{eq:phigamma}
     \Phi(y) &= \frac{\sqrt{y+1}}{8\gamma} \Big(3 \gamma ^2 c_3 \, _2F_1\left(\frac{1}{2},\frac{4}{6-3 \gamma };1+\frac{4}{6-3 \gamma };-y\right)\\
    & +(c_1-3 (\gamma -2) \gamma  c_3) \, _2F_1\left(\frac{3}{2},\frac{4}{6-3 \gamma };1+\frac{4}{6-3 \gamma };-y\right)+8 \gamma  c_4 y^{\frac{4}{3 (\gamma -2)}}\Big).
\end{aleq}
For adiabatic solutions that remain well-defines as $y\rightarrow 0$, we require that $c_1=c_2=c_4=0.$ For a transition to radiation domination ($\gamma=4/3$), we then obtain
\begin{aleq}
    \Phi(y) = \frac{c_3 \left(2 y^2+y-2 \sqrt{y+1}+2\right)}{3 y^2},
\end{aleq}
while for a transition to matter domination ($\gamma=1$), we have
\begin{aleq}
    \Phi(y) = c_3-\frac{1}{4} c_3 (y+1) \, _2F_1\left(1,\frac{11}{6};\frac{7}{3};-y\right).
\end{aleq}
\subsection{Example: $\gamma$-tracker}
The behavior of long-wavelength perturbations within the tracker regime has been discussed in \cite{Malquarti:2002iu} and \cite{Bartolo:2003ad}. These modes can be characterized by the following matrix
\begin{aleq}
M_0 = 
\left(
\begin{array}{cccc}
 -\frac{3 \gamma }{2}-1 & \frac{3 \gamma }{2} & 0 & 0 \\
 0 & 0 & 0 & \frac{3 (2-\gamma )}{\lambda ^2} \\
 0 & 0 & 0 & 3 (2-\gamma ) \left(1-\frac{3 \gamma }{\lambda ^2}\right) \\
 0 & 0 & -\frac{3 \gamma }{2} & 3 \left(\frac{\gamma }{2}-1\right) \\
\end{array}.
\right)
\end{aleq}
This matrix has constant coefficients, allowing us to express the solution as
\begin{aleq}
    \mathbf{v} = \sum_i c_i \mathbf{v}_i a^{\lambda_i},
\end{aleq}
where $\mathbf{v}_i$ and $\lambda_i$ are the eigenvectors and eigenvalues of $M_0$.
The eigenvalues are given by
\begin{aleq}
    & \lambda_1, \lambda_2, \lambda_{3,4} = \left\{0,-\frac{3 \gamma }{2}-1,\frac{3 \left((\gamma -2) \pm \sqrt{(\gamma -2) \left(-24 \gamma ^2+9 \gamma  \lambda ^2-2 \lambda ^2\right)} \lambda \right)}{4 \lambda }\right\}.\\
\end{aleq}
The first two modes are adiabatic ($S(a)=\Gamma(a)=0$), with eigenvectors equal to
\begin{aleq}
    \mathbf{v}_1 = \left(\frac{3 \gamma }{3 \gamma +2},1,0,0 \right), \quad \mathbf{v}_2 = (1,0,0,0).
\end{aleq}
The other two modes are non-adiabatic and the expressions for their eigenvectors are long and not very illuminating. When $\lambda^2>24 \gamma^2/(9\gamma-2)$, the real part of these eigenvalues are negative, and so show an oscillatory behaviour. 
\subsubsection{Radiation tracker}
For the radiation tracker, with $\gamma = 4/3$, this looks
\begin{aleq}
    M_0 = \left(
\begin{array}{cccc}
 -3 & 2 & 0 & 0 \\
 0 & 0 & 0 & \frac{2}{\lambda ^2} \\
 0 & 0 & 0 & 2 \left(1-\frac{4}{\lambda ^2}\right) \\
 0 & 0 & -2 & -1 \\
\end{array}
\right),
\end{aleq}
and we can express the solution as
\begin{aleq}
    \mathbf{v} = \sum_i c_i \mathbf{v}_i a^{\lambda_i},  \quad \mathbf{v} = \left( \Phi, \mathcal{R}, \mathcal{S}, \Gamma\right)^T
\end{aleq}
where $\mathbf{v}_i$ are the eigenvectors of $M_0$, given by
\begin{aleq}\label{evandeigval}
    \mathbf{v}_1,  \mathbf{v}_2,  \mathbf{v}_{3,4} = \left\{ (2,3,0,0), (1,0,0,0), \left(-\frac{4}{5 \lambda ^2 \pm i \lambda  \nu -16},-\frac{4}{\lambda ^2\pm i \lambda  \nu },\frac{-\lambda  \pm i \nu }{4 \lambda },1\right)\right\} ,
\end{aleq}
and $\mathbf{v}_{3,4}$ refers to the plus and minus solutions in Eq.~(\ref{evandeigval}) respectively, with $\nu = \sqrt{64-15 \lambda^2}$ and the values for the eigenvalues $\lambda_i$ are given by 
\begin{aleq}
    \lambda_1 , \lambda_2, \lambda_{3, 4} = \left \{0,-3,\frac{-1\pm \sqrt{\frac{64}{\lambda^2}-15}}{2} \right\} .
\end{aleq}
The first two modes are a leading $\lambda_1 =0$ constant and a subleading $\lambda_2 = -3$ decaying mode, both of which are adiabatic ($\mathcal{S}=0$, $\Gamma=0$). The last two modes are non-adiabatic and decaying (given that $\lvert \lambda \rvert > 2 = \sqrt{3\gamma}$), and these eigenvalues have an imaginary part when $\lambda > \sqrt{64/15} \approx 2.07$.
\subsubsection{Matter tracker}
The discussion of super-horizon modes is qualitatively the same as for the radiation tracker. The large-scale matrix in the matter tracker with $\gamma = 1$ is given by,
\begin{aleq}
    M_0 = \left(
\begin{array}{cccc}
 -\frac{5}{2} & \frac{3}{2} & 0 & 0 \\
 0 & 0 & 0 & \frac{3}{\lambda ^2} \\
 0 & 0 & 0 & 3 \left(1-\frac{3}{\lambda ^2}\right) \\
 0 & 0 & -\frac{3}{2} & -\frac{3}{2} \\
\end{array}
\right).
\end{aleq}
We express the general solution as
\begin{aleq}
    \mathbf{v} = \sum_i c_i \mathbf{v}_i a^{\lambda_i},
\end{aleq}
where $\mathbf{v}_i$ and $\lambda_i$ are the eigenvectors and eigenvalues of $M_0$.
We find that
\begin{aleq}\label{matter_modes}
    & \lambda_1, \lambda_2, \lambda_{3,4} = \left\{0,-\frac{5}{2},\frac{3 \left( \lambda \pm \sqrt{24-7 \lambda ^2} \right)}{4 \lambda }\right\},\\
\end{aleq}
and the respective eigenvectors are given by
\begin{aleq}
    \mathbf{v}_1 = \left( \frac{3}{5},1,0,0 \right), \quad \mathbf{v}_2 = (1,0,0,0), \quad \mathbf{v}_{3,4} = \left(\frac{6}{-7 \lambda ^2\pm i \lambda  \nu +18},\frac{4}{\lambda  (-\lambda \pm i \nu )},-\frac{\lambda \pm i \nu }{2 \lambda },1\right)
\end{aleq}
The last two eigenvalues represent non-adiabatic modes. These modes are subdominant compared to the first mode as they exhibit decay over time as long as $\lvert \lambda \rvert > \sqrt{3} = \sqrt{3\gamma}$, which is a necessary condition for the tracker to exist.


\bibliographystyle{JHEP}
\bibliography{biblio.bib}

\end{document}